\documentclass[11pt,a4paper]{article}

\usepackage[utf8]{inputenc}
\usepackage{amsmath,latexsym}
\usepackage{xcolor}
\usepackage{graphicx}
\usepackage{axodraw2}
\usepackage{slashed}
\usepackage{caption}
\usepackage{subcaption}

\usepackage{tikz}
\usepackage[nomessages]{fp}
\usepackage{jheppub}

\usepackage{makecell}

\def\beq{\begin{equation}}   
\def\eeq{\end{equation}}
\def\bea{\begin{eqnarray}}  
\def\eea{\end{eqnarray}} 
\def\nn{\nonumber}

\def\f21{{}_2F_{1}}
\def\eps{\epsilon}

\def\L{\mathcal{L}}
\def\O{\mathcal{O}}

\newcommand{\equal}{\:\: = \:\:}

\def\as(#1){{\alpha_{\rm s}^{\,#1}}}
\def\ar(#1){{a_{\rm s}^{\,#1}}}

\def\Lt(#1){{L_t^{\:\!#1}}}
\def\lntwo(#1){{\ln^{\:\!#1\!}2}}

\def\Gam(#1,#2){{\gamma_{#1}^{\,#2}}}

\def\B(#1,#2){{\beta_{#1}^{\,#2}}}

\def\L{\mathcal{L}}
\def\eps{\epsilon}
\def\dots{..}

\newcommand{\ed}{\end{document}}

\newcommand{\ice}[1]{\relax}

\def\gluontwoptoneloop{
\raisebox{-20pt}{
  \begin{axopicture}(70,63)
  {\SetColor{Gray}{
  \Gluon(55,20)(70,20){1}{3}
  \Gluon(0,20)(15,20){1}{3}
  }}
  \GluonArc(35,20)(20,0,360){1}{24}
  \BCirc(15,20){5}
  \Line(13,18)(17,22)
  \Line(17,18)(13,22)
  \Line[dash,dsize=1](7,-3)(60,-3)
  \Line[dash,dsize=1](7,-3)(7,43)
  \Line[dash,dsize=1](60,-3)(60,43)
  \Line[dash,dsize=1](7,43)(60,43)
  \end{axopicture}
}
}

\def\gluonthreeptoneloop{
\raisebox{-20pt}{
  \begin{axopicture}(70,63)
  {\SetColor{Gray}{
  \GluonArc(35,20)(20,-90,0){1}{6}
  \GluonArc(35,20)(20,0,90){1}{6}
  \Gluon(55,20)(70,20){1}{3}
  \Gluon(0,20)(15,20){1}{3}
  }}
  \GluonArc(35,20)(20,90,180){1}{6}
  \GluonArc(35,20)(20,180,270){1}{6}
  \Gluon(35,0)(35,40){1}{7}
  \BCirc(15,20){5}
  \Line(13,18)(17,22)
  \Line(17,18)(13,22)
  \Line[dash,dsize=1](7,-3)(40,-3)
  \Line[dash,dsize=1](7,-3)(7,43)
  \Line[dash,dsize=1](40,-3)(40,43)
  \Line[dash,dsize=1](7,43)(40,43)
  \end{axopicture}
  }}
  
\def\gluontwoptwoloop{
\raisebox{-20pt}{  
\begin{axopicture}(70,63)
  {\SetColor{Gray}{
  \Gluon(55,20)(70,20){1}{3}
  \Gluon(0,20)(15,20){1}{3}
  }}
  \GluonArc(35,20)(20,0,90){1}{6}
  \GluonArc(35,20)(20,90,180){1}{6}
  \GluonArc(35,20)(20,180,270){1}{6}
  \GluonArc(35,20)(20,270,360){1}{6}
  \Gluon(35,0)(35,40){1}{7}
  \BCirc(15,20){5}
  \Line(13,18)(17,22)
  \Line(17,18)(13,22)
  \Line[dash,dsize=1](7,-3)(60,-3)
  \Line[dash,dsize=1](7,-3)(7,43)
  \Line[dash,dsize=1](60,-3)(60,43)
  \Line[dash,dsize=1](7,43)(60,43)
  \end{axopicture}  
  }}

  \def\gluonfourptoneloop{
\raisebox{-20pt}{  
  \begin{axopicture}(70,63)
  {\SetColor{Gray}{
  \Gluon(35,20)(55,20){1}{3}
  \GluonArc(35,20)(20,-90,0){1}{6}
  \GluonArc(35,20)(20,0,90){1}{6}
  \Gluon(55,20)(70,20){1}{3}
  \Gluon(0,20)(15,20){1}{3}
  }}
  \GluonArc(35,20)(20,90,180){1}{6}
  \GluonArc(35,20)(20,180,270){1}{6}
  \Gluon(35,0)(35,40){1}{7}
  \BCirc(15,20){5}
  \Line(13,18)(17,22)
  \Line(17,18)(13,22)
  \Line[dash,dsize=1](7,-3)(40,-3)
  \Line[dash,dsize=1](7,-3)(7,43)
  \Line[dash,dsize=1](40,-3)(40,43)
  \Line[dash,dsize=1](7,43)(40,43)
  \end{axopicture}  
  }}

  \def\gluonthreepttwoloop{
\raisebox{-20pt}{    
  \begin{axopicture}(90,63)
  {\SetColor{Gray}{
  \Gluon(0,20)(15,20){1}{3}
  \GluonArc(55,20)(20,-90,0){1}{6}
  \GluonArc(55,20)(20,0,90){1}{6}
  \Gluon(75,20)(90,20){1}{3}
  }}
  \GluonArc(35,20)(20,90,180){1}{6}
  \GluonArc(35,20)(20,180,270){1}{6}
  \Gluon(35,0)(35,40){1}{7}
  \Gluon(55,0)(55,40){1}{7}
  \Gluon(35,0)(55,0){1}{3}
  \Gluon(35,40)(55,40){1}{3}
  \BCirc(15,20){5}
  \Line(13,18)(17,22)
  \Line(17,18)(13,22)
  \Line[dash,dsize=1](7,-3)(60,-3)
  \Line[dash,dsize=1](7,-3)(7,43)
  \Line[dash,dsize=1](60,-3)(60,43)
  \Line[dash,dsize=1](7,43)(60,43)
  \end{axopicture}
  }}
  
  \def\gluontwoptthreeloop{
\raisebox{-20pt}{    
  \begin{axopicture}(90,63)
  {\SetColor{Gray}{
  \Gluon(0,20)(15,20){1}{3}
  \Gluon(75,20)(90,20){1}{3}
  }}
  \GluonArc(55,20)(20,-90,0){1}{6}
  \GluonArc(55,20)(20,0,90){1}{6}
  \GluonArc(35,20)(20,90,180){1}{6}
  \GluonArc(35,20)(20,180,270){1}{6}
  \Gluon(35,0)(35,40){1}{7}
  \Gluon(55,0)(55,40){1}{7}
  \Gluon(35,0)(55,0){1}{3}
  \Gluon(35,40)(55,40){1}{3}
  \BCirc(15,20){5}
  \Line(13,18)(17,22)
  \Line(17,18)(13,22)
  \Line[dash,dsize=1](7,-3)(80,-3)
  \Line[dash,dsize=1](7,-3)(7,43)
  \Line[dash,dsize=1](80,-3)(80,43)
  \Line[dash,dsize=1](7,43)(80,43)
  \end{axopicture}
  }}
  
    \def\gluonfiveptoneloop{
\raisebox{-20pt}
{    
\begin{axopicture}(70,63)
  {\SetColor{Gray}{
  \GluonArc(35,20)(10,-90,0){1}{3}
  \GluonArc(35,20)(10,0,90){1}{3}
  \Gluon(45,20)(55,20){1}{2}
  \GluonArc(35,20)(20,-90,0){1}{6}
  \GluonArc(35,20)(20,0,90){1}{6}
  \Gluon(55,20)(70,20){1}{3}
  \Gluon(0,20)(15,20){1}{3}
  }}
  \GluonArc(35,20)(20,90,180){1}{6}
  \GluonArc(35,20)(20,180,270){1}{6}
  \Gluon(35,0)(35,40){1}{7}
  \BCirc(15,20){5}
  \Line(13,18)(17,22)
  \Line(17,18)(13,22)
  \Line[dash,dsize=1](7,-3)(40,-3)
  \Line[dash,dsize=1](7,-3)(7,43)
  \Line[dash,dsize=1](40,-3)(40,43)
  \Line[dash,dsize=1](7,43)(40,43)
  \end{axopicture}}}

   \def\gluonfourpttwoloop{
   \raisebox{-20pt}{   
  \begin{axopicture}(80,63)
  {\SetColor{Gray}{
  \Gluon(0,20)(15,20){1}{3}
  \Gluon(55,20)(75,20){1}{3}
  \GluonArc(55,20)(20,-90,0){1}{6}
  \GluonArc(55,20)(20,0,90){1}{6}
  \Gluon(75,20)(90,20){1}{3}
  }}
  \GluonArc(35,20)(20,90,180){1}{6}
  \GluonArc(35,20)(20,180,270){1}{6}
  \Gluon(35,0)(35,40){1}{7}
  \Gluon(55,0)(55,40){1}{7}
  \Gluon(35,0)(55,0){1}{3}
  \Gluon(35,40)(55,40){1}{3}
  \BCirc(15,20){5}
  \Line(13,18)(17,22)
  \Line(17,18)(13,22)
  \Line[dash,dsize=1](7,-3)(60,-3)
  \Line[dash,dsize=1](7,-3)(7,43)
  \Line[dash,dsize=1](60,-3)(60,43)
  \Line[dash,dsize=1](7,43)(60,43)
  \end{axopicture}}}
  
  \def\gluonthreeptthreeloop{
     \raisebox{-20pt}{
  \begin{axopicture}(100,63)
  {\SetColor{Gray}{
  \Gluon(0,20)(15,20){1}{3}
  \GluonArc(75,20)(20,-90,0){1}{6}
  \GluonArc(75,20)(20,0,90){1}{6}
  \Gluon(95,20)(110,20){1}{3}
  }}
  \GluonArc(35,20)(20,90,180){1}{6}
  \GluonArc(35,20)(20,180,270){1}{6}
  \Gluon(35,0)(35,40){1}{7}
  \Gluon(55,0)(55,40){1}{7}
  \Gluon(35,0)(55,0){1}{3}
  \Gluon(35,40)(55,40){1}{3}
  \Gluon(75,0)(75,40){1}{7}
  \Gluon(55,0)(75,0){1}{3}
  \Gluon(55,40)(75,40){1}{3}
  \BCirc(15,20){5}
  \Line(13,18)(17,22)
  \Line(17,18)(13,22)
  \Line[dash,dsize=1](7,-3)(80,-3)
  \Line[dash,dsize=1](7,-3)(7,43)
  \Line[dash,dsize=1](80,-3)(80,43)
  \Line[dash,dsize=1](7,43)(80,43)
  \end{axopicture}}}

  \def\gluontwoptfourloop{
     \raisebox{-20pt}{
  \begin{axopicture}(100,63)
  {\SetColor{Gray}{
  \Gluon(0,20)(15,20){1}{3}
  \Gluon(95,20)(110,20){1}{3}
  }}
  \GluonArc(75,20)(20,-90,0){1}{6}
  \GluonArc(75,20)(20,0,90){1}{6}
  \GluonArc(35,20)(20,90,180){1}{6}
  \GluonArc(35,20)(20,180,270){1}{6}
  \Gluon(35,0)(35,40){1}{7}
  \Gluon(55,0)(55,40){1}{7}
  \Gluon(35,0)(55,0){1}{3}
  \Gluon(35,40)(55,40){1}{3}
  \Gluon(75,0)(75,40){1}{7}
  \Gluon(55,0)(75,0){1}{3}
  \Gluon(55,40)(75,40){1}{3}
  \BCirc(15,20){5}
  \Line(13,18)(17,22)
  \Line(17,18)(13,22)
  \Line[dash,dsize=1](7,-3)(100,-3)
  \Line[dash,dsize=1](7,-3)(7,43)
  \Line[dash,dsize=1](100,-3)(100,43)
  \Line[dash,dsize=1](7,43)(100,43)
  \end{axopicture}}}

\def\ghosttwoptoneloop{
\raisebox{-20pt}{
  \begin{axopicture}(70,63)
  \Line[dash](55,20)(70,20)
  \Line[dash](0,20)(15,20)
  \Arc[dash,arrow,clockwise](35,20)(20,180,360)
  \GluonArc(35,20)(20,180,360){1}{12}
  \BCirc(35,0){5}
  \Line(33,-2)(37,2)
  \Line(37,-2)(33,2)
  \end{axopicture}
}
}

\def\ghosttwoptwoloop{
\raisebox{-20pt}{  
\begin{axopicture}(70,63)
  \Line[dash](55,20)(70,20)
  \Line[dash](0,20)(15,20)
  \Arc[dash,arrow,clockwise](35,20)(20,90,360)
  \Arc[dash,arrow,clockwise](35,20)(20,180,90)
  \GluonArc(35,20)(20,180,270){1}{6}
  \GluonArc(35,20)(20,270,360){1}{6}
  \Gluon(35,0)(35,40){1}{7}
  \BCirc(35,0){5}
  \Line(33,-2)(37,2)
  \Line(37,-2)(33,2)
  \end{axopicture}  
  }}

\def\ghostthreeptoneloop{
\raisebox{-20pt}{  
\begin{axopicture}(70,63)
  \Line[dash](0,20)(15,20)
  \Line[dash](35,40)(50,40)
  \Gluon(35,0)(50,0){1}{3}
  \Arc[dash,arrow,clockwise](35,20)(20,180,90)
  \GluonArc(35,20)(20,180,270){1}{6}
  \Gluon(35,0)(35,40){1}{7}
  \BCirc(35,0){5}
  \Line(33,-2)(37,2)
  \Line(37,-2)(33,2)
  \end{axopicture}  
  }}

\def\ghosttwoptthreeloop{
\raisebox{-20pt}{    
  \begin{axopicture}(90,63)
  \Line[dash](0,20)(15,20)
  \Line[dash](75,20)(90,20)  
  \GluonArc(55,20)(20,-90,0){1}{6}
  \Arc[dash,arrow,clockwise](55,20)(20,90,360)
  \Arc[dash,arrow,clockwise](35,20)(20,180,90)
  \GluonArc(35,20)(20,180,270){1}{6}

  \Gluon(35,0)(35,40){1}{7}
  \Gluon(55,0)(55,40){1}{7}
  \Gluon(35,0)(55,0){1}{3}
  \Line[dash](35,40)(55,40)
  \BCirc(35,0){5}
  \Line(33,-2)(37,2)
  \Line(37,-2)(33,2)
  \end{axopicture}
  }}

\def\ghostthreepttwoloop{
\raisebox{-20pt}{    
  \begin{axopicture}(90,63)
  \Line[dash](0,20)(15,20)
  \Arc[dash,arrow,clockwise](35,20)(20,180,90)
  \GluonArc(35,20)(20,180,270){1}{6}
  \Line[dash](55,40)(70,40)
  \Gluon(55,0)(70,0){1}{3}  
  \Gluon(35,0)(35,40){1}{7}
  \Gluon(55,0)(55,40){1}{7}
  \Gluon(35,0)(55,0){1}{3}
  \Line[dash](35,40)(55,40)
  \BCirc(35,0){5}
  \Line(33,-2)(37,2)
  \Line(37,-2)(33,2)
  \end{axopicture}
  }}

\def\ghostfourptoneloop{
\raisebox{-20pt}{  
  \begin{axopicture}(55,63)
  \Line[dash](0,40)(15,40)
  \Gluon(0,0)(15,0){1}{3}  
  \Line[dash](35,40)(50,40)
  \Gluon(35,0)(50,0){1}{3}  
  \Gluon(15,0)(15,40){1}{7}
  \Gluon(35,0)(35,40){1}{7}
  \Gluon(15,0)(35,0){1}{3}
  \Line[dash,arrow](15,40)(35,40)
  \BCirc(15,0){5}
  \Line(13,-2)(17,2)
  \Line(17,-2)(13,2)
  \end{axopicture}
  }}  
  
\def\ghosttwoptfourloop{
\raisebox{-20pt}{    
  \begin{axopicture}(110,63)
  \Line[dash](0,20)(15,20)
  \Line[dash](95,20)(110,20)  
  \GluonArc(75,20)(20,-90,0){1}{6}
  \Arc[dash,arrow,clockwise](75,20)(20,90,360)
  \Arc[dash,arrow,clockwise](35,20)(20,180,90)
  \GluonArc(35,20)(20,180,270){1}{6}
  \Gluon(35,0)(35,40){1}{7}
  \Gluon(55,0)(55,40){1}{7}
  \Gluon(75,0)(75,40){1}{7}
  \Gluon(35,0)(55,0){1}{3}
  \Gluon(55,0)(75,0){1}{3}
  \Line[dash](35,40)(70,40)
  \BCirc(35,0){5}
  \Line(33,-2)(37,2)
  \Line(37,-2)(33,2)
  \end{axopicture}
  }}  
  
\def\ghostthreeptthreeloop{
\raisebox{-20pt}{    
  \begin{axopicture}(95,63)
  \Line[dash](0,20)(15,20)
  \Arc[dash,arrow,clockwise](35,20)(20,180,90)
  \GluonArc(35,20)(20,180,270){1}{6}
  \Gluon(35,0)(35,40){1}{7}
  \Gluon(55,0)(55,40){1}{7}
  \Gluon(75,0)(75,40){1}{7}
  \Gluon(35,0)(55,0){1}{3}
  \Gluon(55,0)(75,0){1}{3}
  \Line[dash](35,40)(75,40)
  \Line[dash](75,40)(90,40)
  \Gluon(75,0)(90,0){1}{3}  
  \BCirc(35,0){5}
  \Line(33,-2)(37,2)
  \Line(37,-2)(33,2)
  \end{axopicture}
  }}

\def\ghostfourpttwoloop{
\raisebox{-20pt}{    
  \begin{axopicture}(75,63)
  \Line[dash](0,40)(15,40)
  \Gluon(0,0)(15,0){1}{3}  
  \Gluon(15,0)(15,40){1}{7}
  \Gluon(35,0)(35,40){1}{7}
  \Gluon(15,0)(35,0){1}{3}
  \Line[dash,arrow](15,40)(35,40)
  \Line[dash](55,40)(70,40)
  \Gluon(55,0)(70,0){1}{3}  
  \Gluon(55,0)(55,40){1}{7}
  \Gluon(35,0)(55,0){1}{3}
  \Line[dash](35,40)(55,40)

  \BCirc(15,0){5}
  \Line(13,-2)(17,2)
  \Line(17,-2)(13,2)
  \end{axopicture}
  }}

\def\ghostfiveptoneloop{
\raisebox{-20pt}{  
  \begin{axopicture}(55,63)
  \Line[dash](0,40)(15,40)
  \Gluon(0,0)(15,0){1}{3}  
  \Line[dash](35,40)(50,40)
  \Gluon(35,0)(50,0){1}{3}  
  \Gluon(35,20)(50,20){1}{3}  
  \Gluon(15,0)(15,40){1}{7}
  \Gluon(35,0)(35,40){1}{7}
  \Gluon(15,0)(35,0){1}{3}
  \Line[dash,arrow](15,40)(35,40)
  \BCirc(15,0){5}
  \Line(13,-2)(17,2)
  \Line(17,-2)(13,2)
  \end{axopicture}
  }}  

\def\BtwolEOMone{
\raisebox{-20pt}{
  \begin{axopicture}(70,63)
  {\SetColor{Gray}{
  \DoubleGluon(55,15)(70,15){1}{3}{1}
  \DoubleGluon(0,15)(15,15){1}{3}{1}
  \GluonArc(30,15)(15,180,270){1}{4}
  \GluonArc(40,15)(15,270,360){1}{4}
  \Gluon(30,0)(40,0){-1}{2}
  \GluonArc(35,15)(20,0,45){1}{3}
  \GluonArc(35,15)(20,135,180){1}{3}
  }}
  \GluonArc(35,15)(20,45,135){1}{6}  
  \GluonArc(35,43.284271247)(20,225,315){1}{6}  
  \BCirc(35,35){5}
  \Line(33,33)(37,37)
  \Line(37,33)(33,37)
  \Line[dash,dsize=1](17,17)(53,17)
  \Line[dash,dsize=1](17,17)(17,43)
  \Line[dash,dsize=1](53,17)(53,43)
  \Line[dash,dsize=1](17,43)(53,43)
  \end{axopicture}
}
}

\def\BthreelEOMone{
\raisebox{-20pt}{
  \begin{axopicture}(70,63)
  {\SetColor{Gray}{
  \DoubleGluon(55,15)(70,15){1}{3}{1}
  \DoubleGluon(0,15)(15,15){1}{3}{1}
  \GluonArc(30,15)(15,180,270){1}{4}
  \GluonArc(40,15)(15,270,360){1}{4}
  \Gluon(30,0)(40,0){-1}{2}
  \GluonArc(35,15)(20,0,45){1}{3}
  \GluonArc(35,15)(20,135,180){1}{3}
  }}
  \GluonArc(35,15)(20,45,135){1}{6}  
  \GluonArc(35,43.284271247)(20,225,270){1}{3}  
  \GluonArc(35,43.284271247)(20,270,315){1}{3}
  \Gluon(35,23.284271247)(35,35){1}{3}
  \BCirc(35,35){5}
  \Line(33,33)(37,37)
  \Line(37,33)(33,37)
  \Line[dash,dsize=1](17,17)(53,17)
  \Line[dash,dsize=1](17,17)(17,43)
  \Line[dash,dsize=1](53,17)(53,43)
  \Line[dash,dsize=1](17,43)(53,43)
  \end{axopicture}
}
}

\def\BfourlEOMone{
\raisebox{-20pt}{
  \begin{axopicture}(70,63)
  {\SetColor{Gray}{
  \DoubleGluon(55,15)(70,15){1}{3}{1}
  \DoubleGluon(0,15)(15,15){1}{3}{1}
  \GluonArc(30,15)(15,180,270){1}{4}
  \GluonArc(40,15)(15,270,360){1}{4}
  \Gluon(30,0)(40,0){-1}{2}
  \GluonArc(35,15)(20,0,45){1}{3}
  \GluonArc(35,15)(20,135,180){1}{3}
  }}
  \GluonArc(35,15)(20,45,135){1}{6}  
  \GluonArc(35,43.284271247)(20,225,255){1}{2}  
  \GluonArc(35,43.284271247)(20,255,285){1}{2}  
  \GluonArc(35,43.284271247)(20,285,315){1}{2}
  \Gluon(35,35)(29.823619098,23.965754722){1}{2}
  \Gluon(35,35)(40.176380902,23.965754722){1}{2}
  \BCirc(35,35){5}
  \Line(33,33)(37,37)
  \Line(37,33)(33,37)
  \Line[dash,dsize=1](17,17)(53,17)
  \Line[dash,dsize=1](17,17)(17,43)
  \Line[dash,dsize=1](53,17)(53,43)
  \Line[dash,dsize=1](17,43)(53,43)
  \end{axopicture}
}
}

\def\Bthreeptoneloop{
\raisebox{-20pt}{
  \begin{axopicture}(70,63)
  {\SetColor{Gray}{
  \GluonArc(35,20)(20,-90,0){1}{6}
  \GluonArc(35,20)(20,0,90){1}{6}
  \DoubleGluon(55,20)(70,20){1}{3}{1}
  \DoubleGluon(0,20)(15,20){1}{3}{1}
  }}
  \GluonArc(35,20)(20,90,180){1}{6}
  \GluonArc(35,20)(20,180,270){1}{6}
  \Gluon(35,0)(35,40){1}{7}
  \BCirc(15,20){5}
  \Line(13,18)(17,22)
  \Line(17,18)(13,22)
  \Line[dash,dsize=1](7,-3)(40,-3)
  \Line[dash,dsize=1](7,-3)(7,43)
  \Line[dash,dsize=1](40,-3)(40,43)
  \Line[dash,dsize=1](7,43)(40,43)
  \end{axopicture}
}}

\def\Bfourptoneloop{
\raisebox{-20pt}{  
  \begin{axopicture}(70,63)
  {\SetColor{Gray}{
  \Gluon(35,20)(55,20){1}{3}
  \GluonArc(35,20)(20,-90,0){1}{6}
  \GluonArc(35,20)(20,0,90){1}{6}
  \DoubleGluon(55,20)(70,20){1}{3}{1}
  \DoubleGluon(0,20)(15,20){1}{3}{1}
  }}
  \GluonArc(35,20)(20,90,180){1}{6}
  \GluonArc(35,20)(20,180,270){1}{6}
  \Gluon(35,0)(35,40){1}{7}
  \BCirc(15,20){5}
  \Line(13,18)(17,22)
  \Line(17,18)(13,22)
  \Line[dash,dsize=1](7,-3)(40,-3)
  \Line[dash,dsize=1](7,-3)(7,43)
  \Line[dash,dsize=1](40,-3)(40,43)
  \Line[dash,dsize=1](7,43)(40,43)
  \end{axopicture}  
  }}

\def\Bthreepttwoloop{
\raisebox{-20pt}{    
  \begin{axopicture}(90,63)
  {\SetColor{Gray}{
  \DoubleGluon(0,20)(15,20){1}{3}{1}
  \GluonArc(55,20)(20,-90,0){1}{6}
  \GluonArc(55,20)(20,0,90){1}{6}
  \DoubleGluon(75,20)(90,20){1}{3}{1}
  }}
  \GluonArc(35,20)(20,90,180){1}{6}
  \GluonArc(35,20)(20,180,270){1}{6}
  \Gluon(35,0)(35,40){1}{7}
  \Gluon(55,0)(55,40){1}{7}
  \Gluon(35,0)(55,0){1}{3}
  \Gluon(35,40)(55,40){1}{3}
  \BCirc(15,20){5}
  \Line(13,18)(17,22)
  \Line(17,18)(13,22)
  \Line[dash,dsize=1](7,-3)(60,-3)
  \Line[dash,dsize=1](7,-3)(7,43)
  \Line[dash,dsize=1](60,-3)(60,43)
  \Line[dash,dsize=1](7,43)(60,43)
  \end{axopicture}
  }}

\def\Bfiveptoneloop{
\raisebox{-20pt}
{    
\begin{axopicture}(70,63)
  {\SetColor{Gray}{
  \GluonArc(35,20)(10,-90,0){1}{3}
  \GluonArc(35,20)(10,0,90){1}{3}
  \Gluon(45,20)(55,20){1}{2}
  \GluonArc(35,20)(20,-90,0){1}{6}
  \GluonArc(35,20)(20,0,90){1}{6}
  \DoubleGluon(55,20)(70,20){1}{3}{1}
  \DoubleGluon(0,20)(15,20){1}{3}{1}
  }}
  \GluonArc(35,20)(20,90,180){1}{6}
  \GluonArc(35,20)(20,180,270){1}{6}
  \Gluon(35,0)(35,40){1}{7}
  \BCirc(15,20){5}
  \Line(13,18)(17,22)
  \Line(17,18)(13,22)
  \Line[dash,dsize=1](7,-3)(40,-3)
  \Line[dash,dsize=1](7,-3)(7,43)
  \Line[dash,dsize=1](40,-3)(40,43)
  \Line[dash,dsize=1](7,43)(40,43)
  \end{axopicture}}}

  \def\Bfourpttwoloop{
   \raisebox{-20pt}{   
  \begin{axopicture}(80,63)
  {\SetColor{Gray}{
  \DoubleGluon(0,20)(15,20){1}{3}{1}
  \Gluon(55,20)(75,20){1}{3}
  \GluonArc(55,20)(20,-90,0){1}{6}
  \GluonArc(55,20)(20,0,90){1}{6}
  \DoubleGluon(75,20)(90,20){1}{3}{1}
  }}
  \GluonArc(35,20)(20,90,180){1}{6}
  \GluonArc(35,20)(20,180,270){1}{6}
  \Gluon(35,0)(35,40){1}{7}
  \Gluon(55,0)(55,40){1}{7}
  \Gluon(35,0)(55,0){1}{3}
  \Gluon(35,40)(55,40){1}{3}
  \BCirc(15,20){5}
  \Line(13,18)(17,22)
  \Line(17,18)(13,22)
  \Line[dash,dsize=1](7,-3)(60,-3)
  \Line[dash,dsize=1](7,-3)(7,43)
  \Line[dash,dsize=1](60,-3)(60,43)
  \Line[dash,dsize=1](7,43)(60,43)
  \end{axopicture}}}
  
  \def\Bthreeptthreeloop{
     \raisebox{-20pt}{
  \begin{axopicture}(100,63)
  {\SetColor{Gray}{
  \DoubleGluon(0,20)(15,20){1}{3}{1}
  \GluonArc(75,20)(20,-90,0){1}{6}
  \GluonArc(75,20)(20,0,90){1}{6}
  \DoubleGluon(95,20)(110,20){1}{3}{1}
  }}
  \GluonArc(35,20)(20,90,180){1}{6}
  \GluonArc(35,20)(20,180,270){1}{6}
  \Gluon(35,0)(35,40){1}{7}
  \Gluon(55,0)(55,40){1}{7}
  \Gluon(35,0)(55,0){1}{3}
  \Gluon(35,40)(55,40){1}{3}
  \Gluon(75,0)(75,40){1}{7}
  \Gluon(55,0)(75,0){1}{3}
  \Gluon(55,40)(75,40){1}{3}
  \BCirc(15,20){5}
  \Line(13,18)(17,22)
  \Line(17,18)(13,22)
  \Line[dash,dsize=1](7,-3)(80,-3)
  \Line[dash,dsize=1](7,-3)(7,43)
  \Line[dash,dsize=1](80,-3)(80,43)
  \Line[dash,dsize=1](7,43)(80,43)
  \end{axopicture}}}

\title{Renormalization of gluonic leading-twist Operators in covariant Gauges}

\author[a]{Giulio Falcioni,}
\author[a]{Franz Herzog}

\affiliation[a]{Higgs Centre for Theoretical Physics, School of Physics and Astronomy, The University of Edinburgh, Edinburgh EH9 3FD, Scotland, UK
}

\emailAdd{giulio.falcioni@ed.ac.uk}
\emailAdd{fherzog@ed.ac.uk}

\abstract{
We provide the all-loop structure of gauge-variant operators required for the renormalisation of Green's functions with insertions of twist-two operators in Yang-Mills theory. Using this structure we work out an explicit basis valid up to 4-loop order for an arbitrary compact simple gauge group. To achieve this we employ a generalised gauge symmetry, originally proposed by Dixon and Taylor, which arises after adding to the Yang-Mills Lagrangian also operators proportional to its equation of motion. Promoting this symmetry to a generalised BRST symmetry allows to generate the ghost operator from a single exact operator in the BRST-generalised sense. We show that our construction complies with the theorems by Joglekar and Lee. We further establish the existence of a generalised anti-BRST symmetry which we employ to derive non-trivial relations among the anomalous dimension matrices of ghost and equation-of-motion operators. For the purpose of demonstration we employ the formalism to compute the $N=2,4$ Mellin moments of the gluonic splitting function up to 4 loops and its $N=6$ Mellin moment up to 3 loops, where we also take advantage of additional simplifications of the background field formalism.}

\begin{document}


\maketitle

\section{Introduction}

The increasing precision with which particle collisions are being measured at the Large Hadron Collider have pushed theoretical predictions in QCD to the next-to-next-to-next-to-leading order (N3LO) in perturbative QCD. First calculations at this order were completed for Higgs boson production \cite{Anastasiou:2015vya,Anastasiou:2016cez,Mistlberger:2018etf,Duhr:2019kwi,Duhr:2020seh,Chen:2021isd} and the Drell-Yan process \cite{Duhr:2019kwi,Duhr:2020sdp}. Even $2\to2$ reactions may become feasible at this order in the not-too-far future as first results for 3-loop dijet production amplitudes have become available \cite{Bargiela:2021wuy,Caola:2021izf}. One of the dominant remaining theoretical uncertainties associated to such N3LO calculations is now related to the lack of knowledge of the 4-loop splitting functions, which determine the evolution of the parton densities at the relevant perturbative order, and which are known completely only up to three loops \cite{Floratos:1978ny,Gonzalez-Arroyo:1979qht,Furmanski:1980cm,Hamberg:1991qt,Vogelsang:1995vh,Mertig:1995ny,Ellis:1996nn,Matiounine:1998ky,Matiounine:1998re,Larin:1993vu,Larin:1996wd,Moch:2004pa,Vogt:2004mw,Ablinger:2014nga,Ablinger:2017tan,Behring:2019tus,Ablinger:2019etw,Blumlein:2021enk,Blumlein:2021ryt}.

Substantial efforts to improve this situation have already been made. In the non-singlet sector, after pioneering calculations of lower moments \cite{Velizhanin:2014fua}, numerical approximations for the 4-loop splitting functions are now known to high accuracy. An analytic reconstruction was achieved in the limit of leading number of colours \cite{Moch:2017uml} and for subleading corrections in the number of quark flavors, $n_f$, \cite{Davies:2016jie}; the leading $n_f$-contributions were known already for some time \cite{Gracey:1996ad} to all orders in perturbation theory. Even some low $N$-moments at five loops \cite{Herzog:2018kwj} are already known. In the singlet sector instead only a handful of lower moments have been calculated so far at the four-loop level \cite{Moch:2021qrk} and this input was found to be insufficient for reliable numerical approximations. 

\begin{figure}[t]
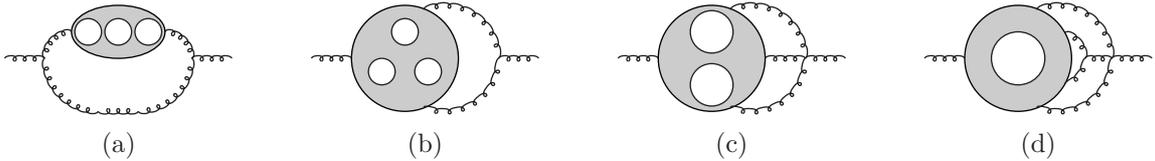

\begin{subfigure}{.2\textwidth}
  \centering
\begin{axopicture}(85,40)
  \Gluon(0,20)(15,20){1}{3}
  \GluonArc(25,20)(10,90,180){1}{4}
  \GluonArc(60,20)(10,0,90){1}{4}
  \GluonArc(35,20)(20,180,270){1}{7}
  \GluonArc(50,20)(20,270,360){1}{7}
  \Gluon(35,0)(50,0){-1}{3}
  \Gluon(70,20)(85,20){1}{3}
  \GOval(42.5,30)(10,17.5)(0){.8}
  \CCirc(31.25,30){5}{Black}{White}
  \CCirc(42.5,30){5}{Black}{White}
  \CCirc(53.75,30){5}{Black}{White}
\end{axopicture}
\caption{\label{fig:mixOEOM1}}
\end{subfigure}
\hfill
\begin{subfigure}{.2\textwidth}
  \centering
\begin{axopicture}(85,40)
  \Gluon(0,20)(15,20){1}{3}
  \GCirc(35,20){20}{.8}
  \CCirc(43.660254038,15){5}{Black}{White}
  \CCirc(26.339745962,15){5}{Black}{White}
  \CCirc(35,30){5}{Black}{White}
  \GluonArc(50,20)(20,0,113.57814878){1}{7}
  \GluonArc(50,20)(20,246.42185122,360){-1}{7}
  \Gluon(70,20)(85,20){1}{3}
\end{axopicture}
\caption{\label{fig:mixOEOM2}}
\end{subfigure}
\hfill
\begin{subfigure}{.2\textwidth}
  \centering
\begin{axopicture}(85,40)
  \Gluon(0,20)(15,20){1}{3}
  \GCirc(35,20){20}{.8}
  \CCirc(35,10){8}{Black}{White}
  \CCirc(35,30){8}{Black}{White}
  \GluonArc(50,20)(20,0,113.57814878){1}{7}
  \GluonArc(50,20)(20,246.42185122,360){-1}{7}
  \Gluon(55,20)(70,20){1}{3}
  \Gluon(70,20)(85,20){1}{3}
\end{axopicture}
\caption{\label{fig:mixOEOM3}}
\end{subfigure}
\hfill
\begin{subfigure}{.2\textwidth}
  \centering
\begin{axopicture}(85,40)
  \Gluon(0,20)(15,20){1}{3}
  \GCirc(35,20){20}{.8}
  \CCirc(35,20){10}{Black}{White}
  \GluonArc(50,20)(20,0,113.57814878){1}{7}
  \GluonArc(50,20)(20,246.42185122,0){-1}{7}
  \GluonArc(50,20)(10,0,75.52248781407009){1}{2}
  \GluonArc(50,20)(10,284.477512186,360){-1}{2}
  \Gluon(60,20)(70,20){1}{2}
  \Gluon(70,20)(85,20){1}{3}
\end{axopicture}
\caption{\label{fig:mixOEOM4}}
\end{subfigure}
\caption{The grey blobs denote examples of multi-loop multi-gluon subgraphs contributing at the four loop order, which contain insertions of the gauge invariant operator, and whose UV-divergences lead to mixing with unphysical operators under renormalisation.}
\label{fig:mixOEOM}
\end{figure}

The reason for why calculations in the non-singlet sector are so much more advanced than those in the singlet sector is not only due to more numerous and complex Feynman diagrams but also due to a more powerful framework. This framework is based on the operator product expansion (OPE)\cite{Gross:1974cs,Georgi:1974wnj}, which allows the extraction of Mellin moments from the anomalous dimensions of leading-twist light-cone operators. While the renormalisation of such operators entering in the non-singlet sector is relatively straight forward, the renormalisation of the corresponding gluonic operators in the off-shell formalism (likely the most promising framework to allow for progress at four loops at this time) is non-trivial due to the mixing into unphysical operators. The mixing into these operators arises from subgraphs with external gluons and ghosts which contain the insertion of the singlet operator; schematic examples appearing at four loops are depicted in figure \ref{fig:mixOEOM}.

An explicit basis for these unphysical operators, valid up to the two-loop level, was worked out by Dixon and Taylor already almost fifty years ago \cite{Dixon:1974ss} and was employed in the computation of one-loop Mellin moments in the Feynman gauge. Hamberg and Van Neerven about twenty years later managed to successfully employ the same framework to perform calculations at the 2-loop order in dimensional regularisation \cite{Hamberg:1991qt}. This calculation was also repeated very recently \cite{Blumlein:2022ndg}. It is interesting to note that the calculation by Hamberg and Van Neerven managed to successfully resolve a number of conflicting results \cite{Floratos:1978ny,Gonzalez-Arroyo:1979qht,Furmanski:1980cm} which were present at that time because of negligence of the mixing into unphysical operators.

Nevertheless the structure of the basis proposed by Dixon and Taylor remained somewhat mysterious. This was pointed out in particular by Collins \cite{Collins:1994ee}, who argued that the unphysical operators appeared to be in conflict with a general theorem  which states that the unphysical operators should be either proportional to the equation of motion (EOM) or BRST-exact. This theorem was in part first conjectured by Kluberg-Stern and Zuber \cite{Kluberg-Stern:1974nmx,Kluberg-Stern:1975ebk} before it was proven by Joglekar and Lee \cite{Joglekar:1975nu,Joglekar:1976eb,Joglekar:1976pe}. Another proof based on cohomology theory was later provided by Henneaux \cite{Henneaux:1993jn}.

In the present paper we revisit the problem. More concretely we reinterpret the basis of Dixon and Taylor as being build up from EOM and ghost operators. The former are proportional to the EOM of the gauge invariant part of the Yang-Mills Lagrangian. Using this notion we are able to write down the general form of the EOM operators at arbitrary loop orders. As was noted already by Dixon and Taylor the combined Lagrangian consisting of the Yang-Mills Lagrangian and their EOM operators is invariant under a generalised gauge symmetry. This generalised gauge symmetry can then be promoted to a generalised BRST symmetry. We prove that the generalised BRST transformation is nilpotent and that the ghost operator can be constructed from a single BRST-exact operator in the generalised BRST sense. While the generalised BRST symmetry was in part pointed out already by Hamberg and Van Neerven in \cite{Hamberg:1991qt}, it was not clearly spelt out how to use it to construct the ghost Lagrangian. We apply the formalism to work out the explicit form of EOM and ghost operators up to four-loop order. We also show that the basis is in accord with the theorem of Joglekar and Lee.

To further simplify calculations in the OPE framework we explore two further symmetry principles. We observe that the ghost term of the unphysical operator can also be generated from a generalised anti-BRST symmetry \cite{Curci:1976bt,Ojima:1980da,Baulieu:1981sb,Binosi:2013cea} and a corresponding generalised anti-BRST-exact operator. In particular we employ this alternative formulation to derive a set of nontrivial identities among the anomalous dimensions of the EOM and ghost operators. We also make use of the background field formalism \cite{DeWitt:1967ub,tHooft:1975uxh,Abbott:1980hw,Abbott:1981ke,Sarkar:1974ni,Kluberg-Stern:1974nmx,Kluberg-Stern:1975ebk,Abbott:1980hw,Abbott:1981ke}. Background field gauge invariance allows one to reduce by one the maximum number of loops at which the anomalous dimensions of the EOM and ghost operators are required - thereby yielding another welcome simplification for the calculation of unphysical counterterms. For the purpose of demonstration we will employ the formalism to re-calculate the $N=2$ and $N=4$ moments of the purely gluonic contributions in the singlet sector up to 4 loops and the $N=6$ moment up to three loops.

In the following we give a brief outline of the paper. In section \ref{sec:background} we summarise our conventions and review some of the relevant background material. The construction of EOM operators and the generalised gauge symmetry is discussed in section \ref{sec:EoMGauge}. The construction of ghost operators and the generalised BRST and anti-BRST invariance is discussed in section \ref{sec:BRST}. There also the compatibility of our construction with the theorems of Joglekar and Lee is shown. The concepts are employed to build an independent basis of gauge-variant operators in section \ref{sec:operatorbases} for various fixed values of $N$. The background-field formulation is presented in section \ref{sec:backgroundfield} and employed in section \ref{sec:Applications} for the computation of Mellin moments up to 4-loop order. We conclude in section \ref{sec:conclusions}.

\section{Background}
\label{sec:background}
\subsection{Yang-Mills Lagrangian}

In the following we summarise our conventions for the Yang Mills Lagrangian.
We define the 
field strength tensor 
as
\begin{equation}
  F_{\mu\nu}^a = \partial_\mu A_\nu^a - \partial_\nu A_\mu^a + g f^{abc} A_\mu^bA_\nu^c\,,
\end{equation}
such that the gauge invariant part of the Yang-Mills action is given by
\begin{equation}
  \label{def:LYM}
  S_{0} = \int d^dx\,\mathcal{L}_{0},\quad\quad\quad
  \mathcal{L}_{0} = -\frac{1}{4}\,F^{\mu\nu}_a\,F_{\mu\nu}^{a}\,.
\end{equation}
Let us further define the covariant derivative in the adjoint representation,
\begin{equation}
  D_\mu^{ac} =\partial_\mu\delta^{ac}+gf^{abc}A_\mu^b\,.
\end{equation}
With this definition the EOM of the Yang-Mills Lagrangian is written compactly as
\begin{equation}
\label{eq:EOM}
\frac{\delta S_0}{\delta A^\mu_a} = \left(D_\nu F^{\nu\mu}\right)_a .
\end{equation}
The action $S_{0}$ is of course invariant under infinitesimal gauge transformations
\begin{equation}
A^a_\mu \to A^a_\mu+\delta_\omega A^a_\mu\,, \quad \text{with} \quad  \delta_\omega A^a_\mu = (D_\mu\omega)^a\,.
\end{equation}

\subsection{Gauge-fixing, Ghosts and BRST}

The gauge invariance is broken by the gauge fixing (GF) and ghost (G) terms, the latter being required to cancel unphysical degrees of freedom of the gauge field.  For the commonly used choice of the linear covariant gauge the gauge-fixing and ghost contributions to the Lagrangian are
\begin{equation}
  \label{def:GF}
  \mathcal{L}_{\text{GF+G}}=-\frac{1}{2\xi}(\partial^\mu A^{a}_\mu)^2-\overline{c}^a\,\partial^\mu D_\mu^{ab}\,c^b\,,
\end{equation}
where $c^a$ and $\bar c^a$ are respectively the ghost and anti-ghost fields. The complete gauge-fixed Yang-Mills action is then given by
\beq
S=\int d^dx\, \L\,, \qquad \text{with}\qquad  \L= \L_0+\L_{\mathrm{GF+G}}\,,
\eeq
and its EOM is given by
\beq
\label{eq:GFEOM}
 \frac{\delta S}{\delta A^\mu_a} =  \left(D_\nu F^{\nu\mu}\right)_a+\frac{1}{\xi}\partial^\mu\partial^\nu A^{a}_{\nu}\
-gf^{abc} (\partial^\mu \overline{c}^b)c^c\,.
\eeq
While eq.~(\ref{def:GF}) breaks gauge invariance it does remain invariant under nilpotent BRST transformations. This feature becomes most transparent after the introduction of an auxiliary field $b_a(x)$, also known as the Nakanishi-Lautrup field \cite{Nakanishi:1966zz,Lautrup:1967zz}. In this formulation the Lagrangian is written as
\begin{equation}
  \label{def:GFB}
  \mathcal{L}_{\text{GF+G}}=    -b^a \partial^\mu A^{a}_\mu       +\frac{\xi}{2}(b^a)^2-\overline{c}^a\,\partial^\mu D_\mu^{ab}\,c^b\,.
\end{equation}
Eq.~(\ref{def:GFB}) can be seen to be equivalent to eq.~(\ref{def:GF}) after substituting the solution of the EOM $b^a=\frac{1}{\xi}\partial^\mu A^a_\mu$. The BRST variation leaving this Lagrangian invariant \cite{Becchi:1975nq,Tyutin:1975qk} is defined as
\beq
\delta_{\mathrm{BRST}}(\bullet)\equiv\theta \,s(\bullet)\,,
\eeq
where $\theta$ is a Grassmann number, $s$ denotes the BRST operator, which being nilpotent satisfies $s^2(\bullet)=0$, and whose action on the fields is given by
\begin{align}
  \label{eq:BRST}
  s b^a =0\,,\quad s A_\mu^a&=(D_\mu\,c)^a\,,\quad s c^a=-\frac{g}{2}\,f^{abc}c^b \,c^c\,,\quad s\overline{c}^a = -b^a\,.
\end{align}
The BRST invariance of eq.~(\ref{def:GFB}) can be made manifest by writing it in BRST-exact form, that is as the BRST variation of an ancestor operator:
\beq
\label{eq:ancestor}
\mathcal{L}_{\text{GF+G}}=  s\O_{\mathrm{ancestor}} \,,  \qquad \O_{\mathrm{ancestor}}=
\overline c^a \big[  \partial^\mu A^a_\mu-\tfrac{1}{2}\xi b^a \big]\,.
\eeq
An interpretation of the BRST symmetry is that it corresponds to a certain subclass of gauge transformations, where the parameter $\omega(x)$ of the gauge transformation is identified with the ghost field times a Grassmann number. There exists in fact a second such symmetry in the gauge-fixed Lagrangian where the role of the ghost field in the BRST variations is replaced with that of the anti-ghost field. This leads to the so-called anti-BRST symmetry \cite{Curci:1976bt,Ojima:1980da,Baulieu:1981sb,Binosi:2013cea}. To discuss this symmetry we first introduce another auxiliary field:
\beq
\bar b^a=-b^a+gf^{abc}\bar c^b c^c
\eeq
The anti-BRST variation is then given by
\beq
\delta_{\overline{\mathrm{BRST}}}(\bullet)\equiv\bar\theta \,\bar s(\bullet)\,,
\eeq
with $\bar\theta$ another Grassmann number and
\begin{align}
  \label{eq:BRST}
  \bar s \bar b^a &=0\,,\quad \bar s A_\mu^a=(D_\mu\,\bar c)^a\,,\quad
  \bar s \bar c^a=-\frac{g}{2}\,f^{abc}\bar c^b \,\bar c^c\,,\quad
  \bar sc^a = -\bar b^a\,.
\end{align}
The BRST and anti-BRST variations fulfil the following consistency condition:
\beq
ss(\bullet)=\bar s \bar s(\bullet) =0=\bar s  s(\bullet)+ s \bar s(\bullet)
\eeq
Similarly to eq.~(\ref{eq:ancestor}) the anti-BRST symmetry of eq.~(\ref{def:GFB}) can be made manifest by writing it in anti-BRST exact form, that is as the anti-BRST variation of another ancestor operator:
\beq
\label{eq:anti-ancestor}
\mathcal{L}_{\text{GF+G}}=  \bar s\,\overline{\O}_{\mathrm{ancestor}} \,,
\qquad \overline{\O}_{\mathrm{ancestor}}=c^a \big[  \tfrac{1}{2}\xi b^a-\partial^\mu A^a_\mu \big]\,.
\eeq

\subsection{Gluonic twist-2 operators}

Let us now consider the extension of the Yang-Mills Lagrangian to include also a general gauge invariant gluonic twist-2 spin-$N$ operator
\begin{equation}
\label{eq:GIoplong}
\O^{(N)}_{\mu_1...\mu_N}(x) = \frac{1}{2} \mathcal{S} \big[F^{a_1}_{\phantom{a_1}\mu\mu_1}\,D_{\mu_{2}}^{a_1a_2}...D_{\mu_{N-1}}^{a_{N-2}a_{N-1}} F^{a_{N-1};\mu}_{\phantom{a_{N-1};\mu}\mu_{N}}\big]+\text{traceless} \,,
\end{equation}
where we have indicated 
\begin{itemize}
\item the sum over all permutations of $\mu_1,...,\mu_N$ via the operation $\mathcal{S}$,
\item and the presence of further terms which make $O^{(N)}_{\mu_1...\mu_N}(x)$ traceless, i.e. the sum vanishes when any two of its Lorentz indices are contracted, by the term `+\text{traceless}'.
\end{itemize}
A well known trick to simplify this expression is to contract it with $N$ identical light-like vectors which we denote by $\Delta^\mu$ and which satisfy $\Delta.\Delta=0$. It is then conventional to introduce the notation
\begin{align}
  F^{\mu;a}=\Delta_\nu\,F^{\mu\nu;a},\qquad  A^a=\Delta_\mu A^{\mu;a},\qquad D=\Delta_\mu\,D^\mu,\qquad \partial=\Delta_\mu\partial^\mu\,.
\end{align}
Using this notation we then define the scalarised version of eq.~(\ref{eq:GIoplong}):
\begin{equation}
\label{def:GIop}
\O_1^{(N)}(x) = \O^{(N)}_{\mu_1...\mu_N}(x) \Delta^{\mu_1}...\Delta^{\mu_N}= \frac{1}{2}\text{Tr}\big[F_{\nu}\,D^{N-2} F^{\nu}\big]\,.
\end{equation}
It is well known that the operator $\O_1^{(N)}(x)$ when inserted into general Green's functions mixes with non-physical operators under renormalisation.
A basis for these non-physical operators will be constructed in the following sections consisting of two kinds of operators, namely operators proportional to the EOM, defined in eq.~(\ref{eq:EOM}), and operators containing ghosts (G). We therefore include besides $\O_1^{(N)}(x)$ also the operators $\O_{\mathrm{EOM}}^{(N)}$ and $\O_{G}^{(N)}(x)$. The complete Lagrangian is then given by
\begin{equation}
  \label{eq:ltilde}
  \tilde{\mathcal{L}}(A,\bar c,c;g,\xi)=\mathcal{L}_{0}+\mathcal{L}_{\text{GF+G}}+\mathcal{C}_1^{(N)} \O^{(N)}_1(x)+\O_{\text{EOM}}^{(N)}+\O_{G}^{(N)}(x)\,,
\end{equation}
where $\mathcal{C}_1^{(N)}$ is the Wilson coefficient associated to $\O^{(N)}_1(x)$. The mass dimension of $\O_1^{(N)}(x)$, $\O_{\text{EOM}}^{(N)}$ and $\O_{G}^{(N)}(x)$ equals the dimension of space-time, $d$, this is achieved by defining $\Delta$ to carry a mass dimension of $2/N-1$.

Let us now briefly discuss the renormalisation of $\tilde{\mathcal{L}}$, which in eq.~(\ref{eq:ltilde}) was defined in terms of \emph{physical} or equivalently \emph{renormalised} fields and couplings. 
The counterterms required to make finite all correlators of the fields $A^a_\mu$, $c^a$ and $\bar{c}^a$ at distinct positions can be readily generated by replacing the fields and couplings with their bare counterparts in $\tilde{\mathcal{L}}(A^b,\bar c^b,c^b;g^b,\xi^b)$ with
\begin{align}
A^{b;\mu}_a(x)&=Z_3^{1/2}A^{\mu}_a(x)\,,&\qquad \bar{c}^{b}_a(x)&=Z_{c}^{1/2}\bar{c}^{b}_a(x)\,,\qquad c^{b}_a(x)=Z_{c}^{1/2}c^{b}_a(x)\,,\nn\\
g^b&=\mu^{\eps} g Z_g\,,&\qquad \xi^b&=\xi Z_3\,.
\end{align}
This replacement is not sufficient to renormalise correlators with an insertion of $\O^{(N)}_1$.
For this purpose it is convenient to introduce the vector notation $\vec{\O}^{(N)}=(\O_1^{(N)},...,\O_n^{(N)})$, with $\O_{1}^{(N)}$ defined in eq.~(\ref{def:GIop}) and the remaining components $\O_{i>1}^{(N)}$, to which we associate Wilson coefficients $\mathcal{C}_i$, forming a basis of operators spanning the space of EOM and ghost operators $\O_{\text{EOM}}^{(N)}$ and $\O_{G}^{(N)}$. The required counterterms are obtained by taking into account the mixing of the operators under renormalisation. This is achieved by making the replacement $$\mathcal{C}_i^{(N),b} = Z_{ji}^{(N)} \mathcal{C}_j^{(N)}\,.$$ The bare Lagrangian then takes the form
\beq
\label{def:OiNren}
\tilde{\mathcal{L}}(A^b,\bar c^b,c^b;\mathcal{C}_i^b,g^b,\xi^b)=\mathcal{L}_{0}(A^b;g^b)+\mathcal{L}_{\text{GF+G}}(A^b,\bar c^b,c^b;g^b,\xi^b)+\sum_{i,j}\mathcal{C}_i^{(N)}Z_{ij}^{(N)} \O_j^{(N),b}
\end{equation}
where the $\O_i^{(N),b}$ denote the operators written in terms of bare couplings and fields. It is well known \cite{Kluberg-Stern:1974nmx,Kluberg-Stern:1975ebk,Joglekar:1975nu,Joglekar:1976eb,Joglekar:1976pe,Henneaux:1993jn} that the structure of $Z_{ij}$ is block triangular in that the physical operator $\O_{1}^{(N)}$ may mix into $\O_{i>1}^{(N)}$ but not vice versa. This is discussed further in section \ref{sec:jogandlee}.

\section{EOM operators and generalised Gauge Symmetry}
\label{sec:EoMGauge}
For the sake of keeping the notation as light as possible we will in the following discuss symmetry properties of the Lagrangian $\tilde \L$ at the level of renormalised fields and parameters. Note that all of these properties can be directly translated to the bare Lagrangian given that it has the same functional form.

\subsection{General formalism}
In this section we will elucidate the general structure of the EOM operator. It is well known that Green's functions are not invariant under field redefinitions 
$$A^a_\mu \to A^a_\mu+\mathcal{G}^a_\mu(A_\alpha,\partial_\alpha A_\beta,\partial_\alpha \partial_\beta  A_\rho,... )\,,$$ where $\mathcal{G}$ is a general local, i.e. polynomial, function of the gauge field $A$ and its derivatives. To leading order in $\mathcal{G}$ the variation of the Yang-Mills action in eq.~(\ref{def:LYM}) can then be written as follows:
\begin{equation}
  \label{eq:varS}
  \delta S_{0} = \int d^dx\, \frac{\delta S_{0}}{\delta A_a^\mu(x)}\, \mathcal{G}^a_\mu(x) =\int d^dx\, \left(D_\nu F^{\nu\mu}\right)^a\,\mathcal{G}^a_\mu(A_\alpha,\partial_\alpha A_\beta,\partial_\alpha \partial_\beta  A_\rho,... )\,.
\end{equation}
For a general form of the function $\mathcal{G}$, as we shall see later, this is actually the most general such EOM operator into which $\O^{(N)}_1$ can mix under renormalisation,  
leading us to write
\begin{equation}
\label{eq:EOMdefG}
 \O_{\text{EOM}}^{(N)}=\left(D_\nu F^{\nu\mu}\right)^a \mathcal{G}^a_\mu(A_\alpha,\partial_\alpha A_\beta,\partial_\alpha \partial_\beta  A_\rho,... )\,.
\end{equation}
A number of constraints on the structure of this EOM operator derive from the overall mass dimension and the twist-2 nature of $\O^{(N)}_1$. This implies that $\mathcal{G}^a_{\mu}$ must be $N$-linear in $\Delta$ and for the mass dimensions to work out the total number of $A$s and $\partial$s entering in every monomial of $\mathcal{G}$ must then equal $N-1$. It follows that $\mathcal{G}$ itself must be proportional to $\Delta$ and that every single $A$ or $\partial$ entering in $\mathcal{G}$ must itself also be contracted with $\Delta$. These considerations therefore pin down the general structure of the EOM operator to be as follows:
\begin{equation}
\label{eq:EOMgeneral1}
\O_{\text{EOM}}^{(N)} =  \left(D .F\right)^a \mathcal{G}^a(A,\partial A,\partial^2 A,... )\,,
\end{equation}
where $\mathcal{G}^a_\mu=\Delta_\mu \mathcal{G}^a$. By expanding $\mathcal{G}$ over all possible monomials which satisfy the power counting constraints we then obtain
\begin{align}
\label{eq:EOMgeneral2}
\O_{\text{EOM}}^{(N)}=\sum_{k=1}^\infty\,\O_{\text{EOM}}^{(N),k}\quad\text{with}\quad\O_{\text{EOM}}^{(N),k} &= g^{k-1}\left(D\cdot F\right)^a\sum_{\substack{i_1+\dots+i_k\\=N-k-1}}\,C^{a;a_1\dots a_k}_{i_1\dots i_k}\,\left(\partial^{i_1}A^{a_1}\right)\dots\left(\partial^{i_k}A^{a_k}\right)\,.
\end{align}
Here the coefficients $C^{a;a_1\dots a_k}_{i_1\dots i_k}$ are in general color-dependent coupling constants which can be further decomposed into some basis of group-invariant color structures. Let us for example consider the case $k=2$, whose general decomposition\footnote{Note we ignore here the fully symmetric rank 3 tensor $d_{abc}$ as it can not appear in Yang-Mills theory.} can be written as
\begin{equation}
\O_{\text{EOM}}^{(N),2} = g\left(D\cdot F\right)^a\sum_{\substack{i_1+i_2\\=N-3}}\,\kappa_{i_1 i_2}\,f^{a\,a_1\,a_2}\,\left(\partial^{i_1}A^{a_1}\right)\left(\partial^{i_2}A^{a_2}\right)\,,
\end{equation}
where the $\kappa_{ij}$s are Wilson coefficients, to be discussed further below. Let us also remark that there exists a general constraint on the $C$-coefficients which derives from the fact that the operators are colour singlets. For general $k$, the coefficients obey the following invariance relation
\begin{align}
\label{eq:colorsingletid}
 C^{b;a_1 \dots a_k}_{i_1\,i_2\dots i_k}f^{b\,a\,x} + C^{a;b \dots a_k}_{i_1\,i_2\dots i_k}f^{b\,a_1\,x}  + \dots + C^{a;a_1 \, a_2 \dots b}_{i_1\,i_2\dots i_k}f^{b\,a_k\,x} = 0.
\end{align}
We now study the symmetry properties of the lagrangian in eq.~(\ref{eq:ltilde}).

\subsection{Generalised Gauge symmetry}
\label{sec:extgauge}
While gauge transformations leave both $\mathcal{L}_{0}$ and $\O_1^{(N)}$ invariant,
the same can not be said about the general EOM operator.
To cancel its variation we will now contruct a generalised gauge transformation,
\beq
\label{Aomegageneralised}
A^a_\mu \to A^a_\mu + \delta_\omega A^a_\mu + \delta_\omega^\Delta A^a_\mu\,,
\eeq
where $\delta_\omega^\Delta$ is multi-linear in $\Delta$ and is such that the gauge variation of $\O_{\text{EOM}}^{(N)}$ is cancelled by the generalised gauge variation of $\L_0$, i.e.
$\delta_\omega^\Delta \L_0$. This leads to
\beq
\label{eq:BRSTidentity}
(D.F)^a_\mu\, \delta_\omega^\Delta A_a^\mu(x)+\delta_\omega \O_{\text{EOM}}^{(N)}=0\,,
\eeq
and combined with eq.~(\ref{eq:EOMgeneral1}) it then follows that the generalised gauge variation satisfies:
\beq
\label{eq:defextgauge}
\delta_\omega^\Delta A^a_\mu + \delta_\omega \mathcal{G}^a_\mu -g\,f^{abc}\,\mathcal{G}^{b}_\mu\omega^c=0\,.
\eeq
Using eqs.~(\ref{eq:EOMgeneral1}) and (\ref{eq:EOMgeneral2}) we then find the following general solution:
\begin{align}
\label{eq:generalisedgaugesolution}
\delta_\omega^\Delta A^a_\mu&=-\Delta_\mu \sum_{k=1}^\infty \sum_{\substack{i_1+\dots+i_k\\N-k-1}}\,\left(\partial^{i_1}A^{a_1}\right)\dots\left(\partial^{i_k+1}\omega^{a_k}\right)\,\sum_{\sigma\in Z_k}\,C^{a;a_{\sigma(1)}\dots a_{\sigma(k)}}_{i_{\sigma(1)}\dots i_{\sigma(k)}}\nonumber\\
 &+g\Delta_\mu \sum_{\substack{i_1+\dots+i_{k+1}\\N-k-2}}\left(\partial^{i_1}A^{a_1}\right)\dots\left(\partial^{i_k}A^{a_k}\right)\left(\partial^{i_{k+1}+1}\omega^{a_{k+1}}\right)\sum_{m=1}^k\,\left(
    \begin{array}{c}
      i_m+i_{k+1}+1\\
      i_m
    \end{array}
    \right)\nonumber\\
&\quad \qquad \qquad \times C^{a;a_1\dots a_{m-1}ba_{m+1}\dots a_k}_{i_1\dots i_m+i_{k+1}+1\dots i_k}f^{b\,a_m\,a_{k+1}}
\end{align}
where we have used eq.~(\ref{eq:colorsingletid}) and symmetry to cancel all terms which contain $\omega$ without derivatives.
By collecting terms of identical field content and powers of $g$ we can bring eq.~(\ref{eq:generalisedgaugesolution}) into the following form
\begin{align}
  \label{eq:deltaAgenk0}
  \delta_\omega^\Delta A^a_\mu &= -\Delta_\mu\; \sum_{k=1}^\infty g^{k-1}\sum_{\substack{i_1+\dots+i_k\\=N-k+1}} \widetilde{C}^{a;a_1\,..\,a_k}_{i_1\,..\,i_k} \left(\partial^{i_1}A^{a_1}\right)..\left(\partial^{i_{k-1}}A^{a_{k-1}}\right)\left(\partial^{i_k+1}\omega^{a_k}\right)\,.
\end{align}
The $\widetilde{C}^{a;a_1\,..\,a_k}_{i_1\,..\,i_k}$ can be extracted from the building blocks of $C^{a;a_1\,..\,a_k}_{i_1\,..\,i_k}$ once a color basis has been specified. We will construct an explicit solution valid up to four loops in the next subsection.

\subsection{Explicit construction up to four loops}
\label{sec:fourloopEOM}
The loop order puts stringent constraints on the type of the EOM operators actually required. The quantity from which we wish to extract the anomalous dimension of $\O^{(N)}_1$ is naturally the gluon 2-point 1PI correlator with an insertion of the operator $\O^{(N)}_1$:
\begin{equation}
  \label{def:GammaNgg}
(\Gamma_{1;gg}^{(N)})^{\mu\nu}_{ab}(Q)=\int d^dx\,d^dz \, e^{iQ.x} \langle 0|T\{A^\mu_a(x)A^\nu_b(0)\O^{(N)}_1(z)\} |0\rangle_{\mathrm{1PI}}.
\end{equation}
At one-loop there are no subdivergences and we only require counterterms with two external gluons. We thus only need the one-loop mixing of $\O^{(N)}_1$ into $\O^{(N),1}_{\mathrm{EOM}}$, as this is the only EOM operator contributing to the two gluon vertex. At two loops we then require two-loop mixing of $\O^{(N)}_1$ into $\O^{(N),1}_{\mathrm{EOM}}$, and one-loop mixing of $\O^{(N)}_1$ into $\O^{(N),2}_{\mathrm{EOM}}$, given that at two loops we can have one-loop subgraphs with three external gluons. This reasoning can be continued at higher loop orders leading to more EOM operators. Diagrams with subgraphs highlighting this pattern are shown in table \ref{tab:EOMmixingdiags} and the corresponding loop numbers, from which we require certain EOMs, are also summarised again in table \ref{tab:EOMmixing}.
\begin{table}[t]
\centering
\begin{tabular}{l c c c c}
\Xhline{3\arrayrulewidth}
$L$ & $\O_{\mathrm{EOM}}^{(N),\le1}$ & $\O_{\mathrm{EOM}}^{(N),\le2}$ & $\O_{\mathrm{EOM}}^{(N),\le3}$ & $\O_{\mathrm{EOM}}^{(N),\le4}$ \phantom{\Bigg\{}\\ [1ex]
\hline
1 & \gluontwoptoneloop & &  &  \\
2 &  \gluontwoptwoloop & \gluonthreeptoneloop &   &   \\
3 & \gluontwoptthreeloop &\gluonthreepttwoloop  &\gluonfourptoneloop &     \\
4 & \gluontwoptfourloop &\gluonthreeptthreeloop  &\gluonfourpttwoloop& \gluonfiveptoneloop          \\
& & & &\\
\Xhline{3\arrayrulewidth}
\end{tabular}
\caption{In the $L$th row the table gives examples of diagrams contributing to the $L$-loop contribution to $\Gamma_{gg}^{(N)}$. Subgraphs whose UV-counterterms require the various EOM operators $\O_{\mathrm{EOM}}^{(N),k}$ are highlighted with dashed boxes.\label{tab:EOMmixingdiags}}
\end{table}
We can therefore ignore terms in eq.~(\ref{eq:EOMgeneral1}) from $k=4$ onwards leading to the following set of EOM operators required up to 4 loops:
\begin{align}
    \label{def:EOMop}
    \O_{\text{EOM}}^{(N),1}&= \eta\;(D.F)^a\;\partial^{N-2}A^a\phantom{\sum_{\substack{i+j=\\N-3}}}\\
    \label{def:EOMop2}
    \O_{\text{EOM}}^{(N),2}&= g(D.F)^a      \sum_{\substack{i+j=\\N-3}}C_{ij}^{abc}(\partial^i A^b)(\partial^j A^c)
\end{align}
\begin{align}
    \label{def:EOMop3}
    \O_{\text{EOM}}^{(N),3}&= g^2(D.F)^a      \sum_{\substack{i+j+k\\=N-4}} C_{ijk}^{abcd}\,  (\partial^{i} A^b)(\partial^{j} A^c)(\partial^{k} A^d)\\
    \label{def:EOMop4}
    \O_{\text{EOM}}^{(N),4}&= g^3(D.F)^a      \sum_{\substack{i+j+k+l\\=N-5}} C_{ijkl}^{abcde}\,  (\partial^{i} A^b)(\partial^{j} A^c)(\partial^{k} A^d)(\partial^{l} A^e)
\end{align}
Let us now discuss the color decomposition of the $C$-coefficients. While at rank two and three possible color structures are limited to $\delta^{ab}$ and $f^{abc}$, color decompositions for operators of higher rank are in general non-trivial, in particular when keeping the color gauge group general as we do here. However the fact that we only require counterterms valid up to certain loop orders imposes strong contstraints and allows us to identify the following color decompositions:
\begin{align}
\label{eq:Cij}
C_{ij}^{abc}&=f^{abc}\kappa_{ij}\\
\label{eq:Cijk}
C_{ijk}^{abcd} &=(ff)^{abcd}\kappa_{ijk}^{(1)}+d_4^{abcd}\kappa_{ijk}^{(2)}+d_{\widehat{4ff}}^{abcd} \kappa_{ijk}^{(3)}\\
\label{eq:Cijkl}
C_{ijkl}^{abcde} &= (fff)^{abcde} \kappa^{(1)}_{ijkl} + d_{4f}^{abcde} \kappa_{ijkl}^{(2)}\,,
\end{align}
where the different color structures are defined as
\begin{table}[t]
\centering
\begin{tabular}{ c c c c c}
\Xhline{3\arrayrulewidth}
\\[-1em]
$L$ & $\O_{EOM}^{(N),1}$ & $\O_{EOM}^{(N),2}$ & $\O_{EOM}^{(N),3}$ & $\O_{EOM}^{(N),4}$ \\
\\[-1em]
\hline
1 & 1 & 0 & 0& 0\\
2 & 2 & 1 & 0& 0\\
3 & 3& 2& 1& 0\\
4 & 4& 3 & 2& 1 \\
\Xhline{3\arrayrulewidth}
\end{tabular}
\caption{The table summarizes the loop orders for which the mixing of $\O_1^{(N)}$ into $\O_{EOM}^{(N),k}$ is required, given a certain loop order of $\Gamma_{gg}^N$.}
\label{tab:EOMmixing}
\end{table}
\begin{align}
(ff)^{abcd}&=f^{abe}f^{cde},& (fff)^{abcde}&=f^{abm}f^{mcn}f^{nde},\quad \nn\\
\label{eq:colordefs}
d_{4f}^{abcde}&=d_4^{abcm}f^{mde},& d_{4ff}^{abcd}&=d_4^{abmn}f^{mce}f^{edn}\,,\\
d_{\widehat{4ff}}^{abcd}&=d_{4ff}^{abcd}-\frac{1}{3}C_A\,d_4^{abcd}\,,& \nn
\end{align}
and the symmetrised trace is defined by
\beq
\label{def:d4abcd}
d_4^{abcd}=\frac{1}{4!}[\mathrm{Tr}(T_A^aT_A^bT_A^cT_A^d)+\text{symmetric permutations}]\,,
\eeq
where $(T_A)^b_{ac}=if^{abc}$. Going beyond four loops would not only require further operators, i.e. $\O_{\mathrm{EOM}}^{(N),k>4}$, but also further color structures in the definitions of $C_{ijk}$ and $C_{ijkl}$. In fact to arbitrary loop-order, there are arbitrarily many independent color structures contributing to $C_{ijk}$ and $C_{ijkl}$. If one was to work in a fixed gauge group this task would be far simpler. For instance in SU($N_c$) we know that the complete basis at 4 and 5  points is expressible in terms of single and double traces of permutations of the generators in the fundamental representation. The penalty for working in an arbitrary gauge group thus becomes increasingly higher at higher loops, but is still mild at the four-loop level.

Let us now come to the definition of the sums appearing in eqs.~(\ref{def:EOMop})-(\ref{def:EOMop4}). These are defined such that we sum over all non-negative integer values of the indices $i,j,k,l$, appearing in the sum which satisfy the respective constraint, e.g. $i+j+k=N-4$. While this sum notation leads to reasonably compact definitions of the EOM operators, it does also lead to overcounting. For instance in the order $g$ term we sum over all indices $i+j=N-3$, but since the associated color tensor, $f^{abc}$, is asymmetric under exchange of $b$ and $c$ the operators appearing in the sum for $j>i$ are related to those with $j<i$. To compensate this over-counting of independent operators we impose relations on the $\kappa$-coefficients. In general there exists a lot of freedom in how to choose these relations. A particularly convenient choice of constraints is obtained by demanding the $\kappa$-coefficients to satisfy the same relations as their respective color factors. That this works can be understood as follows. If we were to use up all the color identities, we would clearly land in an independent basis of operators. By imposing the same identities on the $\kappa$-coefficients, we effectively ensure that solving these identities would lead to the right degrees of freedom - that is the right number of independent $\kappa$-coefficients. This choice is in spirit not dissimilar to the BCJ-choice of numerators \cite{Bern:2008qj} for Feynman diagrams where the numerators of Feynman diagrams are chosen such that they satisfy the same constraints as the corresponding color factors. Here however the motivation is solely to make manipulations with these operators more manageable.

These considerations then finally lead us to impose the following relations on the $\kappa$-coefficients:
\begin{align}
  \label{eq:BCJetaij}
  &\kappa_{ij}+\kappa_{ji}=0,  &(\text{antisymmetry of } f)\\
  \label{eq:BCJeta1ijk}
  &\kappa_{ijk}^{(1)}+\kappa_{ikj}^{(1)}=0,  &(\text{antisymmetry of } f)\\
  \label{eq:BCJetaijkl}
  &\kappa_{ijkl}^{(1)}+\kappa_{ijlk}^{(1)}=0,  &(\text{antisymmetry of } f)\\ 
  \label{eq:BCJeta1ijkJacobi}
  &\kappa_{ijk}^{(1)}+\kappa_{jki}^{(1)}+\kappa_{kij}^{(1)}=0, &(\text{Jacobi}) \\
  \label{eq:BCJetaijklJacobi}
  &\kappa_{ijkl}^{(1)}+\kappa_{iklj}^{(1)}+\kappa_{iljk}^{(1)}=0, &(\text{Jacobi}) \\
  \label{eq:BCJetaijklJacobi2}
  &\kappa_{ijkl}^{(1)}+\kappa_{jilk}^{(1)}+\kappa_{lkji}^{(1)}+\kappa_{klij}^{(1)}=0, &(\text{double Jacobi}) \\
  \label{eq:BCJeta2ijk}
  &\kappa_{ijk}^{(2)}=\kappa_{jik}^{(2)}=\kappa_{ikj}^{(2)}=\kappa_{kji}^{(2)}=\kappa_{jki}^{(2)}=\kappa_{kij}^{(2)},&(\text{symmetry of }d_4)  \\
  \label{eq:BCJeta3ijk}
  &\kappa_{ijk}^{(3)}=\kappa_{ikj}^{(3)}, &(\text{antisymmetry of } f) \\ 
  \label{eq:BCJeta3ijkJacobi}
  &\kappa_{ijk}^{(3)}+\kappa_{jki}^{(3)}+\kappa_{kij}^{(3)}=0, &(\text{generalised Jacobi }) \\
  \label{eq:BCJeta2ijkl}
  &\kappa_{ijkl}^{(2)}+\kappa_{ijlk}^{(2)}=0, &(\text{antisymmetry of } f) \\ 
  \label{eq:BCJeta2ijkl2}
  &\kappa_{ijkl}^{(2)}=\kappa_{jikl}^{(2)}, &(\text{symmetry of }d_4) 
\end{align}
An independent set of operators is then found for any given $N$ by solving these relations. Fixing $N$ this is in principle a straight forward exercise, but is somewat difficult to do keeping $N$ general.

We now give the color identities which lead to eqs.~(\ref{eq:BCJetaij})-(\ref{eq:BCJeta2ijkl2}). The \emph{Jacobi relation} is as usual,
\beq
(ff)^{abcd}+(ff)^{acdb}+(ff)^{adbc}=0\,.
\eeq
By the \emph{double Jacobi relation} we refer to the identity
\begin{equation}
(fff)^{abcde}+(fff)^{acbed}+(fff)^{adebc}+(fff)^{aedcb}=0\,,
\end{equation}
which itself can be derived by repeated use of the Jacobi relation. Another consequence of the Jacobi relation is what is sometimes
refered to as a \emph{generalised Jacobi relation} \cite{vanRitbergen:1998pn}:
\begin{align}
  \label{eq:generalisedJacobi}
d_{4f}^{abcde}+d_{4f}^{bcdae}+d_{4f}^{cdabe}+d_{4f}^{dabce}=0\,. 
\end{align}
This identity does not lead to any relations among the coefficients $\kappa^{(2)}_{ijkl}$, since we fix the position of the index $a$, which contracts the EOM, to be in the $d_4$. The relation would connect it to operators where the $a$ would be attached to the corresponding $f$. However contracting this relation with $f^{deg}$ leads to
\beq
\label{eq:d4ff1}
C_A \,d_4^{abcd}+d_{4ff}^{abcd}+d_{4ff}^{bcad}+d_{4ff}^{cabd}=0\,. 
\eeq
Combining this equation with its permutations, and using the symmetry properties,
\beq
\label{eq:d4ff2}
d_{4ff}^{abcd}=d_{4ff}^{abdc}=d_{4ff}^{bacd}=d_{4ff}^{badc}\,,
\eeq 
which follow directly from the definition in eq.~(\ref{eq:colordefs}), one can further derive the less obvious relation
\begin{equation}
\label{eq:d4ff3}
d_{4ff}^{abcd}=d_{4ff}^{cdab}\,.
\end{equation}
Combining eqs.~(\ref{eq:d4ff1})-(\ref{eq:d4ff3}) we then derive
\begin{equation}
C_A \,d_{4f}^{abcd}+d_{4ff}^{abcd}+d_{4ff}^{acdb}+d_{4ff}^{adbc}=0\,. 
\end{equation}
This relation implies that operators with color structures $d_4$ and $d_{4ff}$ are linearly dependent. To avoid this undesirable feature we introduced the modified color factor $d_{\widehat{4ff}}$ which satisfies
\begin{equation}
d_{\widehat{4ff}}^{abcd}+d_{\widehat{4ff}}^{acdb}+d_{\widehat{4ff}}^{adbc}=0\,,
\end{equation}
and is therefore independent of $d_{4f}^{abcd}$.
Having discussed an explicit basis of the EOM operators and their color structures we can now consider the generalised gauge invariance discussed in section \ref{sec:extgauge}. To 4-loop order the $\Delta$-dependent part of the transformation reads
\begin{align}
  \label{eq:deltaAgenk}
  \delta_w^\Delta A^a_\mu &= -\Delta_\mu\Bigg[\eta\,\left(\partial^{N-1}\omega^a\right)+g\sum_{\substack{i_1+i_2\\=N-3}}\widetilde{C}^{a;a_1\,a_2}_{i_1\,i_2} \left(\partial^{i_1}A^{a_1}\right)\left(\partial^{i_2+1}\omega^{a_2}\right)\nonumber\\
    &+g^2\sum_{\substack{i_1+i_2+i_3\\=N-4}}\widetilde{C}^{a;a_1\,a_2\,a_3}_{i_1\,i_2\,i_3} \left(\partial^{i_1}A^{a_1}\right)\left(\partial^{i_2}A^{a_2}\right)\left(\partial^{i_3+1}\omega^{a_3}\right)\\
    &+g^3\sum_{\substack{i_1+\dots+i_4\\=N-5}}\widetilde{C}^{a;a_1\,a_2\,a_3\,a_4}_{i_1\,i_2\,i_3\,i_4} \left(\partial^{i_1}A^{a_1}\right)\left(\partial^{i_2}A^{a_2}\right)\left(\partial^{i_3}A^{a_3}\right)\left(\partial^{i_4+1}\omega^{a_4}\right)
    +O(g^4)\Bigg],\nn
\end{align}
where $\widetilde{C}^{a;a_1\dots a_n}_{i_1\dots i_n}$ involve the same colour structures which appear in eqs.~(\ref{def:EOMop})-(\ref{def:EOMop4})
\begin{align}
  \label{def:Ctilde2}
  \widetilde{C}^{a;a_1\,a_2}_{i_1\,i_2} &= \eta_{i_1i_2}^{(1)}\,f^{a;a_1\,a_2},\\
  \label{def:Ctilde3}
  \widetilde{C}^{a;a_1a_2a_3}_{i_1i_2i_3} &= \eta^{(1)}_{i_1i_2i_3}\,\left(ff\right)^{aa_1a_2a_3} + \eta^{(2)}_{i_1i_2i_3}\,d^{aa_1a_2a_3} + \eta^{(3)}_{i_1i_2i_3}\,d_{\widehat{4ff}}^{aa_1a_2a_3},\\
  \label{def:Ctilde4}
  \widetilde{C}^{a;a_1a_2a_3a_4}_{i_1i_2i_3i_4} &= \eta^{(1)}_{i_1i_2i_3i_4}\,\left(fff\right)^{aa_1a_2a_3a_4} + \eta^{(2a)}_{i_1i_2i_3i_4}\,d_{4f}^{aa_1a_2a_3a_4} +\eta^{(2b)}_{i_1i_2i_3i_4}\,d_{4f}^{aa_4a_1a_2a_3}\,.
\end{align}
The coefficients $\eta^{(k)}_{i_1\dots i_n}$ are then fixed in terms of the coefficients $\kappa^{(k)}_{i_1\dots i_n}$ of eqs.~(\ref{def:EOMop})-(\ref{def:EOMop4}) by means of eq.~(\ref{eq:generalisedgaugesolution}) and eq.~(\ref{eq:deltaAgenk0}). We obtain the following relations
\allowdisplaybreaks
\begin{align}
  \label{eq:delta_2}
  \eta_{i_1i_2}^{(1)}&=2\,\kappa_{i_1\,i_2} + \eta\Big(
  \begin{array}{c}
    i_1+i_2+1\\
    i_1
  \end{array}
  \Big)\,, \\
  \label{eq:delta1_3}
  \eta^{(1)}_{i_1i_2i_3}&= 2\kappa_{i_1(i_2+i_3+1)}\,\Big(
  \begin{array}{c}
    i_2+i_3+1\\
    i_2
  \end{array}
  \Big) + 2\left[\kappa^{(1)}_{i_1i_2i_3}+\kappa^{(1)}_{i_3i_2i_1}\right],\\
  \label{eq:delta2_3}
  \eta^{(2)}_{i_1i_2i_3} &=3\kappa^{(2)}_{i_1i_2i_3},\\
  \label{eq:delta3_3}
  \eta^{(3)}_{i_1i_2i_3}&=2\left[\kappa^{(3)}_{i_1i_2i_3}-\kappa^{(3)}_{i_3i_2i_1}\right],\\
  \label{eq:delta1_4}
  \eta^{(1)}_{i_1i_2i_3i_4}&= 2\left[\kappa^{(1)}_{i_1i_2(i_3+i_4+1)}+\kappa^{(1)}_{(i_3+i_4+1)i_2i_1}\right]\Big(
  \begin{array}{c}
    i_3+i_4+1\\
    i_3
  \end{array}
  \Big)\nonumber\\
  &+2\left[\kappa^{(1)}_{i_1i_2i_3i_4}+\kappa^{(1)}_{i_1i_4i_3i_2}+\kappa^{(1)}_{i_4i_1i_3i_2}+\kappa^{(1)}_{i_4i_3i_1i_2}\right],\\
  \label{eq:delta2a_4}
  \eta^{(2a)}_{i_1i_2i_3i_4}&= 3\kappa^{(2)}_{i_1i_2(i_3+i_4+1)}\,\Big(
  \begin{array}{c}
    i_3+i_4+1\\
    i_3
  \end{array}
  \Big) + 2\,\kappa^{(2)}_{i_1i_2i_3i_4},\\
  \label{eq:delta2b_4}
  \eta^{(2b)}_{i_1i_2i_3i_4}&=2\,\kappa^{(2)}_{i_4i_1i_2i_3}.
\end{align}
The use and power of these relations will become clear in the next section, where we discuss how the generalised gauge symmetry is
promoted to a generalised BRST symmetry.

\section{Ghost operators and generalised BRST symmetry}
\label{sec:BRST}
\subsection{Generalised BRST symmetry}
The main virtue of the generalised gauge transformation, $\delta_\omega+\delta_\omega^\Delta$, which we established in section \ref{sec:extgauge}, is that we can promote it to a generalised BRST (gBRST) transformation:
\beq
\delta_{\mathrm{BRST}}'(\bullet)\equiv\theta\, s'(\bullet)\,,\qquad  s'= s + s_\Delta.
\eeq
Here $s$ is the action of the usual BRST transformation and $s_\Delta$ is the new $\Delta$-dependent part. To define the action of this symmetry on the fields we follow \cite{Joglekar:1975nu,Mertig:1995ny}. The only non-vanishing action is the variation of the gauge field. It is constructed simply by replacing the gauge parameter $\omega(x)^a$ in eq.~(\ref{eq:deltaAgenk}) with the ghost field $c(x)^a$. We thus obtain
\begin{align}
\label{eq:extbrst}
s_\Delta b^a =0\,,\qquad s_\Delta c^a=0,\qquad s_\Delta\overline{c}^a = 0\,,
\end{align}
and
\beq
\label{eq:extbrstA}
s_\Delta A_\mu^a=-\Delta_\mu \sum_{k=0}^\infty g^{k-1}\sum_{\substack{i_1+\dots+i_k\\=N-k+1}} \widetilde{C}^{a;a_1\,..\,a_4}_{i_1\,..\,i_4}
\left(\partial^{i_1}A^{a_1}\right)..\left(\partial^{i_{k-1}}A^{a_{k-1}}\right)\left(\partial^{i_k+1} c^{a_k}\right)\,.
\eeq
Furthermore eq.~(\ref{eq:defextgauge}) can be promoted to an idenitity for the corresponding BRST variations:
\beq
\label{eq:extbrstrel}
s_\Delta A_a^\mu(x)+s \mathcal{G}_a^\mu-g\,f_{abc} \, \mathcal{G}_b^\mu\,c_c=0\,,
\eeq
This relation is very useful. For instance it allows us to show that $s'$ is nilpotent. In the following we prove this up to terms of order $\Delta^2$, which is all we require for the renormalisation of Green's functions with single insertions of twist-2 operators. First we note that
\beq
\label{eq:nilpotence}
{s'}^2=s^2+ss_\Delta+s_\Delta s+O(\Delta^2)=O(\Delta^2)\,.
\eeq
Given $s^2=0$ we therefore require
\beq
ss_\Delta+s_\Delta s=0\,.
\eeq
To prove this identity it is sufficient to show that it holds for the gauge field. We start with
\begin{align}
 ss_\Delta A_\mu^a&=-s(s \mathcal{G}^a_\mu-g\,f^{abc} \, \mathcal{G}^b_\mu\,c^c)=g\,f^{abc} \, s(\mathcal{G}^b_\mu\,c^c)\nn\\
&=g\,f^{abc} s\mathcal{G}^b_\mu\,c^c-\frac{g^2}{2}\mathcal{G}^b_\mu f^{abc}f^{cde}c^d c^e\\
&=g\,f^{abc} s\mathcal{G}^b_\mu\,c^c-g^2 f^{abc}f^{cde}\mathcal{G}^d_\mu c^b c^e\nn
\end{align}
where we used eq.~(\ref{eq:extbrstrel}) and nilpotence of $s$ in the first line and the Jacobi identity and index relabeling to get to the last line. Next we now consider
\begin{align}
 s_\Delta s A_\mu^a&=s_\Delta (D_\mu^{ac}c^c)=gf^{abc}s_\Delta A_\mu^b c^c\nn\\
 &=-gf^{abc} s \mathcal{G}^b_\mu\, c^c+g^2f^{abc}\,f^{bde} \, \mathcal{G}^d_\mu\,c^e c^c\\
&=-gf^{abc} s \mathcal{G}^b_\mu\, c^c+g^2f^{abc}\,f^{cde} \, \mathcal{G}^d_\mu\,c^b c^e=- ss_\Delta A_\mu^a\nn
 \end{align}
This proves eq.~(\ref{eq:nilpotence}). The nilpotence is thus a direct consequence of the generalised gauge invariance. 

Let us now come to the general form of the gauge-fixing+ghost Lagrangian required to renormalise arbitrary Green's functions with single insertions of $\O_1^{(N)}$. We propose that it can be represented as follows:
\beq
\mathcal{L}_{GF+G}' = s' \O_{\mathrm{ancestor}} \,,
\eeq
where the ancestor is the same one which appears in the usual gauge-fixing and ghost term required for the Yang-Mills Lagrangian; that is the one we defined in eq.~(\ref{eq:ancestor}).  Expanding out $s'$ we thus obtain
\beq
\label{eq:GF+G}
\mathcal{L}_{GF+G}' = \mathcal{L}_{GF+G}+\O_{G}^{(N)} 
\eeq
with 
\beq
\label{eq:gBRSTancestor}
\O_{G}^{(N)}=s_\Delta \O_{\mathrm{ancestor}}=-\overline{c}^a\,\partial^\mu\, s_\Delta A^a_\mu\,.
\eeq
We can then rewrite the complete Lagrangian, introduced in eq.~(\ref{eq:ltilde}), as
\beq
\tilde\L=\L_{\mathrm{EGI}}+\mathcal{L}_{GF+G}'\,,
\eeq
with
\beq
\L_{\mathrm{EGI}}=\L_{\mathrm{YM}}+\O^{(N)}_1+\O_{\mathrm{EOM}}^{(N)}\,.
\eeq
In this formulation the Lagrangian is then manifestly invariant under the generalised BRST transformation $\delta_{\mathrm{BRST}}'$. For $\L_{\mathrm{EGI}}$ this follows immediately from its invariance under generalised gauge transformations. And given the nilpotence of $s'$ it also follows that $\mathcal{L}_{GF+G}'$ is invariant under the symmetry, as it lies in the image of $s'$. Instead $\L_{\mathrm{EGI}}$ lies in the kernel of $s'$. The cohomology of the generalised BRST transformation, defined as the kernel modulo the image of $s'$, is thus unaffected of the details of the gauge fixing function - an important feature which underlies also the usual BRST symmetry.

\subsection{Compatibility with the Theorems of Joglekar and Lee}
\label{sec:jogandlee}
Let us now come to an important issue concerning the mixing between gauge invariant and gauge variant operators. For physics to be independent of the gauge variant operators the renormalisation matrix $Z_{ij}^{(N)}$ should be block triangular. This theorem was proven by Joglekar and Lee \cite{Joglekar:1975nu} and states, essentially, that the block triangular structure is present as long as the unphysical operators belong to two different operator classes\footnote{Note that this is slightly different from the classification into EOM and BRST-exact operators which is often stated and which can be found for instance in \cite{Collins:1994ee}.}:
\begin{itemize}
 \item Class I  operators:
 \beq
 \label{eq:classI}
\O_{\mathrm{I}}= \frac{\delta S}{\delta A^\mu_a} \frac{\delta F[A,c,\bar c]}{\delta(\partial^\mu \bar c^a)}+sF[A,c,\bar c]
 \eeq

 \item Class II operators:
  \beq
\O_{\mathrm{II}}= \frac{\delta S}{\delta c^a} X^a[A,c,\bar c]
 \eeq
\end{itemize}
where $X$ and $F$ are local (polynomial) functionals of the fields.
For our construction this would then imply that $Z_{j1}^{(N)}=0$. However it is not obvious that the ghost and EOM operators presented here fall into these classes. We will now show that they do. Using eq.~(\ref{eq:extbrstrel}) we can write eq.~(\ref{eq:gBRSTancestor}) as follows:
\begin{align}
\label{eq:JoglekarandLee1}
\O_{G}^{(N)}= \overline{c}^a\,\partial^\mu\, s \mathcal{G}^a_\mu
-gf^{abc} \overline{c}^a\,\partial^\mu\, \mathcal{G}^b_\mu c^c\,.
\end{align}
Using now that $s(\bar c^a \partial^\mu\mathcal{G}_\mu^a)=\partial_\mu b^a \mathcal{G}^{\mu;a}+\partial_\mu\bar{c}^a \,s\mathcal{G}^\mu_a$ and the EOM of the $b^a$-field, eq.~(\ref{eq:JoglekarandLee1}) becomes
\begin{align}
\label{eq:JoglekarandLee2}
\O_{G}^{(N)}= -s(\bar c^a \partial\mathcal{G}^a) +\big[\frac{1}{\xi}(\partial\partial_\nu A^{b;\nu})+gf^{abc} (\partial \overline{c}^a)c^c\big] \mathcal{G}^b \,.
\end{align}
Combining this expression with eq.~(\ref{eq:EOMgeneral1}) and eq.~(\ref{eq:GFEOM}) we thus obtain
\beq
\label{eq:jgLeeform}
\O_{G}^{(N)}+\O_{\mathrm{EOM}}^{(N)}=s(\partial \bar c^a \mathcal{G}^a)+ \frac{\delta S}{\delta A_a}\mathcal{G}^a\,.
\eeq
It is thus apparent that our expressions for the ghost and EOM operators are just Class I operators, and therefore comply with the theorems of Joglekar and Lee, if we identify $F=\partial \bar c^a \mathcal{G}^a$ in eq.~(\ref{eq:classI}). In our case the class II operators can not actually contribute due to the leading twist nature. This follows as the ghost EOM $\frac{\delta S}{\delta c^a}$ is already twist $3$. For a similar reason $\mathcal{G}^a$ can also not depend on ghost and anti-ghost fields at twist two.

The structure of the renormalisation matrix $Z_{i\,j}^{(N)}$ is therefore, by the theorem of Joglekar and Lee, expected to be of the form 
\begin{equation}
  \label{eq:Zstructure}
Z^{(N)} = 
\begin{pmatrix}
Z_{1\,1}^{(N)}  & \vline & Z_{1\,2}^{(N)} &..& Z_{1\,n}^{(N)}\\
& \vline& & & \\[-1em]
\hline
& \vline& & & \\[-1em]
0   & \vline & Z_{2\,2}^{(N)} &..& Z_{2\,n}^{(N)}\\
..   & \vline & ... & & .. \\
0   & \vline &Z_{n\,2}^{(N)} &..& Z_{n\,n}^{(N)}\\
\end{pmatrix}\,.
\end{equation}
So while $\O_1^{(N)}$ may mix into the unphysical operators $\O_{i>1}^{(N)}$, the unphysical operators can only mix among themselves.
For calculations of physical quantities, such as S-matrix elements, it is thus fully sufficient to know $Z_{1\,1}^{(N)}$. We require $Z_{1\,i>1}$
only when renormalising Green's functions with insertions of $\O_1^{(N)}$. Instead the $Z_{i>1\,j>1}^{(N)}$ are only required for calculations of Green's functions with insertions of the unphysical operators.

\subsection{Ghost operators up to four loops}
We will now work out the structure of the ghost operator, given in eq.~(\ref{eq:gBRSTancestor}) through four-loop order.
This requires the BRST variation of the gauge field up to 4 loops, which is given by
\begin{align}
  s_\Delta \, A^a_\mu &= -\Delta_\mu\Bigg[\eta\,\left(\partial^{N-1}c^a\right)+g\sum_{\substack{i_1+i_2\\=N-3}}\widetilde{C}^{a;a_1\,a_2}_{i_1\,i_2} \left(\partial^{i_1}A^{a_1}\right)\left(\partial^{i_2+1}c^{a_2}\right)\nonumber\\
    &+g^2\sum_{\substack{i_1+i_2+i_3\\=N-4}}\widetilde{C}^{a;a_1\,a_2\,a_3}_{i_1\,i_2\,i_3} \left(\partial^{i_1}A^{a_1}\right)\left(\partial^{i_2}A^{a_2}\right)\left(\partial^{i_3+1}c^{a_3}\right)\nonumber\\
    &+g^3\sum_{\substack{i_1+\dots+i_4\\=N-5}}\widetilde{C}^{a;a_1\,a_2\,a_3\,a_4}_{i_1\,i_2\,i_3\,i_4} \left(\partial^{i_1}A^{a_1}\right)\left(\partial^{i_2}A^{a_2}\right)\left(\partial^{i_3}A^{a_3}\right)\left(\partial^{i_4+1}c^{a_4}\right)\,
    + O(g^4)\Bigg],
\end{align}
with the coefficients $\widetilde{C}^{a;a_1\,a_2..}_{i_1\,i_2..}$ defined in eqs.~(\ref{def:Ctilde2})-(\ref{def:Ctilde4}) in terms of a range
of $\eta$-coefficients, which in turn are related to the $\kappa$-coefficients, defined in eqs.~(\ref{eq:delta_2})-(\ref{eq:delta2b_4}) and which are attributed to the EOM operators. The ghost operator for arbitrary $N$ as required for calculations up to the four loop level is thus determined to be
\begin{equation}
\label{eq:BRSTop}
    \O_{G}^{(N)}=\sum_k\O_{G}^{(N),k},
\end{equation}
with
\begin{align}
    \label{eq:BRSTop1}
    \O_{G}^{(N),1}&=-\eta\,\left(\partial \overline{c}^a\right)\left(\partial^{N-1}c^a\right),\\
    \label{eq:BRSTop2}
    \O_{G}^{(N),2}&=-g\sum_{\substack{i_1+i_2\\=N-3}}\widetilde{C}^{a;a_1\,a_2}_{i_1\,i_2}\left(\partial \overline{c}^a\right)\, \left(\partial^{i_1}A^{a_1}\right)\left(\partial^{i_2+1}c^{a_2}\right),\\
    \label{eq:BRSTop3}
    \O_{G}^{(N),3}&=-g^2\sum_{\substack{i_1+i_2+i_3\\=N-4}}\widetilde{C}^{a;a_1\,a_2\,a_3}_{i_1\,i_2\,i_3} \left(\partial \overline{c}^a\right)\,\left(\partial^{i_1}A^{a_1}\right)\left(\partial^{i_2}A^{a_2}\right)\left(\partial^{i_3+1}c^{a_3}\right),\\
    \label{eq:BRSTop4}
    \O_{G}^{(N),4}&=-g^3\sum_{\substack{i_1+\dots+i_4\\=N-5}}\widetilde{C}^{a;a_1\,a_2\,a_3\,a_4}_{i_1\,i_2\,i_3\,i_4} \left(\partial \overline{c}^a\right)\,\left(\partial^{i_1}A^{a_1}\right)\left(\partial^{i_2}A^{a_2}\right)\left(\partial^{i_3}A^{a_3}\right)\left(\partial^{i_4+1}c^{a_4}\right).
\end{align}
The ghost operator is therefore completely determined by the generalised BRST symmetry, or, equivalently, by the generalised gauge invariance and the particular form of the gauge-fixing term. Since the couplings appearing in eq.~(\ref{eq:BRSTop}) are determined as linear combinations of an independent set of the $\kappa$-couplings of the EOM operators we can effectively combine the independent parts of the ghost operator with those of the different EOM operators, collecting terms together which share common $\kappa$-coupling coefficients, into what Joglekar and Lee called Class I operators.

One welcome result is thus that the generalised BRST symmetry vastly reduces the independent set of operators one needs to consider. This was of course already oberserved in \cite{Dixon:1974ss} and \cite{Hamberg:1991qt} although it was accounted for in slightly different ways. In \cite{Dixon:1974ss} the relations among the couplings were derived by enforcing the Lie algebra structure on the generalised gauge invariance. Instead in \cite{Hamberg:1991qt} they followed from the generalised BRST invariance of the complete Lagrangian. We like to stress here that in both these references the basis was only considered to two-loop level, and that no connection to EOM operators and BRST exact operators was made. The explicit form of the gauge variant operators and their connection to the ghost operators was thus rather non-trivial and somewhat mysterious. We hope that our presentation finally sheds some light into this long-standing puzzle.

Another advantage of the formalism is that in order to compute the full anomalous dimension mixing matrix we only need to consider the mixing of $\O^{(N)}_1$ into ghost operators, which depending on the method of computation may also require the mixing of the ghost operators among themselves. This is of course much easier to compute then the mixing of $\O^{(N)}_1$ into the EOM operators, whose renormalisation  would naively require the computation of multi-gluon correlators. Instead the anomalous dimensions of ghost operators can be extracted from multi-gluon correlators with a ghost anti-ghost pair; which yields a welcome reduction of complexity. This point will be discussed in more detail in section \ref{sec:Applications} with reference to specific examples.

\subsection{Generalised anti-BRST symmetry}
In the following we will discuss a rather remarkable fact: there exists a second formulation of the generalised gauge-fixing and ghost lagrangian $\mathcal{L}_{GF+G}'$ introduced in eq.~(\ref{eq:GF+G}). Rather than writing it as a gBRST-exact operator  we can write it as an anti-gBRST exact operator with the anti-ancestor operator defined in eq.~(\ref{eq:anti-ancestor}):
\beq
\mathcal{L}_{GF+G}'=\bar{s}'\bar \O_{\mathrm{ancestor}}=\mathcal{L}_{GF+G}+\O_{G}^{(N)}\,.
\eeq
where $\bar{s}'=\bar{s}+\bar{s}_\Delta$ and the anti-gBRST transformation is defined as a generalised gauge transformation with $\omega(x)\to \bar c(x)$, that is
\begin{align}
  \overline{s}_\Delta c^a &= 0,\qquad\overline{s}_\Delta \bar c^a = 0,\qquad \overline{s}_\Delta \bar b^a = 0,\nn\\
  \overline{s}_\Delta A^a_\mu &= -\Delta_\mu\Bigg[\eta\,\left(\partial^{N-1}\overline{c}^a\right)+g\sum_{\substack{i_1+i_2\\=N-3}}\widetilde{C}^{a;a_1\,a_2}_{i_1\,i_2} \left(\partial^{i_1}A^{a_1}\right)\left(\partial^{i_2+1}\overline{c}^{a_2}\right)\nonumber\\
    &+g^2\sum_{\substack{i_1+i_2+i_3\\=N-4}}\widetilde{C}^{a;a_1\,a_2\,a_3}_{i_1\,i_2\,i_3} \left(\partial^{i_1}A^{a_1}\right)\left(\partial^{i_2}A^{a_2}\right)\left(\partial^{i_3+1}\overline{c}^{a_3}\right)\\
    &+g^3\sum_{\substack{i_1+\dots+i_4\\=N-5}}\widetilde{C}^{a;a_1\,a_2\,a_3\,a_4}_{i_1\,i_2\,i_3\,i_4} \left(\partial^{i_1}A^{a_1}\right)\left(\partial^{i_2}A^{a_2}\right)\left(\partial^{i_3}A^{a_3}\right)\left(\partial^{i_4+1}\overline{c}^{a_4}\right)+O(g^4)
    \Bigg]\nn,
\end{align}
The fact that it is possible to define an anti-gBRST transformation and use it to construct the ghost operator may not be too surprising given that this was also possible for the usual renormalisable gauge-fixing+ghost Lagrangian. What is more surprising is that the ghost operator generated by the anti-gBRST exact operator,
\begin{align}
\label{eq:antiBRSTop}
\O_{G}^{(N)} &= -c^a\,\partial^\mu\,\overline{s}_\Delta A^a_\mu,\\
& = \left(\partial c^a\right)\Bigg[\eta\,\left(\partial^{N-1}\overline{c}^a\right)+g\sum_{\substack{i_1+i_2\\=N-3}}\widetilde{C}^{a;a_1\,a_2}_{i_1\,i_2} \left(\partial^{i_1}A^{a_1}\right)\left(\partial^{i_2+1}\overline{c}^{a_2}\right)\nonumber\\
&+g^2\sum_{\substack{i_1+i_2+i_3\\=N-4}}\widetilde{C}^{a;a_1\,a_2\,a_3}_{i_1\,i_2\,i_3} \left(\partial^{i_1}A^{a_1}\right)\left(\partial^{i_2}A^{a_2}\right)\left(\partial^{i_3+1}\overline{c}^{a_3}\right)\nonumber\\
&+g^3\sum_{\substack{i_1+\dots+i_4\\=N-5}}\widetilde{C}^{a;a_1\,a_2\,a_3\,a_4}_{i_1\,i_2\,i_3\,i_4} \left(\partial^{i_1}A^{a_1}\right)\left(\partial^{i_2}A^{a_2}\right)\left(\partial^{i_3}A^{a_3}\right)\left(\partial^{i_4+1}\overline{c}^{a_4}\right)
+O(g^4)    \Bigg],\nn
\end{align}  
is at first sight not equivalent to its gBRST generated cousin. Equating the two with each other,
\beq
c^a\,\partial^\mu\,\overline{s}_\Delta A^a_\mu=\overline{c}^a\,\partial^\mu\, s_\Delta A^a_\mu\,,
\eeq
therefore generates non-trivial identities among the various $\eta$-coefficients. 
Identifying the RHS of eqs.~(\ref{eq:BRSTop}) and (\ref{eq:antiBRSTop}) we then find, after using integration by parts and the product rule, the following relation:
\begin{align}
&0 =  \sum_{\substack{i_1+\dots +i_n\\=N-n-1}}\left(\partial\overline{c}^a\right)\left(\partial^{i_1}A^{a_1}\right)\dots\left(\partial^{i_{n-1}}A^{a_{n-1}}\right)\left(\partial^{i_n+1}c^{a_n}\right)\Bigg\{\widetilde{C}^{a;a_1\dots a_n}_{i_1\dots i_n}-\\
&  \sum_{s_1=0}^{i_1}\dots\sum_{s_{n-1}=0}^{i_{n-1}} \frac{(s_1+\dots+s_{n-1}+i_n)!}{s_1!\dots s_{n-1}!i_n!}
  \times(-1)^{s_1+\dots s_{n-1}+i_n}\widetilde{C}^{a_n;a_1\dots a_{n-1}a}_{(i_1-s_1)\dots(i_{n-1}-s_{n-1})(i_n+s_1+\dots+s_{n-1}})\Bigg\}\,.\nonumber
\end{align}
Let us remark that the summand does not necessarily vanish independently here. It does only as long as the field contents and its derivatives are independent in each term in the sum. One therefore has to be careful when applying this identity.
Using the definitions in eqs.~(\ref{def:Ctilde2})-(\ref{def:Ctilde4}) we then derive the following set of constraints on the couplings of the ghost operators:
\allowdisplaybreaks
\begin{align}
  \label{eq:antiBRSTrel1}
  \eta^{(1)}_{i_1i_2} &=-\sum_{s_1=0}^{i_1}(-1)^{s_1+i_2}\,\Big(
  \begin{array}{c}
    s_1+i_2\\
    s_1
  \end{array}
  \Big)\,\eta^{(1)}_{(i_1-s_1)(i_2+s_1)}, \\
  \label{eq:antiBRSTrel2}
  \eta^{(1)}_{i_1i_2i_3}&=\sum_{s_1=0}^{i_1}\sum_{s_2=0}^{i_2}\frac{(s_1+s_2+i_3)!}{s_1!s_2!i_3!}(-1)^{s_1+s_2+i_3}\,\eta^{(1)}_{(i_2-s_2)(i_1-s_1)(i_3+s_1+s_2)},\\
  \label{eq:antiBRSTrel3}
  \eta^{(2)}_{i_1i_2i_3}&=\sum_{s_1=0}^{i_1}\sum_{s_2=0}^{i_2}\frac{(s_1+s_2+i_3)!}{s_1!s_2!i_3!}(-1)^{s_1+s_2+i_3}\,\eta^{(2)}_{(i_1-s_1)(i_2-s_2)(i_3+s_1+s_2)},\\
  \label{eq:antiBRSTrel4}
  \eta^{(3)}_{i_1i_2i_3}&=\sum_{s_1=0}^{i_1}\sum_{s_2=0}^{i_2}\frac{(s_1+s_2+i_3)!}{s_1!s_2!i_3!}(-1)^{s_1+s_2+i_3}\,\eta^{(3)}_{(i_2-s_2)(i_1-s_1)(i_3+s_1+s_2)},\\ 
  \label{eq:antiBRSTrel5}
  \eta^{(1)}_{i_1i_2i_3i_4}&=-\sum_{s_1=0}^{i_1}\sum_{s_2=0}^{i_2}\sum_{s_3=0}^{i_3}\frac{(s_1+s_2+s_3+i_4)!}{s_1!s_2!s_3!i_4!}\nonumber\\
  &\phantom{-\sum_{s_1=0}^{i_1}\sum_{s_2=0}^{i_2}\sum_{s_3=0}^{i_3}}\times(-1)^{s_1+s_2+s_3+i_4}\,\eta^{(3)}_{(i_3-s_3)(i_2-s_2)(i_1-s_1)(i_4+s_1+s_2+s_3)},\\
  \label{eq:antiBRSTrel6}
  \eta^{(2a)}_{i_1i_2i_3i_4}&=-\sum_{s_1=0}^{i_1}\sum_{s_2=0}^{i_2}\sum_{s_3=0}^{i_3}\frac{(s_1+s_2+s_3+i_4)!}{s_1!s_2!s_3!i_4!}\nonumber\\
  &\phantom{-\sum_{s_1=0}^{i_1}\sum_{s_2=0}^{i_2}\sum_{s_3=0}^{i_3}}\times(-1)^{s_1+s_2+s_3+i_4}\,\eta^{(2a)}_{(i_1-s_1)(i_2-s_2)(i_3-s_3)(i_4+s_1+s_2+s_3)},\\
  \label{eq:antiBRSTrel7}
  \eta^{(2b)}_{i_1i_2i_3i_4}&=\eta^{(2a)}_{i_1i_3i_2i_4}-\eta^{(2a)}_{i_1i_2i_3i_4}+\sum_{s_1=0}^{i_1}\sum_{s_2=0}^{i_2}\sum_{s_3=0}^{i_3}\frac{(s_1+s_2+s_3+i_4)!}{s_1!s_2!s_3!i_4!}\nonumber\\
  &\phantom{\eta^{(2a)}_{i_1i_3i_2i_4}-\eta^{(2a)}_{i_1i_2i_3i_4}}\times(-1)^{s_1+s_2+s_3+i_4}\,\eta^{(2b)}_{(i_1-s_1)(i_2-s_2)(i_3-s_3)(i_4+s_1+s_2+s_4)}.
\end{align}
To the best of our knowledge the existence of these kind of identities was not known by the authors of the previous works \cite{Dixon:1974ss,Hamberg:1991qt}. 
But we can use their one-loop all-$N$ results for $\eta_{ij}^{(1)}$, which in their work was named $\eta_i$ to check eq.~(\ref{eq:antiBRSTrel1}) at this order. Their one-loop result in our notation is given by
\beq
\eps\eta_{ij}^{(1),1}= \frac{1}{2N(N-1)}\Big[(-1)^i- 3\Big(
  \begin{array}{c}
    N-2\\
    i
  \end{array}
  \Big)-\Big(
  \begin{array}{c}
    N-2\\
    i+1
  \end{array}
  \Big)  \Big]\,.
\eeq
Substituting this result into eq.~(\ref{eq:antiBRSTrel1}) we then find:
\beq
\eta^{(1),1}_{ij} +\sum_{s=0}^{i}(-1)^{s+j}\,\Big(
\begin{array}{c}
    s+j\\
    s
\end{array}
\Big)\,\eta^{(1),1}_{(i-s)(j+s)}
= 
 \frac{3}{2\eps}\frac{(-1)^N-1 } {N(N-1)}\Big(
  \begin{array}{c}
    N-2\\
    1+i
  \end{array}
  \Big)
\eeq
The right hand side thus indeed vanishes for all positive even values of $N$, as required.

We initially found these identities after inspecting the results of explicit calculations. We could explain the extra relations by imposing a ghost-antighost exchange symmetry; whose origin we then finally derived as a consequence of the anti-gBRST symmetry. Since the $\eta$-coefficients can be written in terms of the $\kappa$-coefficients it then follows that the set of $\kappa$-coefficients associated to the different EOM operators is not actually independent. That is there are nontrivial relations among the EOM-operators. To solve these relations in closed form is in general difficult but it is not too hard to solve them for fixed $N$ on a case-by-case basis. We will give examples and demonstrate the use of these relations in section \ref{sec:operatorbases} where we construct minimal bases of operators for the lowest values of $N$, and study the size of the basis for higher $N$.

\section{Operator Bases Construction for fixed $N$}
\label{sec:operatorbases}
In this section we will construct explicit bases of the unphysical (EOM and ghost) operators which can mix with the gauge-invariant operator $\O^{(N)}_1$ for fixed $N$, valid up to the four-loop level. The structure of the EOM operator $\O_{\text{EOM}}^{(N),k}$ up to four loops was discussed in section \ref{sec:fourloopEOM}.
As explained there, we only require $k\leq4$ for general $N$ when working up to four loops. The corresponding EOM operators were presented in eqs.~(\ref{def:EOMop})- (\ref{def:EOMop4}). Another constraint on $k$ arises for fixed $N$ since we only have a total budget of $N-1$ $\partial$s and $A$s to spend in the $\mathcal{G}$-function multiplying the EOM in eq.~(\ref{eq:EOMgeneral2}). Since $\O_{\text{EOM}}^{(k)}$ requires at least $k$ $As$, this leads to $k\leq N-1$.

Being determined by the gBRST symmetry explicit expressions for all ghost operators required up to four loops are given in eqs.~(\ref{eq:BRSTop1})-(\ref{eq:BRSTop4}). Their color decompositions are given in eqs.(\ref{def:Ctilde2})-(\ref{def:Ctilde4}), with the Wilson coefficients being related to those of the EOM operators via eqs.(\ref{eq:delta_2})-(\ref{eq:delta2b_4}). Finally, the generalised anti-BRST symmetry imposes further constraints given in eqs.(\ref{eq:antiBRSTrel1})-(\ref{eq:antiBRSTrel7}), reducing unphysical operators to a yet smaller basis. Having all these definitions at our deposal we are now in a position to construct explicit and minimal bases. In the remaining part of this section we provide explicitly the bases that are relevant for the renormalisation of $\O^{(N)}_1$, with $N=2,4$ and $N=6$, and we describe the space of independent operators for higher $N$.

\subsection{$N=2$ operators}
The construction of a basis of unphysical operators mixing with $\O_1^{(2)}$, which has dimension 4, is straightforward. There is a single EOM operator, $\O_{\text{EOM}}^{(2),1}$, defined in eq.~(\ref{def:EOMop}). The corresponding ghost operator, $\O_{G}^{(2),1}$, is given in eq.~(\ref{eq:BRSTop1}). They read
\begin{equation}
  \O_{\text{EOM}}^{(2),1} = \eta\,(D.F)^a\,A^a,\qquad  \O_{G}^{(2),1} = -\eta\,\partial\bar{c}^a\partial c^a.
\end{equation}
We note that both operators feature the same coupling constant, $\eta$, which follows from the generalised BRST symmetry. In practice, this fact has important consequences for renormalisation, because it implies that the $\O_{\text{EOM}}^{(2)}$ and $\O_{G}^{(2)}$ mix with $\O^{(N)}_1$ with the {\textit{same}} counterterm. In other words, we find only one unphysical operator mixing with the gauge invariant operator of $N=2$. Following the vector notation introduced in sec.\ref{sec:background}, the twist-2 operators of dimension 4 are written as $\vec{\O}^{(2)}=(\O_1^{(2)},\O_2^{(2)})$ with
\begin{align}
  \label{eq:O2basis1}
  \O_1^{(2)}&=\frac{1}{2}\text{Tr}\big[F_{\nu} F^{\nu}\big],\\
  \label{eq:Oeta(N=2)}
    \O_{2}^{(2)}&=(D.F)^a\,A^a+\overline{c}^a\partial^2c^a.
\end{align}

\subsection{$N=4$ operators}
The mass dimension-6 operator $\O_1^{(4)}$ undergoes a less trivial mixing pattern. All EOM operators in $\O_{\text{EOM}}^{(4),k}$ with $k\leq3$ are relevant and each sector generates associated ghost operators. As for the case $N=2$, we can readily write down the EOM operator $\O_{\text{EOM}}^{(N),1}$ and its associated ghost operator
\begin{align}
  \label{eq:O4EOM1}
  \O_{\text{EOM}}^{(4),1} &= \eta\,(D.F)^a\,\partial^2A^a,\\
  \label{eq:O4G1}
  \O_{G}^{(4),1}&=-\eta\,\partial \bar{c}^a\,\partial^3c^a.
\end{align}
Next we consider $\O_{\text{EOM}}^{(4),2}$, eq.~(\ref{def:EOMop2}), which involves only one operator, due to the antisymmetry of the coefficients $\kappa_{ij}$, eq.~(\ref{eq:BCJetaij}). It reads
\begin{align}
  \label{eq:O4EOM2}
  \O_{\text{EOM}}^{(4),2} &= 2\,g\kappa_{01}\,f^{aa_1a_2}\,(D.F)^a\,A^{a_1}\,\partial A^{a_2}.
\end{align}
The ghost operator, $\O_{G}^{(4),2}$, is defined in eq.~(\ref{eq:BRSTop2}) in terms of the coefficients $\tilde{C}^{aa_1a_2}_{i_1i_2}$ of eq.~(\ref{def:Ctilde2}) as
\begin{align}
  \O_{G}^{(4),2}&= -gf^{aa_1a_2}\,\partial \bar{c}^a\Big[\eta^{(1)}_{01}\,A^{a_1}\,\partial^2 c^{a_2} + \eta^{(1)}_{10} \partial A^{a_1}\,\partial c^{a_2}\Big].
\end{align}
The generalised BRST symmetry imposes that the coefficients $\eta^{(1)}_{01}$ and $\eta^{(1)}_{10}$ are related to the parameters in $\O_{\text{EOM}}^{(4),1}$ and $\O_{\text{EOM}}^{(4),2}$, respectively $\eta$ and $\kappa_{01}$. These relations are given in eq.~(\ref{eq:delta_2}) and lead to
\begin{align}
  \label{eq:O4G2}
 \O_{G}^{(4),2} &=- gf^{aa_1a_2} \Big[\eta\,\partial\bar{c}^a\Big(2\partial A^{a_1}\, \partial c^{a_2} + A^{a_1}\partial^2c^{a_2}\Big)   + 2\kappa_{01}\,\partial\bar{c}^a\Big(A^{a_1}\,\partial^2c^{a_2}-\partial A^{a_1}\, \partial c^{a_2}\Big) \Big].
\end{align}
The unphysical operators in eqs.~(\ref{eq:O4EOM1}), (\ref{eq:O4G1}), (\ref{eq:O4EOM2}) and (\ref{eq:O4G2}) contribute to the renormalisation of $\O_1^{(4)}$ starting from two loops \cite{Dixon:1974ss,Hamberg:1991qt}. From three loops onwards, we must take into account also the EOM operator $\O_{\text{EOM}}^{(4),3}$, eq.~(\ref{def:EOMop3}), which reads
\begin{align}
  \label{eq:O4EOM3}
  \O_{\text{EOM}}^{(4),3} &= g^2\kappa^{(2)}_{000}\,d^{aa_1a_2a_3}\,(D.F)^a\,A^{a_1}A^{a_2}A^{a_3},
\end{align}
where we applied eqs.~(\ref{eq:BCJeta1ijk}), (\ref{eq:BCJeta1ijkJacobi}), (\ref{eq:BCJeta2ijk}), (\ref{eq:BCJeta3ijk}) and (\ref{eq:BCJeta3ijkJacobi}) to restrict the independent couplings to a single operator at mass dimension $6$. Due to the fully symmetric nature of the colour structure $d^{aa_1a_2a_3}$ of eq.~(\ref{eq:O4EOM3}),  we find that two-point correlators with an insertion of $\O_{\text{EOM}}^{(4),3}$ vanish automatically at one and at two loops. This implies that $\O_{\text{EOM}}^{(4),3}$ enters the renormalisation of $\O_1^{(N)}$ only at four loops. We derive the ghost operator $\O_{G}^{(4),3}$, eq.~(\ref{eq:BRSTop3}), by computing the coefficients (\ref{eq:delta1_3})-(\ref{eq:delta3_3}), which enter $\widetilde{C}^{aa_1a_2a_3}_{i_1i_2i_3}$ in eq.~(\ref{def:Ctilde3}). We get
\begin{align}
  \O_{G}^{(4),3}&=-g^2\,\partial\bar{c}^a\,A^{a_1}A^{a_2}\,\partial c^{a_3}\Big[\eta^{(1)}_{000}(ff)^{aa_1a_2a_3}+\eta^{(2)}_{000}d^{aa_1a_2a_3}+\eta^{(3)}_{000}d^{aa_1a_2a_3}_{\widehat{4ff}}\Big]\nonumber\\
  \label{eq:O4G3}
    &=-g^2\,\partial\bar{c}^a\,A^{a_1}A^{a_2}\,\partial c^{a_3}\,\Big[2\kappa_{01}\,(ff)^{aa_1a_2a_3}+3\kappa^{(2)}_{000}\,d^{aa_1a_2a_3}\Big].
\end{align}
By taking into account only the relations deriving from the generalised BRST symmetry, we obtained a set of three unphysical operators, each of them corresponding to the terms in eqs.~(\ref{eq:O4EOM1})-(\ref{eq:O4EOM2}) and (\ref{eq:O4G2})-(\ref{eq:O4G3}) that are proportional to the coefficients $\eta$, $\kappa_{01}$ and $\kappa^{(2)}_{000}$, respectively.
However, the generalised anti-BRST symmetry introduces an additional constraint on these coefficients and reduces the set of independent operators further. By specialising $i_1,i_2=0,1$ in eq.~(\ref{eq:antiBRSTrel1}) we find
\begin{equation}
  \label{eq:O4antiBRSTrel}
  2\eta^{(1)}_{10} = \eta^{(1)}_{01} \iff 2\kappa_{01} = \eta.
\end{equation}
This identity is surprising, because it relates the couplings of different EOM operators, which are free a priori. Therefore the EOM and ghost Lagrangian feature only two independent parameters, e.g. $\eta$ and $\kappa^{(2)}_{000}$, which are chosen as coupling constants of two independent unphysical operators. In conclusion, we obtain a basis of operators with spin 4 (and dimension 6) $\vec{\O}^{(4)}=(\O_1^{(4)},\O_2^{(4)},\O_3^{(4)})$ with
\begin{align}
  \label{eq:basisO41}
  \O_1^{(4)}&=\frac{1}{2}\text{Tr}\big[F_{\nu}D^2F^{\nu}\big],\\
  \label{eq:basisO42}
  \O_2^{(4)}&=(D.F)^a\Big[ \partial^2A^a + gf^{aa_1a_2}A^{a_1}\partial A^{a_2}\Big]-\partial \bar{c}^a\,\partial^3c^a - gf^{aa_1a_2}\,\partial \bar{c}^a\Big[2A^{a_1}\partial^2 c^{a_2}+\partial A^{a_1}\,\partial c^{a_2}\Big]\nonumber\\
  &-g^2(ff)^{aa_1a_2a_3}\,\partial \bar{c}^a\,A^{a_1}A^{a_2}\,\partial c^{a_3},\\
  \label{eq:basisO43}
  \O_3^{(4)}&=d^{aa_1a_2a_3}\Big[(D.F)^aA^{a_1}A^{a_2}A^{a_3}-3\,\partial \bar{c}^a\,A^{a_1}A^{a_2}\,\partial c^{a_3}\Big].
\end{align}

\subsection{$N=6$ operators}
The basis of operators at mass dimension 8, which includes the gauge invariant operator $\O_1^{(6)}$, is generated by the full set of operators $\O_{\text{EOM}}^{(6),k}$, with $k\leq 4$. For $k=1$ we get immediately the EOM and ghost operators
\begin{align}
  \label{eq:O6EOM1G1}
  \O_{\text{EOM}}^{(6),1}&=\eta\,(D.F)^a\partial^4 A^a, &\O_{G}^{(6),1}=-\eta\,\partial\bar{c}^a\,\partial^5c^a.
\end{align}
The definition in eq.~(\ref{def:EOMop2}) and antisymmetry of the coefficents $\kappa_{ij}$, eq.~(\ref{eq:BCJetaij}), imply that $\O_{\text{EOM}}^{(6),2}$ includes two independent terms
\begin{align}
  \label{eq:O6EOM2}
  \O_{\text{EOM}}^{(6),2}&=gf^{aa_1a_2}\,(D.F)^a\Big[2\kappa_{03} A^{a_1}\partial^3A^{a_2}+2\kappa_{12}(\partial A^{a_1})\partial^2A^{a_2}\Big].
\end{align}
To get the ghost sector $\O_{G}^{(6),2}$, we expand out eq.~(\ref{eq:BRSTop2}) with $i_1,i_2=0,\dots,3$  and we use the definitions in eqs.~(\ref{def:Ctilde2}) and (\ref{eq:delta_2}), to get
\begin{align}
  \label{eq:O6G2}
  \O_{G}^{(6),2}=&-gf^{aa_1a_2}\,\partial \bar{c}^a\Big\{\eta\Big[A^{a_1}\partial^4c^{a_2}+4\,\partial A^{a_1}\,\partial^3c^{a_2}+6\,\partial^2 A^{a_1}\partial^2c^{a_2}+4\,\partial^3A^{a_1}\partial c^{a_2}\Big]\nonumber\\&+2\kappa_{03}\Big[A^{a_1}\partial^4 c^{a_2}-\partial^3 A^{a_1}\,\partial c^{a_2}\Big] +2\kappa_{12}\Big[\partial A^{a_1}\,\partial^3c^{a_2}-\partial^2 A^{a_1}\,\partial^2c^{a_2}\Big]  \Big\}.
\end{align}
Similarly, we write down the operator $\O_{\text{EOM}}^{(6),3}$, following the definition in eq.~(\ref{def:EOMop3}) and the relations eq.~(\ref{eq:BCJeta1ijk}), (\ref{eq:BCJeta1ijkJacobi}) and (\ref{eq:BCJeta2ijk})-(\ref{eq:BCJeta3ijkJacobi}) on the coefficients, to obtain
\begin{align}
  \label{eq:O6EOM3}
  \O_{\text{EOM}}^{(6),3}=&+2g^2\,(ff)^{aa_1a_2a_3}(D.F)^a\Big[\kappa^{(1)}_{002}\,A^{a_1}A^{a_2}\,\partial^2A^{a_3} + \kappa^{(1)}_{101}\,\partial A^{a_1}\,A^{a_2}\partial A^{a_3}\Big]\nonumber\\
  &+3g^2\,d^{aa_1a_2a_3}(D.F)^a\Big[\kappa^{(2)}_{002}A^{a_1}A^{a_2}\partial^2A^{a_3}+\kappa^{(2)}_{011}A^{a_1}\,\partial A^{a_2}\partial A^{a_3}\Big]\nonumber\\
  &+2g^2\,d^{aa_1a_2a_3}_{\widehat{4ff}}(D.F)^a\Big[\kappa^{(3)}_{002}(A^{a1}A^{a_2}\partial^2A^{a_3}-\partial^2 A^{a_1}\,A^{a_2}A^{a_3})\nonumber\\
  &\phantom{2g^2\,d^{aa_1a_2a_3}_{\widehat{4ff}}(D.F)^a\Big[\;}+\kappa^{(3)}_{101}(\partial A^{a_1}\,A^{a_2}\partial A^{a_3}-A^{a_1}\partial A^{a_2}\,\partial A^{a_3})\Big].
\end{align}
By expanding out eq.~(\ref{eq:BRSTop3}) for $i_1,i_2,i_3=0\dots2$ and by using the definitions in eqs.~(\ref{def:Ctilde3}), (\ref{eq:delta1_3}), (\ref{eq:delta2_3}) and (\ref{eq:delta3_3}) we obtain the related ghost operator
\begin{align}
  \label{eq:O6G3}
  \O_{G}^{(6),3}=&-2g^2\kappa^{(1)}_{03}\,(ff)^{aa_1a_2a_3}\,\partial\bar{c}^a\Big[A^{a_1}A^{a_2}\partial^3c^{a_3}+3A^{a_1}\partial A^{a_2}\,\partial^2c^{a_3}+3A^{a_1}\partial^2A^{a_2}\,\partial c^{a_3}\Big]\nonumber\\
  &-2g^2\kappa^{(1)}_{12}\,(ff)^{aa_1a_2a_3}\,\partial\bar{c}^a\Big[\partial A^{a_1}\,A^{a_2}\partial^2c^{a_3}-\partial^2A^{a_1}\,A^{a_2}\partial c^{a_3}+2\,\partial A^{a_1}\,\partial A^{a_2}\,\partial c^{a_3} \Big]\nonumber\\
  &-2g^2\kappa^{(1)}_{002}\,(ff)^{aa_1a_2a_3}\,\partial\bar{c}^a\Big[A^{a_1}A^{a_2}\partial^3c^{a_3}+\partial^2 A^{a_1}\,A^{a_2}\partial c^{a_3}-2A^{a_1}\partial^2A^{a_2}\,\partial c^{a_3}\Big]\nonumber\\
  &-2g^2\kappa^{(1)}_{101}\,(ff)^{aa_1a_2a_3}\,\partial\bar{c}^a\Big[2\,\partial A^{a_1}\,A^{a_2}\partial^2c^{a_3}-A^{a_1}\partial A^{a_2}\,\partial^2c^{a_3}-\partial A^{a_1}\,\partial A^{a_2}\,\partial c^{a_3}\Big]\nonumber\\
  &-3g^2\kappa^{(2)}_{002}\,d^{aa_1a_2a_3}\,\partial\bar{c}^a\Big[A^{a_1}A^{a_2}\partial^3c^{a_3}+2A^{a_1}\partial^2A^{a_2}\,\partial c^{a_3}\Big]\nonumber\\
  &-3g^2\kappa^{(2)}_{011}\,d^{aa_1a_2a_3}\,\partial\bar{c}^a\Big[\partial A^{a_1}\,\partial A^{a_2}\,\partial c^{a_3}+2A^{a_1}\partial A^{a_2}\,\partial^2c^{a_3} \Big]\nonumber\\
  &-6g^2\kappa^{(3)}_{002}\,d^{aa_1a_2a_3}_{\widehat{4ff}}\,\partial\bar{c}^a\Big[A^{a_1}A^{a_2}\partial^3c^{a_3}-\partial^2A^{a_1}\,A^{a_2}\partial c^{a_3}\Big]\nonumber\\
  &-6g^2\kappa^{(3)}_{101}\,d^{aa_1a_2a_3}_{\widehat{4ff}}\,\partial\bar{c}^a\Big[\partial A^{a_1}\,\partial A^{a_2}\,\partial c^{a_3}-A^{a_1}\partial A^{a_2}\,\partial^2c^{a_3}\Big].
\end{align}
We construct $\O_{\text{EOM}}^{(6),4}$ and $\O_{G}^{(6),4}$, by expanding out eqs.~(\ref{def:EOMop4}) and (\ref{eq:BRSTop4}) with $i_1\dots i_4=0,1$. After imposing the relations in eqs.(\ref{eq:BCJetaijkl}), (\ref{eq:BCJetaijklJacobi}), (\ref{eq:BCJetaijklJacobi2}), (\ref{eq:BCJeta2ijkl}) and (\ref{eq:BCJeta2ijkl2}), which constrain the coefficients of $C^{a;a_1\dots a_4}_{i_1\dots i_4}$, defined in eq.~(\ref{eq:Cijkl}), we choose $\kappa^{(1)}_{0001}$ and $\kappa^{(2)}_{0001}$ as independent parameters in $\O_{\text{EOM}}^{(6),4}$. At this point, $\O_{G}^{(6),4}$ is written in terms of the coefficients appearing in $\O_{\text{EOM}}^{(6),4}$ and $\O_{\text{EOM}}^{(6),3}$, by means of eqs.(\ref{def:Ctilde4}), (\ref{eq:delta1_4}), (\ref{eq:delta2a_4}) and (\ref{eq:delta2b_4}), which give 
\begin{align}
  \label{eq:O6EOM4}
  \O_{\text{EOM}}^{(6),4}=&+2g^3\kappa^{(1)}_{0001}\,(fff)^{aa_1a_2a_3a_4}\,(D.F)^aA^{a_1}A^{a_2}A^{a_3}\partial A^{a_4}\nonumber\\
  &+2g^3\kappa^{(2)}_{0001}\,d_{4f}^{aa_1a_2a_3a_4}\,(D.F)^aA^{a_1}A^{a_2}A^{a_3}\partial A^{a_4},
\end{align}
\begin{align}
  \label{eq:O6G4}
  \O_{G}^{(6),4}=&-2g^3\kappa^{(1)}_{002}\,(fff)^{aa_1a_2a_3a_4a_5}\,\partial\bar{c}^a\Big[A^{a_1}A^{a_2}A^{a_3}\partial^2c^{a_4}+2A^{a_1}A^{a_2}(\partial A^{a_3})\partial c^{a_4}\Big]\nonumber\\
  &+2g^3\kappa^{(1)}_{101}\,(fff)^{aa_1a_2a_3a_4a_5}\,\partial\bar{c}^a\Big[A^{a_1}\partial A^{a_2}\,A^{a_3}\partial c^{a_4}-\partial A^{a_1}\,A^{a_2}A^{a_3}\partial c^{a_4}\Big]\nonumber\\
  &-3g^3\kappa^{(2)}_{002}\,d_{4f}^{aa_1a_2a_3a_4}\,\partial\bar{c}^a\Big[A^{a_1}A^{a_2}A^{a_3}\partial^2c^{a_4}+2A^{a_1}A^{a_2}\partial A^{a_3}\,\partial c^{a_4}\Big]\nonumber\\
  &-6g^3\kappa^{(2)}_{011}\,d_{4f}^{aa_1a_2a_3a_4}\,\partial\bar{c}^a\partial A^{a_1}\,A^{a_2}A^{a_3}\partial c^{a_4}\nonumber\\
  &-2g^3\kappa^{(1)}_{0001}\,(fff)^{aa_1a_2a_3a_4a_5}\,\partial\bar{c}^a\Big[A^{a_1}A^{a_2}A^{a_3}\partial^2c^{a_4}+3A^{a_1}\partial A^{a_2}\,A^{a_3}\partial c^{a_4}\nonumber\\
    &\phantom{-2g^3\kappa^{(1)}_{0001}\,(fff)^{aa_1a_2a_3a_4a_5}\,(\partial\bar{c}^a)\Big[}-3A^{a_1}A^{a_2}\partial A^{a_3}\,\partial c^{a_4}-\partial A^{a_1}\,A^{a_2}A^{a_3}\partial c^{a_4}\Big]\nonumber\\
    &-2g^3\kappa^{(2)}_{0001}\,\partial\bar{c}^a\Big[d_{4f}^{aa_1a_2a_3a_4}(A^{a_1}A^{a_2}A^{a_3}\partial^2c^{a_4}-A^{a_1}A^{a_2}\partial A^{a_3}\,\partial c^{a_4})\nonumber\\
      &\phantom{-2g^3\kappa^{(2)}_{0001}\,\partial\bar{c}^a\Big[}+2d_{4f}^{aa_4a_1a_2a_3}\partial c^{a_4}\,A^{a_1}A^{a_3}\partial A^{a_3}\Big].
\end{align}
At mass dimension 8, we found a total of eleven unphysical operators, parameterised by an equal number of free coefficients $\eta$, $\kappa_{03}$, $\kappa_{12}\dots$ , that are required to renormalise $\O_1^{(6)}$ up to four loops. This picture simplifies significantly by taking into account the anti-BRST relations. For instance, by evaluating eq.~(\ref{eq:antiBRSTrel1}) for $i_1,i_2=0\dots 3$, we obtain 
\begin{equation}
  \left\{
  \begin{array}{l}
    2\eta^{(1)}_{12}-3\eta^{(1)}_{03}=0\\
    \eta^{(1)}_{03}+\eta^{(1)}_{12}-\eta^{(1)}_{21}+2\eta^{(1)}_{30}=0
  \end{array}
  \right.
\end{equation}
where the $\eta^{(1)}_{ij}$ depend on $\eta$, $\kappa_{12}$ and $\kappa_{03}$, as in eq.~(\ref{eq:delta_2}). The equations above are both solved simultanously by imposing
\begin{equation}
  \label{eq:antiBRSTrelO61}
  5\eta+4\kappa_{12}-6\kappa_{03}=0.
\end{equation}
Similarly, we derive further constraints by expanding eqs.~(\ref{eq:antiBRSTrel2}) - (\ref{eq:antiBRSTrel7}), which lead to
\begin{align}
  \label{eq:antiBRSTrelO62}
  &\kappa^{(1)}_{101}-2\kappa^{(1)}_{002}+\frac{5}{6}\eta+\frac{5}{3}\kappa_{12} = 0,\\
  \label{eq:antiBRSTrelO63}
  &\kappa^{(2)}_{011}-\kappa^{(2)}_{002}=0,\\
  \label{eq:antiBRSTrelO64}
  &\kappa^{(3)}_{101}+2\kappa^{(3)}_{002}=0,\\
  \label{eq:antiBRSTrelO65}
  &3\kappa^{(1)}_{0001}+\eta+2\kappa_{12}-3\kappa^{(1)}_{002}=0,\\
  \label{eq:antiBRSTrelO66}
  &2\kappa^{(2)}_{0001}-3\kappa^{(2)}_{002}=0.
\end{align}
In conclusion, by imposing the relations on the coefficients of eqs.~(\ref{eq:O6EOM1G1})-(\ref{eq:O6G4}), which are given in eqs.(\ref{eq:antiBRSTrelO61})-(\ref{eq:antiBRSTrelO66}), we obtain a minimal basis of only five independent unphysical operators at dimension 8. For instance, we might solve eqs.(\ref{eq:antiBRSTrelO61})-(\ref{eq:antiBRSTrelO66}) in terms of $\eta$, $\kappa_{12}$, $\kappa^{(1)}_{002}$, $\kappa^{(2)}_{002}$, $\kappa^{(3)}_{002}$ and pick the following basis of independent operators
\begin{align}
  \label{eq:basisO61}
  \O_1^{(6)}=\frac{1}{2}\text{Tr}\Big[F_\nu\,D^4\,F^\nu\Big],
\end{align}
\begin{align}
  \O_2^{(6)}&=(D.F)^a\partial^4A^a-\partial\bar{c}^a\,\partial^5c^a+gf^{a_1a_2a_3}\Big[\frac{5}{3}(D.F)^{a_1}A^{a_2}\partial^3A^{a_3}-\partial\bar{c}^{a_1}\Big(\frac{8}{3}A^{a_2}\partial^4c^{a_3}\nonumber\\
    &+4\partial A^{a_2}\partial^3c^{a_2}+6\partial^2A^{a_2}\partial^2c^{a_3}+\frac{7}{3}\partial^3A^{a_2}\partial c^{a_3}\Big)\!\Big]+g^2(ff)^{aa_1a_2a_3}\Big[\frac{5}{3}(D.F)^a\partial A^{a_1}\partial A^{a_2}A^{a_3}\nonumber\\
    &-\frac{5}{3}\partial\bar{c}^a\Big(A^{a_1}A^{a_2}\partial^3c^{a_3}+4A^{a_1}\partial A^{a_2}\partial^2c^{a_3}+3A^{a_1}\partial^2A^{a_2}\partial c^{a_3}+\partial A^{a_1}\partial A^{a_2}\partial c^{a_3}\nonumber\\
    &-2\partial A^{a_1}A^{a_2}\partial^2c^{a_3}\Big)\Big]-g^3(fff)^{aa_1a_2a_3a_4}\Big[\frac{2}{3}(D.F)^aA^{a_1}A^{a_2}A^{a_3}\partial A^{a_4}\nonumber\\
    &+\frac{1}{3}\partial\bar{c}^a\Big(2A^{a_1}A^{a_2}A^{a_3}\partial^2 c^{a_4}+6A^{a_1}A^{a_2}\partial A^{a_3}\partial c^{a_4}-A^{a_1}\partial A^{a_2}A^{a_3}\partial c^{a_4}\nonumber\\
    &-8\partial A^{a_1}A^{a_2}A^{a_3}\partial c^{a_4}\Big)\Big]
  \label{eq:basisO62}
\end{align}

\begin{align}
  \O_3^{(6)}&=gf^{aa_1a_2}\Big[(D.F)^a\Big(2\partial A^{a_1}\,\partial^2A^{a_2}+\frac{4}{3}A^{a_1}\,\partial^3 A^{a_2}\Big)-\frac{4}{3}\partial\bar{c}^a\Big(a^{a_1}\partial^4c^{a_2}+\frac{3}{2}\partial A^{a_1}\,\partial^2c^{a_2}\nonumber\\
    &-\frac{3}{2}\partial^2A^{a_1}\,\partial^2c^{a_2}-\partial^3A^{a_1}\partial c^{a_2}\Big)\Big]+g^2(ff)^{aa_1a_2a_3}\Big[-\frac{10}{3}(D.F)^a\partial A^{a_1}A^{a_2}\partial A^{a_3}\nonumber\\
    &-\partial\bar{c}^a\Big(\frac{4}{3}A^{a_1}A^{a_2}\partial^3c^{a_3}+\frac{22}{3}A^{a_1}\partial A^{a_2}\partial^2c^{a_3}+A^{a_1}\partial^2A^{a_2}\partial c^{a_3}-\frac{14}{3}\partial A^{a_1}A^{a_2}\partial^2c^{a_3}\nonumber\\
    &-2\partial^2A^{a_1}A^{a_2}\partial c^{a_3}+\frac{22}{3}\partial A^{a_1}\partial A^{a_2}\partial c^{a_3}\Big)\Big] -\frac{4}{3} g^3(fff)^{aa_1a_2a_3a_4}\Big[(D.F)^aA^{a_1}A^{a_2}A^{a_3}\partial A^{a_4}\nonumber\\
    &-\partial\bar{c}^a\!\Big(A^{a_1}A^{a_2}A^{a_3}\partial^2c^{a_4}-3A^{a_1}A^{a_2}\partial A^{a_3}\partial c^{a_4}\!+\frac{A^{a_1}\partial A^{a_2}A^{a_3}\partial c^{a_4}}{2}+4\partial A^{a_1}A^{a_2}A^{a_3}\partial c^{a_4}\!\Big)\!\Big]
  \label{eq:basisO63}
\end{align}

\begin{align}
  \O_4^{(6)}&=2g^2(ff)^{aa_1a_2a_3}\Big[(D.F)^a\Big(A^{a_1}A^{a_2}\partial^2A^{a_3}+2\partial A^{a_1}A^{a_2}\partial A^{a_3}\Big)-\partial\bar{c}^a\Big(A^{a_1}A^{a_2}\partial^3c^{a_3}\nonumber\\
    &-2A^{a_1}\partial A^{a_2}\partial^2c^{a_3}-2A^{a_1}\partial^2A^{a_2}\partial c^{a_3}+4\partial A^{a_1}A^{a_2}\partial^2c^{a_3}+\partial^2A^{a_1}A^{a_2}\partial c^{a_3}\nonumber\\
    &-2\partial A^{a_1}\partial A^{a_2}\partial c^{a_3}\Big)\Big]+2g^3(fff)^{aa_1a_2a_3a_4}\Big[(D.F)^aA^{a_1}A^{a_2}A^{a_3}\partial A^{a_4}\nonumber\\
    &-\partial\bar{c}^a\Big(\!2A^{a_1}A^{a_2}A^{a_3}\partial^2c^{a_4}-A^{a_1}A^{a_2}\partial A^{a_3}\partial c^{a_4}+A^{a_1}\partial A^{a_2}A^{a_3}\partial c^{a_4}+3\partial A^{a_1}A^{a_2}A^{a_3}\partial c^{a_4}\Big)\!\Big]
  \label{eq:basisO64}
\end{align}

\begin{align}
  \O_5^{(6)}&=3g^2d^{aa_1a_2a_3}\Big[(D.F)^a\Big(A^{a_1}A^{a_2}\partial^2A^{a_3}+A^{a_1}\partial A^{a_2}\partial A^{a_3}\Big)-\partial\bar{c}^a\Big(A^{a_1}A^{a_2}\partial^3c^{a_3}\nonumber\\
    &+2A^{a_1}\partial A^{a_2}\partial^2c^{a_3}+2A^{a_1}\partial^2A^{a_2}\partial c^{a_3}+\partial A^{a_1}\partial A^{a_2}\partial c^{a_3}\Big)\Big]\nonumber\\
  &+3g^3d_{4f}^{aa_1a_2a_3a_4}\Big[(D.F)^aA^{a_1}A^{a_2}A^{a_3}\partial A^{a_4}-\partial\bar{c}^a\Big(2A^{A_1}A^{a_2}A^{a_3}\partial^2c^{a_4}+A^{a_1}A^{a_2}\partial A^{a_3}\partial c^{a_4}\nonumber\\
    &+2A^{a_1}\partial A^{a_2}A^{a_3}\partial c^{a_4}\Big)\Big]-6g^3d_{4f}^{aa_4a_1a_2a_3}\partial\bar{c}^aA^{a_1}A^{a_2}\partial A^{a_3}\partial c^{a_4}
  \label{eq:basisO65}
\end{align}

\begin{align}
  \O_6^{(6)}&=2g^2d_{4ff}^{aa_1a_2a_3}\Big[(D.F)^a\Big(A^{a_1}A^{a_2}\partial^2A^{a_3}+2A^{a_1}\partial A^{a_2}\partial A^{a_3}-\partial^2A^{a_1}A^{a_2}A^{a_3}\nonumber\\
    &-2\partial A^{a_1}A^{a_2}\partial A^{a_3}\Big)-6\partial\bar{c}^a\Big(A^{a_1}A^{a_2}\partial^3c^{a_3}+2A^{a_1}\partial A^{a_2}\partial^2c^{a_3}-\partial^2A^{a_1}A^{a_2}\partial c^{a_3}\nonumber\\
    &-2\partial A^{a_1}\partial A^{a_2}\partial c^{a_3}\Big)\Big]
  \label{eq:basisO66}
\end{align}

\subsection{Operators of higher $N$}
The construction of an operator basis to renormalise twist-2 operators of higher spin $N$ is summarised by the following steps.
\begin{itemize}
\item[1] List all the EOM operators, $\O_{\text{EOM}}^{(N),k}$, defined in eq.~(\ref{eq:EOMgeneral2}). Up to four loops, only the terms with $k\leq 4$, given in eqs.~(\ref{def:EOMop})-(\ref{def:EOMop4}) are relevant. All these operators have been written in terms of the colour structures in eqs.(\ref{eq:Cij})-(\ref{eq:Cijkl}) and associated parameters. The latter obey the relations in eqs.(\ref{eq:BCJetaij})-(\ref{eq:BCJeta2ijkl2}), which define an independent set of EOM operators, considering Bose symmetry only.
\item[2] The structure of ghost operators is dictated by the generalised BRST symmetry, eq.~(\ref{eq:gBRSTancestor}). The operators that mix with $\O_1^{(N)}$ up to four loops are given, for every value of $N$, in eqs.~(\ref{eq:BRSTop1})-(\ref{eq:BRSTop4}). They involve the colour structures given in eqs.(\ref{def:Ctilde2})-(\ref{def:Ctilde4}). Eqs.~(\ref{eq:delta_2})-(\ref{eq:delta2b_4}) uniquely determine {\textbf{all}} parameters of the ghost Lagrangian, in terms of the parameters of the EOM operators.
\item[3] Impose the anti-BRST symmetry, eq.~(\ref{eq:antiBRSTop}). The latter implies relations among the coefficients of the EOM operators via eqs.~(\ref{eq:antiBRSTrel1})-(\ref{eq:antiBRSTrel7}). These reduce the number of independent operators to a minimal set.
\end{itemize}
The steps above allow to automate easily the construction of the operators, e.g. in \verb+FORM+ \cite{Ruijl:2017dtg}. Finding independent operators boils down to finding a set of coefficients which solves the linear relations\footnote{We notice that both eqs.~(\ref{eq:antiBRSTrel6}) and (\ref{eq:antiBRSTrel7}) originate from the structure associated to the coefficients $\kappa_{i_1i_2i_3i_4}^{(2)}$ in $\O_{\text{EOM}}^{(N),4}$. We checked explicitly up to $N=10$ that eq.~(\ref{eq:antiBRSTrel7}) is automatically satisfied by the solutions of eq.~(\ref{eq:antiBRSTrel6}), which rely also on eq.~(\ref{eq:antiBRSTrel3}), and therefore it doesn't provide further simplifications of the basis.} in eqs.(\ref{eq:BCJetaij})-(\ref{eq:BCJeta2ijkl2}) and (\ref{eq:antiBRSTrel1})-(\ref{eq:antiBRSTrel6}), using the definitions in eqs.(\ref{eq:delta_2})-(\ref{eq:delta2b_4}). By solving these, we determine the number of independent unphysical operators of higher spin $N$.
For up to $N=16$ the size of the basis is given in table \ref{table:opnops}. The second line in table \ref{table:opnops} gives the size of the basis without using anti-BRST relations, while the first line includes them.
\begin{table}[h]
\begin{center}
\begin{tabular}{l c c c c c c c c c}
\Xhline{3\arrayrulewidth}
\\[-1em]
  Spin $N$ &2& 4 & 6 & 8 & 10 & 12 & 14 & 16 \\
\\[-1em]
 \hline
\\[-1em]
  w aBRST&1& 2 & 5 & 12 & 25 & 50 & 87 & 140\\
  w/o aBRST  &1& 3 & 11 & 30 & 66 & 126 & 215 & 339\\
\Xhline{3\arrayrulewidth}
\end{tabular}
\end{center}
\caption{Table showing the number of independent operators with and without the use of anti-BRST (aBRST) relations.}
\label{table:opnops}
\end{table}
While the basis grows significantly with the spin $N$, we find that most of the free parameters are associated to the operators $\O_{\text{EOM}}^{(N),4}$. For instance, $\O_{\text{EOM}}^{(N),4}$ generates $112$ out of the $140$ unphysical operators at $N=16$. Since mixing with $\O_{\text{EOM}}^{(N),4}$ is only relevant at one loop, see table \ref{tab:EOMmixing}, these operators do not introduce prohibitive obstacles.

\section{Background-field formulation}
\label{sec:backgroundfield}

A powerful trick to simplify calculations of anomalous dimensions is to use the background field method.
The basic idea is to split the gauge field into a classical (non-propagating) background field component $B$ and
a purely Quantum field component $Q$ as follows:
\beq
A^\mu_a(x)=B^\mu_a(x)+Q^\mu_a(x)\,.
\eeq
One can then consider Green's functions with external background fields. By using a clever gauge fixing and ghost ghost Lagrangian, for the Quantum field \cite{DeWitt:1967ub,tHooft:1975uxh,Abbott:1980hw,Abbott:1981ke}
\beq
\L_{\mathrm{BGF+BG}}(Q,B,\bar c ,c)=-\frac{1}{2\xi}(\bar D^\mu A_\mu)^2-\bar c^a \bar D_\mu^{ab} D^{\mu;bc}c^c\,,
\eeq
where the background- and background+quantum-field covariant derivatives are defined as
\beq
\bar D_\mu^{ac}=\partial_\mu \delta^{ac}+gf^{abc}B^{b}_\mu\,,\qquad D_\mu^{ac}=\partial_\mu \delta^{ac}+gf^{abc}(B+Q)^{b}_\mu\,,
\eeq
it then follows that the quantum gauge-fixed Lagrangian,
\beq
\L_B(Q,B,\bar c,c)=\L_0(Q+B)+\L_{\mathrm{BGF+BG}}(Q,B,\bar c ,c)\,,
\eeq
stays invariant under background-field gauge transformations
\begin{align}
\delta_w^B B_\mu^a(x)&=\bar D_\mu^{ac} \omega(x)^c\,,\nn\\
\delta_w^B Q_\mu^a(x)&=gf^{abc}Q^{b}_\mu \omega(x)^c\,.
\end{align}
We now wish to discuss the form of the complete Lagrangian $\tilde \L(A,\bar c,c)$, introduced in eq.~(\ref{eq:ltilde}), which contains besides the Yang-Mills, gauge fixing and ghost terms also the twist-2 gauge invariant gluonic operator $\O_N(A)$, the EOM operator $\O_{EOM}^{(N)}(A)$ and the ghost operator $\O_{G}^{(N)}(A)$. Here we have
purposefully included a dependence on $A$, although we will not write out explicitly the dependence on its derivatives.

The lifting of $\tilde \L(A,\bar c,c)$ into the background field formalism is straight forward for the gauge invariant part but requires some minor modifications to the EOM and ghost operator. We therefore introduce their background field versions $\O_{\mathrm{BEOM}}^{(N)}(Q,B,\bar c ,c)$ and $\O_{\mathrm{BG}}^{(N)}(Q,B)$. Before giving a detailed derivation of the form of the Lagrangian we will state their form below. The complete Lagrangian then reads
\begin{align}
\tilde \L_B(A,B,\bar c,c)=&\L_0(Q+B)+\L_{\mathrm{BGF+BG}}(Q,B,\bar c ,c)+\O^{(N)}_1(Q+B)\nn\\
&+\O_{\mathrm{BEOM}}^{(N)}(Q,B,\bar c ,c)+\O_{\mathrm{BG}}^{(N)}(Q,B)\,,
\end{align}
where
\begin{equation}
\label{eq:BOG}
\O_{BG}^{(N)}=\sum_k\O_{BG}^{(N),k},\qquad \O_{\mathrm{BEOM}}^{(N)}=\sum_k\O_{\mathrm{BEOM}}^{(N),k},
\end{equation}
\begin{align}
     \label{def:BEOMop1}
\O_{\text{BEOM}}^{(N),k} &= g^{k-1}\left(D\cdot F(B+Q)\right)^a\sum_{\substack{i_1+\dots+i_k\\=N-k-1}}\,C^{a;a_1\dots a_k}_{i_1\dots i_k}\,\left(\bar D^{i_1}Q^{a_1}\right)\dots\left(\bar D^{i_k}Q^{a_k}\right)\,,
\end{align}
\begin{align}
\O_{BG}^{(N),k}&=- g^{k-1}\sum_{\substack{i_1+\dots+i_k\\=N-k+1}} \widetilde{C}^{a;a_1\,..\,a_4}_{i_1\,..\,i_4}
\left(\bar D \overline{c}^a\right) \left(\bar D^{i_1}Q^{a_1}\right)..\left(\bar D^{i_{k-1}}Q^{a_{k-1}}\right)\left(\bar D^{i_k+1} c^{a_k}\right)\,.
\label{eq:BBRSTop1}
\end{align}
Note in particular that the coefficients $C_{ijk..}^{abc..}$ and $\widetilde C_{ijk..}^{abc..}$ are identical in their definitions to those defined respectively in eqs.~(\ref{eq:Cij})-(\ref{eq:Cijkl}) and (\ref{def:Ctilde2})-(\ref{def:Ctilde4}). The set of EOM and ghost operators in the background gauge formalism is thus directly related to those in the standard formulation.

To understand the structure of the EOM operator note that it should be generated from an infinitesimal field transformation of the kind $Q\to Q+\mathcal{G}^B(Q,B,\partial Q,\partial B,...)$, since the Quantum effective action contains a path integral only over the field $Q$ being a functional of $B$. This fixes the form of the EOM operator as follows:
\begin{align}
\O_{BEOM}^{(N)}=&\int d^Dx \frac{\delta S_0(A+Q)}{\delta Q^\mu_a(x)}\,\mathcal{G}^{B;a}_\mu(Q^\mu,B^\mu,\partial^\mu Q,\partial^\mu B,...)\nn \\
=&\big(D.F(Q+B)\big)^a\,\mathcal{G}^{B;a}(Q,B,\partial Q,\partial B,...)
\end{align}
where we have used also our earlier considerations about the mass dimension and counting of $\Delta$-contractions. Finally we make the assertion that
\beq
\label{eq:GB}
\mathcal{G}^{B;a}_\mu(Q,B,\partial Q,\partial B,...)=\mathcal{G}^{a}_\mu(Q,\bar D Q,...)\,.
\eeq
with $\mathcal{G}^{a}_\mu$ defined in eqs.~(\ref{eq:EOMdefG}) and (\ref{eq:EOMgeneral1}). There are a number of considerations which fix this relation.
First we require $\mathcal{G}^{B;a}$ to be background-field gauge covariant - thus it can only depend on $Q$ or on its background-field covariant derivatives. However this fixes only its dependence on $Q$ and $B$ but not its functional form, $\mathcal{G}^B=\mathcal{G}$. To fix this form we set $B=0,Q=A$ in the complete Lagrangian, i.e. we consider $\tilde \L_B(A,0,\bar c,c)$. For this Lagrangian to generate the same Green's functions as $\tilde \L(A,\bar c,c)$ (note their gauge-invariant parts are now identical) we therefore require:
\beq
\tilde \L_B(A,0,\bar c,c)=\tilde\L(A,\bar c,c)
\eeq
From this it immediately follows that
\beq
\mathcal{G}^{B;a}_\mu(Q,\bar D Q,...)\Big|_{B=0,Q= A}=\mathcal{G}^{B;a}_\mu(A,\partial A,...)=\mathcal{G}^{a}_\mu(A,\partial A,...)
\eeq
and we see that eq.~(\ref{eq:GB}) satisfies these constraints uniquely.

Let us now turn our attention to the ghost operator in the background formalism. Again we need to satisfy the constraints that it should coincide with $\O_{G}^{(N)}$ when we set $B=0,Q=A$ and that it should be background gauge invariant. A simple recipe which satisfies all these constraints is to  to make the replacements $A\to Q,\partial\to \bar D$ in $\O_{G}^{(N)}$. A more thorough path to arrive at the same answer would involve working out the generalised gauge invariance and its associated generalised BRST symmetry. In turn one could write the ghost operator in gBRST exact form, in the background field formalism.

\subsection{Bases of operators up to four loops}
In the background field method, we determine the renormalisation constants of $\O_1^{(N)}$ by computing the counterterms of correlators of the background field 
\begin{equation}
  \label{eq:GammaOBBmunu}
  \left(\Gamma_{\O_i;BB}^{(N)}\right)^{a_1 a_2}_{\nu_1\nu_2}(g,\xi;p^\mu) = \int d^dx_1 d^dx_2\,e^{ip \cdot( x_1-x_2)}\,\langle 0 | T\left[B^{a_1}_{\nu_1}(x_1)B^{a_2}_{\nu_2}(x_2)\O_i^{(N)}(0)\right]|0\rangle_{\mathrm{1PI}},
\end{equation}
where the subscript $\mathrm{1PI}$ indicates one-particle-irreducibe, amputated Green's functions. In the equation above, the operator $\O_i^{(N)}=\O_i^{(N)}(B+Q)$ is inserted with zero momentum. Counterterms proportional to $\O_{\text{BEOM}}^{(N)}$ and $\O_{BG}^{(N)}$ are required in order to cancel divergences of the diagrams that contribute to eq.~(\ref{eq:GammaOBBmunu}). Notably, these unphysical operators always involve at least one quantum gluon or a ghost-antighost pair, as it follows from the definitions in eqs.~(\ref{def:BEOMop1}) and (\ref{eq:BBRSTop1}). Therefore, EOM and ghost operators are only required from the two-loop level onwards, in order to cancel the {\textit{subdivergences}} of the correlator in eq.~(\ref{eq:GammaOBBmunu}); and no unphysical counterterm can arise at one loop \cite{Sarkar:1974ni,Pascual:1984zb}. In table \ref{tab:BEOMmixingdiags} we report example diagrams showing subdivergences of $\Gamma_{\O_i;BB}^{(N)}$, which are renormalised by each term $\O_{\mathrm{EOM}}^{(N),k}$. Table \ref{tab:BEOMmixing} summarises the maximal loop order at which each operator $\O^{(N),k}_{\mathrm{EOM}}$ enters the renormalisation of eq.~(\ref{eq:GammaOBBmunu}). By comparing the last line of tables \ref{tab:EOMmixing} and \ref{tab:BEOMmixing} we find that there is an advantage in renormalising correlators of background fields, in that unphysical counterterms are needed only up to 3 loops. In contrast without background-field invariance the counterterm $\O_{\text{EOM}}^{(N),1}$ would be required up to 4 loops, as in table \ref{tab:EOMmixing}.
\begin{table}[h]
\centering
\begin{tabular}{l c c c c}
\Xhline{3\arrayrulewidth}
$L$ & $\O_{\mathrm{BEOM}}^{(N),\le1}$ & $\O_{\mathrm{BEOM}}^{(N),\le2}$ & $\O_{\mathrm{BEOM}}^{(N),\le3}$ & $\O_{\mathrm{BEOM}}^{(N),\le4}$ \phantom{\Bigg\{}\\ [1ex]
\hline
2 & \BtwolEOMone & \Bthreeptoneloop &   &   \\
3 & \BthreelEOMone &\Bthreepttwoloop  &\Bfourptoneloop &     \\
4 & \BfourlEOMone &\Bthreeptthreeloop  &\Bfourpttwoloop& \Bfiveptoneloop          \\
& & & &\\
\Xhline{3\arrayrulewidth}
\end{tabular}
\caption{In the $L$th row the table gives examples of diagrams contributing to the $L$-loop contribution to $\Gamma_{\O_1;BB}^{(N)}$. Subgraphs whose UV-counterterms require the various EOM operators $\O_{\mathrm{EOM}}^{(N),k}$ are highlighted with dashed boxes.\label{tab:BEOMmixingdiags}}
\end{table}

\begin{table}[h]
\centering
\begin{tabular}{c c c c c}
\Xhline{3\arrayrulewidth}
\\[-1em]
$\Gamma_{\O_1;BB}^{(N)}$ & $\O_{BEOM}^{(N),1}$ & $\O_{BEOM}^{(N),2}$ & $\O_{BEOM}^{(N),3}$ & $\O_{BEOM}^{(N),4}$ \\ [1ex]
\hline
1 & 0 & 0 & 0& 0\\
2 & 1 & 1 & 0& 0\\
3 & 2& 2& 1& 0\\
4 & 3& 3 & 2& 1 \\
\Xhline{3\arrayrulewidth}
\end{tabular}
\caption{The table summarizes the loop orders for which the mixing of $\O_1^{(N)}$ into $\O_{EOM}^{(N),k}$ is required, given a certain loop order of $\Gamma_{\O_1;BB}^{(N)}$.}
\label{tab:BEOMmixing}
\end{table}
In the next section we will compute the counterterms required to renormalise these subdivergences. To this end, it is convenient to reduce to a basis of independent operators.
In the background-field method a basis for a given fixed value of $N$ is obtained by modifying the corresponding basis obtained without background field, according to the replacements:
\begin{align}
  \label{eq:Bkguplift}
  (D.F)^a &\longrightarrow (D.F(Q+B))^a, &  (\partial^i A^a) &\longrightarrow (\bar{D}^i Q)^a,\\
  \label{eq:Bkguplift2}
  (\partial\bar{c}^a) &\longrightarrow (\bar{D}\bar{c})^a, &(\partial^i c^a) &\longrightarrow (\bar{D}^i c)^a.
\end{align}
For instance, the basis for $N=2$ can be directly read off eq.~(\ref{eq:Oeta(N=2)}), giving
\begin{align}
  \O_1^{(2)}&= F_\nu^a(Q+B) F^{\nu;a}(Q+B)\,,\\
  \O_2^{(2)}&= (D.F(Q+B))^a\,Q^a+\overline{c}^a\bar{D}^{a a_1}\bar{D}^{a_1 a_2}c^{a_2}.
\end{align}
Similarly, bases for $N=4$ and $N=6$ are obtained by applying eqs.~(\ref{eq:Bkguplift}) and (\ref{eq:Bkguplift2}) to eqs.(\ref{eq:basisO41})-(\ref{eq:basisO43}) and to eqs.(\ref{eq:basisO61})-(\ref{eq:basisO66}), respectively.

\section{Calculations and results}
\label{sec:Applications}
In this section we renormalise gauge invariant operators of spin $N=2$, $4$ and $6$, using the bases in eqs.~(\ref{eq:O2basis1})-(\ref{eq:Oeta(N=2)}), (\ref{eq:basisO41})-(\ref{eq:basisO43}) and (\ref{eq:basisO61})-(\ref{eq:basisO66}), respectively. In these bases, we proceed to calculate the associated renormalisation constants $Z^{(N)}_{i,j}$, defined in eq.~(\ref{def:OiNren}), which in the $\overline{\mathrm{MS}}$ scheme can be expanded as follows,
\begin{equation}
  \label{def:ZNijelement}
  Z_{i\,j}^{(N)}=\delta_{i\,j} + \delta Z^{(N)}_{i\,j},\quad\text{with}\quad\delta Z^{(N)}_{i\,j}=\sum_{r=1}^{\infty}\frac{1}{\epsilon^r}\;Z^{(N),r}_{i\,j}(\alpha_s).
\end{equation}
The renormalisation matrix is block triangular with $Z^{(N)}_{j>1\,1}=0$, as described in eq.~(\ref{eq:Zstructure}), and only $Z_{1\,1}^{(N)}$ is required to describe the scale evolution of the gauge invariant operator $\O_1^{N}$ in physical matrix elements. In particular, from the definition of the anomalous dimension matrix,
        \begin{equation}
          \gamma_{ij}^{(N)} = -\mu^2\frac{d^2}{d\mu^2}Z^{(N)}_{ik}\,(Z^{-1})_{kj}\,,
        \end{equation}
one can obtain
\begin{equation}
  \label{def:physAD}
  \gamma^{(N)}\equiv\gamma^{(N)}_{1\,1}=\alpha_s\frac{\partial}{\partial\alpha_s}\, Z^{(N),1}_{1\,1}(\alpha_s)\,.
\end{equation}
Off-diagonal elements of the renormalisation matrix do not contribute to the anomalous dimension of the physical operators. However, the computational method that we adopt to determine $Z^{(N)}_{1\,1}$ requires the knowledge of a set of mixing contributions $Z^{(N)}_{1\,i>1}$. Below we describe the calculation of these renormalisation constants and that of the physical anomalous dimension.

\subsection{Mixing with EOM and Ghost Operators}
The renormalisation constants $Z_{1\,i}^{(N)}$, with $i>1$, are determined by the counterterms of one-particle-irreducible, amputated Green functions, with one insertion of $\O_1^{(N)}$ and external ghost and gluon fields. We list examples of diagrams contributing to such Green's functions in table \ref{tab:ghostmixingdiags} for general $N$. In practice, if we work at fixed values of $N$, not all these contributions enter. In table \ref{tab:ghostmixingdiagsfixedN} we show the structure of the relevant counterterms for $N=2,4$ and $6$.
\begin{table}[t]
\centering
\begin{tabular}{l c c c c}
\Xhline{3\arrayrulewidth}
\\[-1em]
$L$ & $\O_{\mathrm{EOM+G}}^{(N),\le1}$ & $\O_{\mathrm{EOM+G}}^{(N),\le2}$ & $\O_{\mathrm{EOM+G}}^{(N),\le3}$ & $\O_{\mathrm{EOM+G}}^{(N),\le4}$ \\
\\[-1em]
\hline
2 & \ghosttwoptoneloop & \ghostthreeptoneloop &   &   \\
3 & \ghosttwoptwoloop & \ghostthreepttwoloop   &\ghostfourptoneloop &     \\
4 & \ghosttwoptthreeloop &\ghostthreeptthreeloop  &\ghostfourpttwoloop& \ghostfiveptoneloop          \\
& & & &\\
\Xhline{3\arrayrulewidth}
\end{tabular}
\caption{In the $L$th row the table gives examples of diagrams containing a ghost-anti-ghost pair and gluons whose UV-counterterms determine the mixing of $\O^{(N)}_1$ into both the EOM - and ghost - operators as required for the computation of the $L$-loop contribution to $\Gamma_{1;BB}^{(N)}$.}
\label{tab:ghostmixingdiags}
\end{table}
%
%
Specifically, we consider the following correlators with an operator insertion at zero momentum,
\begin{align}
  \label{def:ghostOicor}
  (\Gamma_{i;c\overline{c}}^{(N)})^{ab}(g,\xi,p) &= \int d^dx_1 d^dx_2\, e^{ip\cdot(x_1-x_2)}\,\langle 0|T\big[c^a(x_1)\overline{c}^b(x_2)\O_i^{(N)}(0)\big]|0\rangle_{\mathrm{1PI}}.
\end{align}
Examples of Feynman diagrams contributing to eq.~(\ref{def:ghostOicor}) are depicted in the first column of table \ref{tab:ghostmixingdiags}. For every value of $N$, $\Gamma_{i;c\overline{c}}^{(N)}$ vanishes at tree level, unless $\O_i^{(N)}=\O_2^{(N)}$, as it can be seen by inspecting the operators bases in eqs.~(\ref{eq:Oeta(N=2)}), (\ref{eq:basisO41})-(\ref{eq:basisO43}) and (\ref{eq:basisO61})-(\ref{eq:basisO66}). Therefore we write
\begin{equation}
  \label{def:Gammaccbtree}
  (\Gamma_{i;c\overline{c}}^{(N)})^{ab}(g,\xi,p)=\left\{
  \begin{array}{lcc}
    \delta^{ab}\,\left[\Gamma^{(N),0}_{i;c\bar{c}}(p) \, + \,\delta\Gamma_{i;c\overline{c}}^{(N)}(g,\xi,p)\right] & & \text{if}\quad i = 2\\
    \\
    \delta^{ab}\,\delta \Gamma_{i;c\overline{c}}^{(N)}(g,\xi,p) & & if \quad i \neq 2
  \end{array}
  \right.
\end{equation}
in order to separate the tree level contribution from the term $\delta\Gamma_{i;c\bar{c}}^{(N)}$, which represents the sum of loop corrections to all orders, namely
\begin{equation}
  \label{def:deltaGammaccbLoop}
  \delta\Gamma_{i;c\overline{c}}^{(N)}(g,\xi,p) = \sum_{r=1}^{\infty}\,\Gamma_{i;c\overline{c}}^{(N),r}(\xi,p)\,\left(\frac{\alpha_s}{4\pi}\right)^r,
\end{equation}
with $\alpha_s=\frac{g^2}{4\pi}$. Counterterms of $\delta\Gamma_{i;c\bar{c}}^{(N)}$ must therefore be proportional to $\O_2^{(N)}$. In particular, inserting $\O_1^{(N)}$ into eq.~(\ref{def:ghostOicor}), we get
\begin{equation}
  \label{def:ccRstarToZ}
  \mathcal{Z}\left[\delta\Gamma_{1;c\bar{c}}^{(N)}\right] = Z_c\,\delta Z_{1\,2}^{(N)}\,\Gamma^{(N),0}_{2,c\bar{c}},\qquad \forall N
\end{equation}
where $\mathcal{Z}$ extracts the local counterterm of each Feynman diagram contributing to eq.~(\ref{def:ghostOicor}). To this end, we apply the $R^*$ operation \cite{Chetyrkin:1982nn,Chetyrkin:1984xa,Smirnov:1986me,Chetyrkin:2017ppe}, using a formulation that is valid for a general Feynman rule of the inserted operator \cite{Herzog:2017bjx,deVries:2019nsu,Beekveldt:2020kzk,Cao:2021cdt}
\begin{equation}
  \label{def:Rstar}
\mathcal{Z}\left[\delta\Gamma_{1;c\bar{c}}^{(N)}\right] = - K_\epsilon\bar{R}^*\left[\mathcal{T}^{(N)}_{p}\delta\Gamma_{1;c\bar{c}}^{(N)}\left|_{p=0}\right.\right].
\end{equation}
Here $\mathcal{T}^{(N)}_{p}$ denotes a Taylor expansion operator which extracts the term of order $p^N$. The operation $K_\epsilon$ extracts the singular terms of Laurent series in $\epsilon$
\begin{equation}
  K_\epsilon\left[\sum_{k=-n}^\infty\,f_{(k)}\,\epsilon^k\right] = \sum_{k=-n}^{-1}\,f_{(k)}\,\epsilon^k\,,
\end{equation}
and the operation $\bar{R}^*$  isolates the local counterterm by subtracting all UV subdivergences and IR divergences.
\begin{table}[t]
\centering
\begin{tabular}{l c c c c}
\Xhline{3\arrayrulewidth}
\\[-1em]
$L$ & $\O_{2}^{(N\ge 2)}$   & $\O_{3}^{(4)}$ & $\O_{3}^{(6)}$ & $\O_{i\in\{3,4,5,6\}}^{(6)}$\\ 
\\[-1em]
\hline
2 & \ghosttwoptoneloop & &\ghostthreeptoneloop &      \\
3 & \ghosttwoptwoloop  & &\ghostthreepttwoloop   &\ghostfourptoneloop      \\
4 & \ghosttwoptthreeloop & \ghostfourptoneloop &\ghostthreeptthreeloop  &\ghostfourpttwoloop \\
  & \\
\Xhline{3\arrayrulewidth}
\end{tabular}
\caption{In the $L$th row the table gives examples of diagrams containing a ghost-anti-ghost pair and gluons whose UV-counterterms determine the mixing of $\O^{(N)}$ into unphysical operators $\O_i^{(N\in \{2,4,6\})}$ as required for the computation of the $L$-loop contribution to $Z_{11}^{(N\in \{2,4,6\}),L}$ in the background field method.}
\label{tab:ghostmixingdiagsfixedN}
\end{table}
In addition to eq.~(\ref{def:ccRstarToZ}), we determined the elements $Z_{1\,2}^{(N)}$ of the mixing matrix using an alternative approach, described in appendix \ref{app:ADQCD}. In this way we obtain
\begin{align}
    \label{eq:Z(O2,Oeta)}
    \delta Z_{1\,2}^{(2)}&=-\frac{\alpha_s}{4\pi}\,\frac{C_A}{2\epsilon}+\left(\frac{\alpha_s}{4\pi}\right)^2\,C_A^2\left[\frac{19}{24\epsilon^2}+\frac{5}{48}\frac{\xi_F}{\epsilon}-\frac{35}{48\epsilon}\right]+\left(\frac{\alpha_s}{4\pi}\right)^3\,C_A^3\Big[ - \frac{779}{432\epsilon^3}\nonumber\\
      &+\frac{1}{\epsilon^2}\Big(\frac{2807}{864}-\frac{35\xi_F}{216}+\frac{5\xi_F^2}{288}\Big)+\frac{1}{\epsilon}\Big(-\frac{16759}{7776}-\frac{11\zeta_3}{72}+\frac{377\xi_F}{1728}+\frac{5\zeta_3\,\xi_F}{72}-\frac{65\xi_F^2}{1728}\Big)\Big]\nn\\
    &+O(\alpha_s^4),\\
    \label{eq:Z(O4,2)}
    \delta Z_{1\,2}^{(4)}&=-\frac{\alpha_s}{4\pi}\,\frac{C_A}{12\epsilon}-\left(\frac{\alpha_s}{4\pi}\right)^2\,C_A^2\left[\frac{97}{1440\epsilon^2}-\frac{\xi_F}{320\epsilon}+\frac{8641}{86400\epsilon}\right]+\left(\frac{\alpha_s}{4\pi}\right)^3\,C_A^3\Big[\frac{9437}{86400\epsilon^3}\nonumber\\
      &+\frac{1}{\epsilon^2}\Big(-\frac{1520341}{15552000}+\frac{853\xi_F}{86400}\Big)+\frac{1}{\epsilon}\Big(-\frac{166178237}{466560000}-\frac{\zeta_3}{2400}+\frac{37199\xi_F}{648000}+\frac{37\zeta_3\,\xi_F}{9600}\Big)\Big]\nn\\
    &+O(\alpha_s^4),\\
    \label{eq:Z(O6,2)}
  \delta Z_{1\,2}^{(6)}&=-\frac{\alpha_s}{4\pi}\,\frac{C_A}{30\epsilon}-\left(\frac{\alpha_s}{4\pi}\right)^2\,C_A^2\left[\frac{653}{10080\epsilon^2}+\frac{19\xi_F}{20160\epsilon}+\frac{185093}{4233600\epsilon}\right] + O(\alpha_s^3).
\end{align}
Here $\xi_F=1-\xi$ is the gauge fixing parameter, such that $\xi_F=0$ recovers the result in Feynman gauge.

In order to extract the terms $Z_{1\,i>2}^{(N>2)}$, we compute the counterterms of three- and four-point correlators, depicted in the second and in the third columns of table \ref{tab:ghostmixingdiags}, respectively. For this purpose we consider the three-point Green's function
\begin{align}
  \label{def:ccbgOicor}
  (\Gamma_{i;c\overline{c}g}^{(N)})^{abc}_\mu(g,\xi,p_1,p_2) &= \int d^dx_1 d^dx_2d^dx_3 \;e^{ip_1\cdot(x_1-x_3)}e^{ip_2\cdot(x_2-x_3)}\;\nonumber\\
  &\phantom{\int d^dx_1 d^dx_2d^dx_3}\times\langle 0|T\left[c^a(x_1)\overline{c}^b(x_2)A^c_\mu(x_3)\O_i^{(N)}(0)\right]|0\rangle,
\end{align}
which is expanded as follows:
\begin{equation}
\left(\Gamma_{i;c\bar{c}g}^{(N)}\right)^{abc}_\mu(g,\xi,p_1,p_2)= g\Big[\left(\Gamma_{i;c\bar{c}g}^{(N),0}\right)^{abc}_\mu(p_1,p_2) + \underbrace{\sum_{r=1}^{\infty}\left(\Gamma_{i;c\bar{c}g}^{(N),r}\right)^{abc}_\mu(\xi,p_1,p_2)\left(\frac{\alpha_s}{4\pi}\right)^r}_{\left(\delta\Gamma_{i;c\bar{c}g}^{(N)}\right)_\mu^{abc}}\Big],
\end{equation}
where we separated the tree-level contribution from the loop corrections, similarly to eqs.~(\ref{def:Gammaccbtree}) and (\ref{def:deltaGammaccbLoop}). The counterterm of eq.~(\ref{def:ccbgOicor}), with an insertion of $\O_1^{(N)}$, reads
\begin{equation}
  \label{def:ccbgRstarToZ}
  \mathcal{Z}\left[\left(\delta\Gamma_{1;c\bar{c}g}^{(N)}\right)^{abc}_\mu\right] = g Z_gZ_c\sqrt{Z_3}\,\sum_{k>1}\delta Z_{1\,k}^{(N)}\,\left(\Gamma^{(N),0}_{k,c\bar{c}g}\right)^{abc}_\mu\,.
\end{equation}
The terms $g(\Gamma^{(N),0}_{k,c\bar{c}g})^{abc}_\mu$ are ghost-antighost-gluon vertices generated by each operator $\O^{(N)}_{k}$, with $k>1$. Notably, there is no such counterterm for $N=2$, as it can be seen by inspecting $\O_2^{(2)}$ in eq.~(\ref{eq:Oeta(N=2)}). For $N=4$, the operator $\O_2^{(4)}$, given in eq.~(\ref{eq:basisO42}), generates both the ghost-antighost vertex and the ghost-antighost-gluon vertex. Therefore, the same counterterm $\delta Z^{(4)}_{1\,2}$ will suffice to renormalise both eqs.~(\ref{def:ccRstarToZ}) and (\ref{def:ccbgRstarToZ}). For consistency, we verified that $\delta Z^{(4)}_{1\,2}$ extracted from eq.~(\ref{def:ccbgRstarToZ}) agrees with the result in eq.~(\ref{eq:Z(O4,2)}). For $N=6$ we find
\begin{align}
  \label{eq:Z(O6,3)a}
  \mathcal{Z}\left[\left(\delta\Gamma_{1;c\bar{c}g}^{(6)}\right)^{abc}_\mu\right] =g Z_gZ_c\sqrt{Z_3}\,\left[\delta Z^{(6)}_{1\,2}\left(\Gamma^{(6),0}_{2,c\bar{c}g}\right)^{abc}_\mu + \delta Z^{(6)}_{1\,3}\left(\Gamma^{(6),0}_{3,c\bar{c}g}\right)^{abc}_\mu\right],
\end{align}
which can be solved for $\delta Z_{1\,3}^{(6)}$, upon computing the left hand-side, by means of the $R^*$ operation, and by replacing the result for $\delta Z^{(6)}_{1\,2}$, given in eq.~(\ref{eq:Z(O6,2)}). We get
\begin{align}
  \label{eq:Z(O6,3)}
  \delta Z_{1\,3}^{(6)}&=-\frac{\alpha_s}{4\pi}\,\frac{C_A}{48\epsilon}-\left(\frac{\alpha_s}{4\pi}\right)^2\,C_A^2\left[\frac{2021}{40320\epsilon^2}+\frac{235813}{8467200\epsilon}+O(\xi_F)\right] + O(\alpha_s^3),
\end{align}
where we performed the calculation in Feynman gauge, dropping terms proportional to $\xi_F$.

Finally, we determine the remaining elements of the mixing matrices for operators $\O_1^{(4)}$ and $\O_1^{(6)}$, by computing the counterterms of the four-point functions
\begin{align}
  \label{def:ccbggOicor}
  (\Gamma_{i;c\overline{c}gg}^{(N)})^{abcd}_{\mu\nu}(g,\xi,p_1,p_2,p_3) &= \int d^dx_1 d^dx_2d^dx_3d^dx_4 \;e^{ip_1\cdot(x_1-x_4)}e^{ip_2\cdot(x_2-x_4)}e^{ip_3\cdot(x_3-x_4)}\;\nonumber\\
  &\phantom{\int}\times\langle 0|T\left[c^a(x_1)\overline{c}^b(x_2)A^c_\mu(x_3)A^d_\nu(x_4)\O_i^{(N)}(0)\right]|0\rangle,\\
  &\hspace{-2.5cm}\equiv g^2\Big[\left(\Gamma_{i;c\bar{c}gg}^{(N),0}\right)^{abcd}_{\mu\nu}(p_1,p_2,p_3) + \underbrace{\sum_{r=1}^{\infty}\,\left(\Gamma_{i;c\bar{c}gg}^{(N),r}\right)^{abcd}_{\mu\nu}(\xi,p_1,p_2,p_3)\,\left(\frac{\alpha_s}{4\pi}\right)^r}_{(\delta\Gamma_{i;c\bar{c}gg}^{(N)})^{abcd}_{\mu\nu}}\Big].
\end{align}
The counterterms of eq.~(\ref{def:ccbggOicor}) are given by
\begin{align}
  \label{def:ccbggRstarToZ}
  \mathcal{Z}\left[\left(\delta\Gamma_{1;c\bar{c}gg}^{(N)}\right)^{abcd}_{\mu\nu}\right] &= g^2Z_g^2 Z_cZ_3\,\sum_{k>1}\delta Z_{1\,k}^{(N)}\,\left(\Gamma^{(N),0}_{k,c\bar{c}gg}\right)^{abcd}_{\mu\nu}.
\end{align}
By specialising the equation above to $N=4$, we find that it receives only one contribution from the vertex associated to the operator $\O_3^{(4)}$, written in eq.~(\ref{eq:basisO43}). We get
\begin{align}
  \mathcal{Z}\left[\left(\delta\Gamma_{1;c\bar{c}gg}^{(4)}\right)^{abcd}_{\mu\nu}\right] &= g^2Z_g^2Z_cZ_3\,\delta Z_{1\,3}^{(4)}\,\left(\Gamma^{(4),0}_{3,c\bar{c}gg}\right)^{abcd}_{\mu\nu},
\end{align}
which leads to
\begin{equation}
  \label{eq:Z(O4,3)}
  \delta Z_{1\,3}^{(4)} = \frac{\alpha_s}{4\pi}\,\frac{C_A}{24\epsilon} + O(\alpha_s^2).
\end{equation}
For $N=6$, all operators $\O^{(6)}_{i>1}$ contribute to eq.~(\ref{def:ccbggRstarToZ}). By plugging the known results for $\delta Z^{(6)}_{1\,2}$ and $\delta Z^{(6)}_{1\,3}$, given in eqs.~(\ref{eq:Z(O6,2)}) and (\ref{eq:Z(O6,3)}) respectively, into eq.~(\ref{def:ccbggRstarToZ}), we get
\begin{align}
  \label{eq:Z(O6,4)}
  \delta Z_{1\,4}^{(6)} &= -\frac{\alpha_s}{4\pi}\frac{C_A}{32\epsilon} + O\left(\alpha_s^2\right),  &\delta Z_{1\,5}^{(6)} = \frac{\alpha_s}{4\pi}\frac{C_A}{24\epsilon} + O\left(\alpha_s^2\right),  &\quad\delta Z_{1\,6}^{(6)} = O\left(\alpha_s^2\right).
\end{align}
The terms of $O(\alpha_s^2)$ contribute to renormalise $\O_1^{(6)}$ only at four loops, because they arise from divergent four-point subdiagrams at two loops, such as the one depicted in the botton right entry of table \ref{tab:ghostmixingdiags}. In this work we renormalise the gauge invariant operator of spin $N=6$ up to three loops and therefore we don't need to compute such contributions.

Eqs.~(\ref{eq:Z(O2,Oeta)})-(\ref{eq:Z(O6,2)}), (\ref{eq:Z(O6,3)}), (\ref{eq:Z(O4,3)}) and (\ref{eq:Z(O6,4)}) include all off-diagonal terms of the mixing matrix $\delta Z_{1\,i}^{(N)}$, which are required to renormalise the gauge invariant operators $\O_1^{(N)}$ at $N=2$ and $4$ up to 4 loops and $N=6$ up to three loops. The calculation of the physical contribution $Z_{1\,1}^{(N)}$ is described in the remaining part of this section.

\subsection{Renormalisation of physical operators}
The renormalisation constants $Z^{(N)}_{1\,1}$, which determine the anomalous dimension of the gauge invariant operator via eq.~(\ref{def:physAD}), are best extracted from correlators of the background field $B$. Using the definition in eq.~(\ref{eq:GammaOBBmunu}) and the definition of the gauge invariant operators in eq.~(\ref{eq:GIoplong}) we have
\begin{equation}
  \label{def:GammaBBmu1muN}
  \left(\Gamma^{(N)}_{1;BB}\right)^{ab}_{\nu_1\nu_2}(g,\xi,p) \equiv \left(\Gamma^{(N)}_{1;BB}\right)^{ab}_{\nu_1\nu_2;\mu_1\dots\mu_N}(g,\xi,p)\Delta^{\mu_1}\dots\Delta^{\mu_N}.
\end{equation}
The renormalisation of eq.~(\ref{def:GammaBBmu1muN}) requires a single counterterm
\begin{equation}
  \mathcal{Z}\left[\left(\Gamma^{(N)}_{1;BB}\right)^{ab}_{\nu_1\nu_2}\right] = Z_B\,Z^{(N)}_{1\,1}\,\left(\Gamma^{(N),0}_{1;BB}\right)^{ab}_{\nu_1\nu_2},
\end{equation}
where $(\Gamma^{(N)}_{1;BB})^{ab}_{\nu_1\nu_2}$ is the tree-level contribution to eq.~(\ref{def:GammaBBmu1muN}). In practice, applying the $R^*$ operation becomes computationally challenging at higher loop orders or higher $N$-values\footnote{The mass dimension of the operator increases with the spin, as $d=N+2$, and therefore also the degree of divergence of the Feynman diagrams of eq.~(\ref{def:GammaBBmu1muN}). This requires to compute high order terms in the Taylor expansion of the diagrams, see eq.~(\ref{def:Rstar}), which can generate large numbers of terms.}. Instead, we renormalise the {\textit{bare}} Green's functions, which are defined by using bare fields (including $\O^{(N),b}_i$) and bare parameters in eq.~(\ref{eq:GammaOBBmunu}). We compute the scalar quantities
\begin{equation}
 \label{eq:GammaOBB}
 \Gamma_{i;BB}^{(N)}(g_B,\xi_B,p^2) = \frac{\delta^{a_1a_2}}{N_A}\;\frac{g^{\nu_1\nu_2}H_N^{\mu_1\dots\mu_N}(p)}{(d-1)}\;\left(\Gamma_{i;BB}^{(N)}\right)^{a_1 a_2}_{\nu_1\nu_2;\mu_1\dots\mu_N}(g_B,\xi_B,p),
\end{equation}
where $d=4-2\epsilon$ is the dimension of spacetime and $N_A$ the dimension of the adjoint representation of the gauge group. $H_N^{\mu_1\dots\mu_N}(p)$ are the harmonic tensors introduced in refs. \cite{Gorishnii:1983su,Gorishnii:1986gn,Larin:1996wd}, which project the Green's function on its symmetric and traceless component. The harmonic projectors are defined to satisfy
\begin{align}
  H_N^{\mu_1\dots\mu_N}(p)\,g_{\mu_i\mu_j}&=0\quad\quad\text{and}\quad\quad H_N^{\mu_1\dots\mu_i\dots\mu_j\dots\mu_N}(p)=H_N^{\mu_1\dots\mu_j\dots\mu_i\dots\mu_N}(p)\quad\forall i,j\nn\\
  H_N^{\mu_1\dots\mu_N}(p)p_{\mu_N}&=H_{N-1}^{\mu_1\dots\mu_{N-1}}(p)\,p^2,
\end{align}
and they are explicitly constructed in \cite{Larin:1996wd}. We generated all the Feynman diagrams that contribute to eq.~(\ref{eq:GammaOBB}) with \verb+QGRAF+ \cite{Nogueira:1991ex}, we performed the color and Lorentz algebra with inhouse code, which is written in \verb+FORM+ \cite{Ruijl:2017dtg} and makes use of the package \verb+COLOR+ \cite{vanRitbergen:1998pn}. All the Feynman integrals that contribute to eq.~(\ref{eq:GammaOBB}) are massless two-point functions, also called {\textit{p-integrals}} \cite{Baikov:2010hf,Lee:2011jt,Georgoudis:2018olj,Georgoudis:2021onj}, which we computed with the code \verb+Forcer+ \cite{Ruijl:2017cxj}.
In order to renormalise eq.~(\ref{eq:GammaOBB}), we separate the tree-level from loop contributions
\begin{equation}
  \label{def:GammaiBBtree}
  \Gamma_{i;BB}^{(N)}(g_B,\xi_B) = \left\{
  \begin{array}{lc}
    \Gamma_{i;BB}^{(N),0} + \delta\Gamma_{i;BB}^{(N)}(g_B,\xi_B),&\quad\text{for}\;i=1\\
    \\
  \delta\Gamma_{i;BB}^{(N)}(g_B,\xi_B), &\quad\text{for}\; i\neq 1.
  \end{array}
  \right.
\end{equation}
where we omit the dependence on $p^2$, which can be reconstructed via dimensional analysis, and with
\begin{equation}
    \delta\Gamma_{i;BB}^{(N)}(g_B,\xi_B) = \sum_{r=1}^{\infty}\Gamma_{i;BB}^{(N),r}(\xi_B)\,\left(\frac{\alpha_{s,B}}{4\pi}\right)^r,
\end{equation}
where $\alpha_{s,B}=g_B^2/(4\pi)$. Upon considering $\Gamma_{1;BB}^{(N)}(g_B,\xi_B)$ in eq.~(\ref{eq:GammaOBB}), we find the renormalised correlator to obey
\begin{equation}
  \label{eq:ren}
K_\epsilon\left[Z_B\,\Big[Z_{1\,1}^{(N)} \Gamma_{1;BB}^{(N)}(g_B,\xi_B)+\sum_{i>1}\,\delta Z_{1\,i}^{(N)}\,\delta\Gamma_{i;BB}^{(N)}(g_B,\xi_B)\Big]\right] = 0,
\end{equation}
where $Z_B=\frac{1}{Z_g^2}$ is the renormalisation constant of the background field \cite{Abbott:1980hw,Abbott:1981ke}. Eq.~(\ref{eq:ren}) can be solved in terms of the renormalisation constant $Z_{1\,1}^{(N)}$ of the gauge invariant operator. Using identities eqs.~(\ref{def:ZNijelement}) and (\ref{def:GammaiBBtree}), we then get
\begin{equation}
  \label{eq:renExpl}
  \delta Z_{1\,1}^{(N)}\,\Gamma_{1;BB}^{(N),0} = -\frac{1}{Z_B}\,K_\epsilon\left[Z_B\sum_{i\geq1}Z_{1\,i}^{(N)}\delta\Gamma_{i;BB}^{(N)}(g_B,\xi_B)\right].
\end{equation}
The equation above holds to all loop orders. The renormalisation constants $Z_{1\,i>1}^{(N)}$, on the right hand-side of eq.~(\ref{eq:renExpl}), are required to renormalise sub-divergences of $\Gamma_{1;BB}^{(N)}$, which involve quantum gluons and/or a ghost-antighost pair. Each sub-divergence is proportional to one of the unphysical operators $\O^{(N)}_{i>1}$. This determines the maximal loop order at which $Z_{1\,i}^{(N)}$ has been computed, as shown in table \ref{tab:ghostmixingdiagsfixedN}.
The {\textit{diagonal}} renormalisation constant, $\delta Z_{1\,1}^{(N)}$, appears on both sides of eq.~(\ref{eq:renExpl}). However, we notice that the $Z_{1\,1}^{(N)}$ appearing on the right hand-side is multiplied by $\delta\Gamma_{1;BB}^{(N)}(g_B,\xi_B)$, which starts at $O(\alpha_s)$. Therefore, eq.~(\ref{eq:renExpl}) allows us to compute $Z_{1\,1}^{(N)}$ at $L$-loops, given the knowledge of $Z^{(N)}_{1\,i}$ at $l<L$ loops as discussed before. We plug the $l$-loop values of $Z^{(N)}_{1\,i}$, given in eqs.~(\ref{eq:Z(O2,Oeta)})-(\ref{eq:Z(O6,2)}), (\ref{eq:Z(O6,3)}), (\ref{eq:Z(O4,3)}) and (\ref{eq:Z(O6,4)}) respectively, into eq.~(\ref{eq:renExpl}). After computing the relevant correlators $\delta\Gamma_{i;BB}^{(N)}(g_B,\xi_B)$ at the required $L-l$ loop order, we find
\begin{align}
  \label{eq:Z11^2}
  Z_{1\,1}^{(2)}&=1 + O\left(\alpha_s^5\right),\\
  Z_{1\,1}^{(4)}&=1 + \frac{\alpha_s}{4\pi}\,\frac{21C_A}{5\epsilon} + \left(\frac{\alpha_s}{4\pi}\right)^2C_A^2\,\left(\frac{28}{25\epsilon^2}+\frac{7121}{1000\epsilon}\right) + \left(\frac{\alpha_s}{4\pi}\right)^3C_A^3\,\left(-\frac{1316}{1125\epsilon^3}\right.\nonumber\\
  &\left.-\frac{151441}{45000\epsilon^2}+\frac{103309639}{4050000\epsilon}\right)+ \left(\frac{\alpha_s}{4\pi}\right)^4\left\{C_A^4\left[\frac{11186}{5625\epsilon^4}+ \frac{1512989}{450000\epsilon^3}-\frac{5437269017}{162000000\epsilon^2}\right.\right.\nonumber\\
    &\left.\left.+\frac{1}{\epsilon}\Big(\frac{1502628149}{13500000} + \frac{1146397\zeta_3}{45000}-\frac{126\zeta_5}{5}\Big)\right]  + \frac{d_{AA}}{N_A}\,\left(\frac{21623}{600\epsilon}+\frac{3899\,\zeta_3}{15\epsilon}-\frac{1512\,\zeta_5}{5\epsilon}\right)\right\}\nonumber\\
  \label{eq:Z11^4}
  &+O(\alpha_s^5),\\
  Z_{1\,1}^{(6)}&=1 + \frac{\alpha_s}{4\pi}\,\frac{83\,C_A}{14\epsilon} + \left(\frac{\alpha_s}{4\pi}\right)^2C_A^2\,\left(\frac{7885}{1176\epsilon^2}+\frac{1506899}{148176\epsilon}\right) + \left(\frac{\alpha_s}{4\pi}\right)^3C_A^3\,\left(-\frac{465215}{148176\epsilon^3} \right.\nonumber\\
  &\left.+ \frac{243375989}{18670176\epsilon^2} + \frac{96390174479}{2613824640\epsilon}\right) + O\left(\alpha_s^4\right),
  \label{eq:Z11^6}
\end{align}
where $d_{AA}=d_4^{abcd}d_4^{abcd}$, with $d_4^{abcd}$ defined in eq.~(\ref{def:d4abcd}). As a check on our calculation, we verified that all non-local divergences of the form $1/\epsilon^p\,\log^q(\mu^2/p^2)$, which appear in the bare correlators, cancel upon combining the required counterterms. Furthermore, we verified that the dependence on the gauge parameter $\xi$ cancels up to three loops in eqs.~(\ref{eq:Z11^2}) and (\ref{eq:Z11^4}). The $O(\alpha_s^4)$-terms in those equations were computed only in Feynman gauge. Similarly, the calculation of the $O(\alpha^3_s)$-terms in eq.~(\ref{eq:Z11^6}) was performed in Feynman gauge and the cancellation of $\xi$ was verified to two loops.
The result $Z_{1\,1}^{(2)}=1$, in eq.~(\ref{eq:Z11^2}), agrees with the findings of refs. \cite{Freedman:1974gs,Freedman:1974ze}, which imply that $\O_1^{(2)}$ does not renormalise to all orders. Finally, by extracting the anomalous dimension $\gamma^{(N)}$, as written in eq.~(\ref{def:physAD}), we find agreement with the results at three and at four loops given in refs. \cite{Larin:1996wd} and \cite{Moch:2021qrk}.

\section{Conclusions}
\label{sec:conclusions}

In this paper we generalised a method, originally by Dixon and Taylor \cite{Dixon:1974ss},
for the construction of unphysical operators which are required for the renormalisation of Green's functions with insertions of twist-two gluonic operators. As one increases the loop order of the Green's function more unphysical operators are in general required for its renormalisation. The previously known basis was restricted to two-loop calculations, and it was unclear how to systematically extend it to higher loop order, thereby preventing the OPE method to be used for calculations of the singlet splitting functions. We have uncovered a general and systematic formalism for extending the basis to arbitrary loop order. Using this formalism we then worked out the explicit basis for calculations up to four-loop order and used it to perform calculations of the $N=2,4$ Mellin moments at four loops and the $N=6$ Mellin moment at three loops, obtaining the correct known results.

The formalism we developed can essentially be broken down to a few key concepts. The first is that we identified the gluonic gauge-variant operators in the Dixon-Taylor basis with EOM operators, these are not EOM operators of the gauge-fixed Lagrangian, but EOM operators of the gauge invariant part of the Lagrangian. With this identification we could easily write down the all-loop structure of the EOM operator. The second concept is that of a generalised gauge transformation which leaves invariant the Lagrangian made up of the gauge invariant and EOM operators. Following the works of Hamberg and Van Neerven \cite{Hamberg:1991qt} and Joglekar and Lee \cite{Joglekar:1975nu} this generalised gauge invariance is promoted to a generalised BRST symmetry. We then propose that the most general ghost operator can be written as the generalised BRST action acting on a single BRST ancestor operator. The ghost operator is therefore identified as an BRST-exact operator in the BRST generalised sense. This proposition not only reproduces the previously known ghost operators required at two loops, but we also confirmed that it complies with the theorems of Joglekar and Lee \cite{Joglekar:1975nu}. Indeed we show that the operators generated with our procedure can be always written as a sum of a BRST-exact term (in the sense of the original, not generalised, BRST transformations) and a term that vanishes on the equation of motion of the complete Yang-Mills lagrangian, as required by \cite{Joglekar:1975nu}.

We explored two further symmetry principles to simplify calculations of unphysical counterterms. The first is the anti-BRST symmetry which can be used to derive a ghost anti-ghost exchange symmetry of the ghost operators. This symmetry allows one to drastically reduce the number of independent unphysical operators. Another symmetry principle is background field gauge invariance, which we employed in our calculations. Background field invariance allows to do calculations without unphysical operators at the one-loop level, beyond one-loop counterterms a number of unphysical operators are however still required to perform calculations.

The task of computing unphysical counterterms requires the extraction of local renormalisation counterterms of Green's functions containing a ghost anti-ghost pair and multiple gluons. For instance to determine the anomalous dimension of the gauge invariant operator at the four-loop level generally requires, among others, the counterterms associated to Green's functions containing a ghost anti-ghost pair with two gluons at two loops and with one gluon at three loops. These quantities can thus not be extracted through a naive calculation of a self energy diagram. In this work we employed a fully automated implementation of the local $R^*$-operation, an operation which allows to extract the counterterms of Greens's functions of arbitrary many external particles from self energy diagrams, via the technique of IR rearrangement and IR subtractions. However the $R^*$-operation becomes very expensive for higher moments, due to the many derivatives and many counterterms one requires. Already at $N=6$ we found that the calculations were becoming prohibitively time-consuming even with substantial computing resources. It may be possible with further optimisation to push the $R^*$-approach to higher $N$, however we believe that a more streamlined approach could be more promising. We leave further improvements of this task to the future.

Assuming that the problem of calculating these UV counterterms can be solved efficiently one can expect that the methods presented here should allow for a much more efficient approach to computing Mellin moments of gluonic splitting functions at N3LO than the brute force approach which was currently used \cite{Moch:2021qrk}. To extend the methods presented here to singlet splitting functions containing also quarks will require further extensions of the formalism. We do not believe these to give major complications.

\section*{Acknowledgements}
We would like to thank Sven Moch, Jos Vermaseren and Andreas Vogt for many insightful discussions and their continuous encouragement. G.F. would like to thank Arnd Behring and Mattia Dalla Brida for numerous discussions on related topics.
F.H. is supported by the NWO Vidi grant 680-47-551 and the UKRI FLF Mr/S03479x/1.
G.F. is supported by the ERC Starting Grant 715049 ‘QCDforfuture’ with Principal Investigator Jennifer Smillie and by the STFC Consolidated Grant ‘Particle Physics at the Higgs Centre’.

\appendix
\section{Computing anomalous dimension in QCD}
\label{app:ADQCD}
It is convenient to spell out also a procedure to compute anomalous dimensions which does not rely on the background field method, but involves instead only the calculation of bare Green's functions with external gluons or ghosts. These were defined in eqs.~(\ref{def:GammaNgg}) and (\ref{def:ghostOicor}), respectively, and they read
\begin{align}
  \label{def:GammaNggmu1muN2}
  \left(\Gamma^{(N)}_{i;gg}\right)^{a_1a_2}_{\nu_1\nu_2} &\equiv \Delta^{\mu_1}\dots\Delta^{\mu_N}\;\left(\Gamma^{(N)}_{i;gg}\right)^{a_1a_2}_{\nu_1\nu_2;\mu_1\dots\mu_N},\\
  \label{def:ghostOicormu1muN2}
  \left(\Gamma^{(N)}_{i;c\bar{c}}\right)^{a_1a_2}&\equiv \Delta^{\mu_1}\dots\Delta^{\mu_N}\;\left(\Gamma^{(N)}_{i;c\bar{c}}\right)^{a_1a_2}_{\mu_1\dots\mu_N}.
\end{align}
We compute these correlators with the help of \verb+FORCER+, after applying harmonic and colour projectors to reduce eqs.~(\ref{def:GammaNggmu1muN2}) and (\ref{def:ghostOicormu1muN2}), as described below eq.~(\ref{eq:GammaOBB})
\begin{align}
  \label{eq:GammaNggSCAL}
  \Gamma^{(N)}_{i;gg}(g_B,\xi_B,p^2)&=\frac{\delta^{a_1a_2}}{N_A}\frac{g^{\nu_1\nu_2}\,H^{\mu_1\dots\mu_N}(p)}{(d-1)}\,\left(\Gamma^{(N)}_{i;gg}\right)^{a_1a_2}_{\nu_1\nu_2;\mu_1\dots\mu_N},\\
  \label{eq:GammaNccSCAL}
  \Gamma^{(N)}_{i;c\bar{c}}(g_B,\xi_B,p^2)&=\frac{\delta^{a_1a_2}}{N_A} \,H^{\mu_1\dots\mu_N}(p)\, \left(\Gamma^{(N)}_{i;c\bar{c}}\right)^{a_1a_2}_{\mu_1\dots\mu_N}.
\end{align}
By definition, the ghost correlator $\Gamma_{i;c\bar{c}}$ doesn't vanish at tree level, only if we consider insertion of the operator $\O_2^{(N)}$, which is chosen to contain the term $\O_G^{(N),1}$ eq.~(\ref{eq:BRSTop1}), as we have done in the construction of operator bases for $N=2,4$ and $6$ in eqs.~(\ref{eq:Oeta(N=2)}), (\ref{eq:basisO42}) and (\ref{eq:basisO62}). The gluon correlator $\Gamma^{(N)}_{i;gg}$ receives contributions at tree level from both the gauge invariant operator $\O_1^{(N)}$ and from $\O_2^{(N)}$, which includes the term $\O_{\text{EOM}}^{(N),1}$, eq.~(\ref{def:EOMop}), related to $\O_G^{(N),1}$ by (generalised) BRST symmetry. We get
\begin{subequations}
  \begin{align}
  \label{eq:GammaNggtree}
  \Gamma^{(N)}_{i;gg}(g_B,\xi_B,p^2)&=\left\{
  \begin{array}{lr}
    \Gamma^{(N),0}_{i;gg}(p^2) + \delta\Gamma^{(N)}_{i;gg}(g_B,\xi_B,p^2) & \quad \text{for}\;i=1,2\\\\
    \delta\Gamma^{(N)}_{i;gg}(g_B,\xi_B,p^2) &\quad \text{for}\;i>2
  \end{array}
  \right.\\
  \label{eq:GammaNcctree}  
  \Gamma^{(N)}_{i;c\bar{c}}(g_B,\xi_B,p^2)&=\left\{
  \begin{array}{lr}
    \Gamma^{(N),0}_{i;c\bar{c}}(p^2) + \delta\Gamma^{(N)}_{i;c\bar c}(g_B,\xi_B,p^2) & \quad \text{for}\;i=2\\\\
    \delta\Gamma^{(N)}_{i;c\bar{c}}(g_B,\xi_B,p^2) &\quad \text{for}\;i\neq2
  \end{array}
  \right.
  \end{align}
\end{subequations}
In order to compute the renormalisation constant $Z_{1\,1}^{(N)}$, we renormalise the bare correlators $\Gamma_{1;gg}^{(N)}$ and $\Gamma_{1;c\bar{c}}^{(N)}$, where we inserted the gauge invariant operator $\O_1^{(N),b}$
\begin{subequations}
\begin{align}
  K_\epsilon\Big[Z_3\sum_{i\geq1}Z_{1\,i}\;\Gamma^{(N)}_{i;gg}(g_B,\xi_B,p^2)\Big] &= 0,\\
  K_\epsilon\Big[Z_c\sum_{i\geq1}Z_{1\,i}\;\Gamma^{(N)}_{i;c\bar{c}}(g_B,\xi_B,p^2)\Big] &= 0.
\end{align}
\end{subequations}
We separate the contributions of the tree-level terms and of the loop corrections to the Green's functions, according to eqs.~(\ref{eq:GammaNggtree}) and (\ref{eq:GammaNcctree}) and we solve the equations above for $Z_{1\,1}^{(N)}$ and $Z_{1\,2}^{(N)}$. We find
\begin{align}
  \label{eq:Z11QCD}
  Z_{1\,1}\Gamma^{(N),0}_{1;gg} + \delta Z_{1\,2}\Gamma^{(N),0}_{2;gg} &=-\frac{1}{Z_3}\,K_\epsilon\Big[Z_3\sum_{i\geq1}Z_{1\,i}\;\delta\Gamma^{(N)}_{i;gg}(g_B,\xi_B,p^2)\Big],\\
  \label{eq:Z12QCD}
  \delta Z_{1\,2}\Gamma^{(N),0}_{2;c\bar{c}} &=-\frac{1}{Z_c}\,K_\epsilon\Big[Z_c\sum_{i\geq1}Z_{1\,i}\;\delta\Gamma^{(N)}_{i;c\bar{c}}(g_B,\xi_B,p^2)\Big].
\end{align}
We solve the equations above order-by-order is $\alpha_s$. Provided we have knowledge of the renormalisation constants $Z_{1\,i}^{(N)}$ up to $L-1$ loops, which enter the right hand-side of both eqs.~(\ref{eq:Z11QCD}) and (\ref{eq:Z12QCD}), we determine $\delta Z_{1\,2}^{(N)}$ to $L$ loops by means of eq.~(\ref{eq:Z12QCD}). We applied this method to compute $Z_{1\,2}^{(N)}$ in eqs.(\ref{eq:Z(O2,Oeta)})-(\ref{eq:Z(O6,2)}) with complete dependence on the gauge parameter $\xi$. Finally, by replacing $\delta Z_{1\,2}^{(N)}$ at $L$-loop in the left hand-side of eq.~(\ref{eq:Z11QCD}), we determine the renormalisation $Z_{1\,1}^{(N)}$ to $L$ loops.

\bibliographystyle{JHEP}
\bibliography{refs}

\providecommand{\href}[2]{#2}\begingroup\raggedright\begin{thebibliography}{10}

\bibitem{Anastasiou:2015vya}
C.~Anastasiou, C.~Duhr, F.~Dulat, F.~Herzog and B.~Mistlberger, \emph{{Higgs
  Boson Gluon-Fusion Production in QCD at Three Loops}},
  \href{http://dx.doi.org/10.1103/PhysRevLett.114.212001}{\emph{Phys. Rev.
  Lett.} {\bf 114} (2015) 212001} [\href{https://arxiv.org/abs/1503.06056}{{\tt
  arXiv:1503.06056}}].

\bibitem{Anastasiou:2016cez}
C.~Anastasiou, C.~Duhr, F.~Dulat, E.~Furlan, T.~Gehrmann, F.~Herzog et~al.,
  \emph{{High precision determination of the gluon fusion Higgs boson
  cross-section at the LHC}},
  \href{http://dx.doi.org/10.1007/JHEP05(2016)058}{\emph{JHEP} {\bf 05} (2016)
  058} [\href{https://arxiv.org/abs/1602.00695}{{\tt arXiv:1602.00695}}].

\bibitem{Mistlberger:2018etf}
B.~Mistlberger, \emph{{Higgs boson production at hadron colliders at N$^{3}$LO
  in QCD}}, \href{http://dx.doi.org/10.1007/JHEP05(2018)028}{\emph{JHEP} {\bf
  05} (2018) 028} [\href{https://arxiv.org/abs/1802.00833}{{\tt
  arXiv:1802.00833}}].

\bibitem{Duhr:2019kwi}
C.~Duhr, F.~Dulat and B.~Mistlberger, \emph{{Higgs Boson Production in
  Bottom-Quark Fusion to Third Order in the Strong Coupling}},
  \href{http://dx.doi.org/10.1103/PhysRevLett.125.051804}{\emph{Phys. Rev.
  Lett.} {\bf 125} (2020) 051804} [\href{https://arxiv.org/abs/1904.09990}{{\tt
  arXiv:1904.09990}}].

\bibitem{Duhr:2020seh}
C.~Duhr, F.~Dulat and B.~Mistlberger, \emph{{Drell-Yan Cross Section to Third
  Order in the Strong Coupling Constant}},
  \href{http://dx.doi.org/10.1103/PhysRevLett.125.172001}{\emph{Phys. Rev.
  Lett.} {\bf 125} (2020) 172001} [\href{https://arxiv.org/abs/2001.07717}{{\tt
  arXiv:2001.07717}}].

\bibitem{Chen:2021isd}
X.~Chen, T.~Gehrmann, E.W.N.~Glover, A.~Huss, B.~Mistlberger and A.~Pelloni,
  \emph{{Fully Differential Higgs Boson Production to Third Order in QCD}},
  \href{http://dx.doi.org/10.1103/PhysRevLett.127.072002}{\emph{Phys. Rev.
  Lett.} {\bf 127} (2021) 072002} [\href{https://arxiv.org/abs/2102.07607}{{\tt
  arXiv:2102.07607}}].

\bibitem{Duhr:2020sdp}
C.~Duhr, F.~Dulat and B.~Mistlberger, \emph{{Charged current Drell-Yan
  production at N$^{3}$LO}},
  \href{http://dx.doi.org/10.1007/JHEP11(2020)143}{\emph{JHEP} {\bf 11} (2020)
  143} [\href{https://arxiv.org/abs/2007.13313}{{\tt arXiv:2007.13313}}].

\bibitem{Bargiela:2021wuy}
P.~Bargiela, F.~Caola, A.~von~Manteuffel and L.~Tancredi, \emph{{Three-loop
  helicity amplitudes for diphoton production in gluon fusion}},
  \href{https://arxiv.org/abs/2111.13595}{{\tt arXiv:2111.13595}}.

\bibitem{Caola:2021izf}
F.~Caola, A.~Chakraborty, G.~Gambuti, A.~von~Manteuffel and L.~Tancredi,
  \emph{{Three-loop gluon scattering in QCD and the gluon Regge trajectory}},
  \href{https://arxiv.org/abs/2112.11097}{{\tt arXiv:2112.11097}}.

\bibitem{Floratos:1978ny}
E.G.~Floratos, D.A.~Ross and C.T.~Sachrajda, \emph{{Higher Order Effects in
  Asymptotically Free Gauge Theories. 2. Flavor Singlet Wilson Operators and
  Coefficient Functions}},
  \href{http://dx.doi.org/10.1016/0550-3213(79)90094-4}{\emph{Nucl. Phys. B}
  {\bf 152} (1979) 493}.

\bibitem{Gonzalez-Arroyo:1979qht}
A.~Gonzalez-Arroyo and C.~Lopez, \emph{{Second Order Contributions to the
  Structure Functions in Deep Inelastic Scattering. 3. The Singlet Case}},
  \href{http://dx.doi.org/10.1016/0550-3213(80)90207-2}{\emph{Nucl. Phys. B}
  {\bf 166} (1980) 429}.

\bibitem{Furmanski:1980cm}
W.~Furmanski and R.~Petronzio, \emph{{Singlet Parton Densities Beyond Leading
  Order}}, \href{http://dx.doi.org/10.1016/0370-2693(80)90636-X}{\emph{Phys.
  Lett. B} {\bf 97} (1980) 437}.

\bibitem{Hamberg:1991qt}
R.~Hamberg and W.L.~van~Neerven, \emph{{The Correct renormalization of the
  gluon operator in a covariant gauge}},
  \href{http://dx.doi.org/10.1016/0550-3213(92)90593-Z}{\emph{Nucl. Phys. B}
  {\bf 379} (1992) 143}.

\bibitem{Vogelsang:1995vh}
W.~Vogelsang, \emph{{A Rederivation of the spin dependent next-to-leading order
  splitting functions}},
  \href{http://dx.doi.org/10.1103/PhysRevD.54.2023}{\emph{Phys. Rev. D} {\bf
  54} (1996) 2023} [\href{https://arxiv.org/abs/hep-ph/9512218}{{\tt
  hep-ph/9512218}}].

\bibitem{Mertig:1995ny}
R.~Mertig and W.L.~van~Neerven, \emph{{The Calculation of the two loop spin
  splitting functions P(ij)(1)(x)}},
  \href{http://dx.doi.org/10.1007/s002880050138}{\emph{Z. Phys. C} {\bf 70}
  (1996) 637} [\href{https://arxiv.org/abs/hep-ph/9506451}{{\tt
  hep-ph/9506451}}].

\bibitem{Ellis:1996nn}
R.K.~Ellis and W.~Vogelsang, \emph{{The Evolution of parton distributions
  beyond leading order: The Singlet case}},
  \href{https://arxiv.org/abs/hep-ph/9602356}{{\tt hep-ph/9602356}}.

\bibitem{Matiounine:1998ky}
Y.~Matiounine, J.~Smith and W.L.~van~Neerven, \emph{{Two loop operator matrix
  elements calculated up to finite terms}},
  \href{http://dx.doi.org/10.1103/PhysRevD.57.6701}{\emph{Phys. Rev. D} {\bf
  57} (1998) 6701} [\href{https://arxiv.org/abs/hep-ph/9801224}{{\tt
  hep-ph/9801224}}].

\bibitem{Matiounine:1998re}
Y.~Matiounine, J.~Smith and W.L.~van~Neerven, \emph{{Two loop operator matrix
  elements calculated up to finite terms for polarized deep inelastic lepton -
  hadron scattering}},
  \href{http://dx.doi.org/10.1103/PhysRevD.58.076002}{\emph{Phys. Rev. D} {\bf
  58} (1998) 076002} [\href{https://arxiv.org/abs/hep-ph/9803439}{{\tt
  hep-ph/9803439}}].

\bibitem{Larin:1993vu}
S.A.~Larin, T.~van~Ritbergen and J.A.M.~Vermaseren, \emph{{The Next
  next-to-leading QCD approximation for nonsinglet moments of deep inelastic
  structure functions}},
  \href{http://dx.doi.org/10.1016/0550-3213(94)90268-2}{\emph{Nucl. Phys. B}
  {\bf 427} (1994) 41}.

\bibitem{Larin:1996wd}
S.A.~Larin, P.~Nogueira, T.~van~Ritbergen and J.A.M.~Vermaseren, \emph{{The
  Three loop QCD calculation of the moments of deep inelastic structure
  functions}},
  \href{http://dx.doi.org/10.1016/S0550-3213(97)80038-7}{\emph{Nucl. Phys. B}
  {\bf 492} (1997) 338} [\href{https://arxiv.org/abs/hep-ph/9605317}{{\tt
  hep-ph/9605317}}].

\bibitem{Moch:2004pa}
S.~Moch, J.A.M.~Vermaseren and A.~Vogt, \emph{{The Three loop splitting
  functions in QCD: The Nonsinglet case}},
  \href{http://dx.doi.org/10.1016/j.nuclphysb.2004.03.030}{\emph{Nucl. Phys. B}
  {\bf 688} (2004) 101} [\href{https://arxiv.org/abs/hep-ph/0403192}{{\tt
  hep-ph/0403192}}].

\bibitem{Vogt:2004mw}
A.~Vogt, S.~Moch and J.A.M.~Vermaseren, \emph{{The Three-loop splitting
  functions in QCD: The Singlet case}},
  \href{http://dx.doi.org/10.1016/j.nuclphysb.2004.04.024}{\emph{Nucl. Phys. B}
  {\bf 691} (2004) 129} [\href{https://arxiv.org/abs/hep-ph/0404111}{{\tt
  hep-ph/0404111}}].

\bibitem{Ablinger:2014nga}
J.~Ablinger, A.~Behring, J.~Bl\"umlein, A.~De~Freitas, A.~von~Manteuffel and
  C.~Schneider, \emph{{The 3-loop pure singlet heavy flavor contributions to
  the structure function $F_2(x,Q^2)$ and the anomalous dimension}},
  \href{http://dx.doi.org/10.1016/j.nuclphysb.2014.10.008}{\emph{Nucl. Phys. B}
  {\bf 890} (2014) 48} [\href{https://arxiv.org/abs/1409.1135}{{\tt
  arXiv:1409.1135}}].

\bibitem{Ablinger:2017tan}
J.~Ablinger, A.~Behring, J.~Bl\"umlein, A.~De~Freitas, A.~von~Manteuffel and
  C.~Schneider, \emph{{The three-loop splitting functions $P_{qg}^{(2)}$ and
  $P_{gg}^{(2, N_F)}$}},
  \href{http://dx.doi.org/10.1016/j.nuclphysb.2017.06.004}{\emph{Nucl. Phys. B}
  {\bf 922} (2017) 1} [\href{https://arxiv.org/abs/1705.01508}{{\tt
  arXiv:1705.01508}}].

\bibitem{Behring:2019tus}
A.~Behring, J.~Bl\"umlein, A.~De~Freitas, A.~Goedicke, S.~Klein,
  A.~von~Manteuffel et~al., \emph{{The Polarized Three-Loop Anomalous
  Dimensions from On-Shell Massive Operator Matrix Elements}},
  \href{http://dx.doi.org/10.1016/j.nuclphysb.2019.114753}{\emph{Nucl. Phys. B}
  {\bf 948} (2019) 114753} [\href{https://arxiv.org/abs/1908.03779}{{\tt
  arXiv:1908.03779}}].

\bibitem{Ablinger:2019etw}
J.~Ablinger, A.~Behring, J.~Bl\"umlein, A.~De~Freitas, A.~von~Manteuffel,
  C.~Schneider et~al., \emph{{The three-loop single mass polarized pure singlet
  operator matrix element}},
  \href{http://dx.doi.org/10.1016/j.nuclphysb.2020.114945}{\emph{Nucl. Phys. B}
  {\bf 953} (2020) 114945} [\href{https://arxiv.org/abs/1912.02536}{{\tt
  arXiv:1912.02536}}].

\bibitem{Blumlein:2021enk}
J.~Bl\"umlein, P.~Marquard, C.~Schneider and K.~Sch\"onwald, \emph{{The
  three-loop unpolarized and polarized non-singlet anomalous dimensions from
  off shell operator matrix elements}},
  \href{http://dx.doi.org/10.1016/j.nuclphysb.2021.115542}{\emph{Nucl. Phys. B}
  {\bf 971} (2021) 115542} [\href{https://arxiv.org/abs/2107.06267}{{\tt
  arXiv:2107.06267}}].

\bibitem{Blumlein:2021ryt}
J.~Bl\"umlein, P.~Marquard, C.~Schneider and K.~Sch\"onwald, \emph{{The
  three-loop polarized singlet anomalous dimensions from off-shell operator
  matrix elements}},
  \href{http://dx.doi.org/10.1007/JHEP01(2022)193}{\emph{JHEP} {\bf 01} (2022)
  193} [\href{https://arxiv.org/abs/2111.12401}{{\tt arXiv:2111.12401}}].

\bibitem{Velizhanin:2014fua}
V.N.~Velizhanin, \emph{{Four-loop anomalous dimension of the third and fourth
  moments of the nonsinglet twist-2 operator in QCD}},
  \href{http://dx.doi.org/10.1142/S0217751X20501997}{\emph{Int. J. Mod. Phys.
  A} {\bf 35} (2020) 2050199} [\href{https://arxiv.org/abs/1411.1331}{{\tt
  arXiv:1411.1331}}].

\bibitem{Moch:2017uml}
S.~Moch, B.~Ruijl, T.~Ueda, J.A.M.~Vermaseren and A.~Vogt, \emph{{Four-Loop
  Non-Singlet Splitting Functions in the Planar Limit and Beyond}},
  \href{http://dx.doi.org/10.1007/JHEP10(2017)041}{\emph{JHEP} {\bf 10} (2017)
  041} [\href{https://arxiv.org/abs/1707.08315}{{\tt arXiv:1707.08315}}].

\bibitem{Davies:2016jie}
J.~Davies, A.~Vogt, B.~Ruijl, T.~Ueda and J.A.M.~Vermaseren, \emph{{Large-$n_f$
  contributions to the four-loop splitting functions in QCD}},
  \href{http://dx.doi.org/10.1016/j.nuclphysb.2016.12.012}{\emph{Nucl. Phys. B}
  {\bf 915} (2017) 335} [\href{https://arxiv.org/abs/1610.07477}{{\tt
  arXiv:1610.07477}}].

\bibitem{Gracey:1996ad}
J.A.~Gracey, \emph{{Anomalous dimensions of operators in polarized deep
  inelastic scattering at O(1/N(f))}},
  \href{http://dx.doi.org/10.1016/S0550-3213(96)00485-3}{\emph{Nucl. Phys. B}
  {\bf 480} (1996) 73} [\href{https://arxiv.org/abs/hep-ph/9609301}{{\tt
  hep-ph/9609301}}].

\bibitem{Herzog:2018kwj}
F.~Herzog, S.~Moch, B.~Ruijl, T.~Ueda, J.A.M.~Vermaseren and A.~Vogt,
  \emph{{Five-loop contributions to low-N non-singlet anomalous dimensions in
  QCD}}, \href{http://dx.doi.org/10.1016/j.physletb.2019.01.060}{\emph{Phys.
  Lett. B} {\bf 790} (2019) 436} [\href{https://arxiv.org/abs/1812.11818}{{\tt
  arXiv:1812.11818}}].

\bibitem{Moch:2021qrk}
S.~Moch, B.~Ruijl, T.~Ueda, J.A.M.~Vermaseren and A.~Vogt, \emph{{Low moments
  of the four-loop splitting functions in QCD}},
  \href{http://dx.doi.org/10.1016/j.physletb.2021.136853}{\emph{Phys. Lett. B}
  {\bf 825} (2022) 136853} [\href{https://arxiv.org/abs/2111.15561}{{\tt
  arXiv:2111.15561}}].

\bibitem{Gross:1974cs}
D.J.~Gross and F.~Wilczek, \emph{{Asymptotically free gauge theories. 2.}},
  \href{http://dx.doi.org/10.1103/PhysRevD.9.980}{\emph{Phys. Rev. D} {\bf 9}
  (1974) 980}.

\bibitem{Georgi:1974wnj}
H.~Georgi and H.D.~Politzer, \emph{{Electroproduction scaling in an
  asymptotically free theory of strong interactions}},
  \href{http://dx.doi.org/10.1103/PhysRevD.9.416}{\emph{Phys. Rev. D} {\bf 9}
  (1974) 416}.

\bibitem{Dixon:1974ss}
J.A.~Dixon and J.C.~Taylor, \emph{{Renormalization of wilson operators in gauge
  theories}}, \href{http://dx.doi.org/10.1016/0550-3213(74)90598-7}{\emph{Nucl.
  Phys. B} {\bf 78} (1974) 552}.

\bibitem{Blumlein:2022ndg}
J.~Bl\"umlein, P.~Marquard, C.~Schneider and K.~Sch\"onwald, \emph{{The
  Two-Loop Massless Off-Shell QCD Operator Matrix Elements to Finite Terms}},
  \href{https://arxiv.org/abs/2202.03216}{{\tt arXiv:2202.03216}}.

\bibitem{Collins:1994ee}
J.C.~Collins and R.J.~Scalise, \emph{{The Renormalization of composite
  operators in Yang-Mills theories using general covariant gauge}},
  \href{http://dx.doi.org/10.1103/PhysRevD.50.4117}{\emph{Phys. Rev. D} {\bf
  50} (1994) 4117} [\href{https://arxiv.org/abs/hep-ph/9403231}{{\tt
  hep-ph/9403231}}].

\bibitem{Kluberg-Stern:1974nmx}
H.~Kluberg-Stern and J.B.~Zuber, \emph{{Renormalization of Nonabelian Gauge
  Theories in a Background Field Gauge. 1. Green Functions}},
  \href{http://dx.doi.org/10.1103/PhysRevD.12.482}{\emph{Phys. Rev. D} {\bf 12}
  (1975) 482}.

\bibitem{Kluberg-Stern:1975ebk}
H.~Kluberg-Stern and J.B.~Zuber, \emph{{Renormalization of Nonabelian Gauge
  Theories in a Background Field Gauge. 2. Gauge Invariant Operators}},
  \href{http://dx.doi.org/10.1103/PhysRevD.12.3159}{\emph{Phys. Rev. D} {\bf
  12} (1975) 3159}.

\bibitem{Joglekar:1975nu}
S.D.~Joglekar and B.W.~Lee, \emph{{General Theory of Renormalization of Gauge
  Invariant Operators}},
  \href{http://dx.doi.org/10.1016/0003-4916(76)90225-6}{\emph{Annals Phys.}
  {\bf 97} (1976) 160}.

\bibitem{Joglekar:1976eb}
S.D.~Joglekar, \emph{{Local Operator Products in Gauge Theories. 1.}},
  \href{http://dx.doi.org/10.1016/0003-4916(77)90014-8}{\emph{Annals Phys.}
  {\bf 108} (1977) 233}.

\bibitem{Joglekar:1976pe}
S.D.~Joglekar, \emph{{Local Operator Products in Gauge Theories. 2.}},
  \href{http://dx.doi.org/10.1016/0003-4916(77)90170-1}{\emph{Annals Phys.}
  {\bf 109} (1977) 210}.

\bibitem{Henneaux:1993jn}
M.~Henneaux, \emph{{Remarks on the renormalization of gauge invariant operators
  in Yang-Mills theory}},
  \href{http://dx.doi.org/10.1016/0370-2693(93)91187-R}{\emph{Phys. Lett. B}
  {\bf 313} (1993) 35} [\href{https://arxiv.org/abs/hep-th/9306101}{{\tt
  hep-th/9306101}}].

\bibitem{Curci:1976bt}
G.~Curci and R.~Ferrari, \emph{{On a Class of Lagrangian Models for Massive and
  Massless Yang-Mills Fields}},
  \href{http://dx.doi.org/10.1007/BF02729999}{\emph{Nuovo Cim. A} {\bf 32}
  (1976) 151}.

\bibitem{Ojima:1980da}
I.~Ojima, \emph{{Another BRS Transformation}},
  \href{http://dx.doi.org/10.1143/PTP.64.625}{\emph{Prog. Theor. Phys.} {\bf
  64} (1980) 625}.

\bibitem{Baulieu:1981sb}
L.~Baulieu and J.~Thierry-Mieg, \emph{{The Principle of BRS Symmetry: An
  Alternative Approach to Yang-Mills Theories}},
  \href{http://dx.doi.org/10.1016/0550-3213(82)90454-0}{\emph{Nucl. Phys. B}
  {\bf 197} (1982) 477}.

\bibitem{Binosi:2013cea}
D.~Binosi and A.~Quadri, \emph{{Anti-BRST symmetry and background field
  method}}, \href{http://dx.doi.org/10.1103/PhysRevD.88.085036}{\emph{Phys.
  Rev. D} {\bf 88} (2013) 085036} [\href{https://arxiv.org/abs/1309.1021}{{\tt
  arXiv:1309.1021}}].

\bibitem{DeWitt:1967ub}
B.S.~DeWitt, \emph{{Quantum Theory of Gravity. 2. The Manifestly Covariant
  Theory}}, \href{http://dx.doi.org/10.1103/PhysRev.162.1195}{\emph{Phys. Rev.}
  {\bf 162} (1967) 1195}.

\bibitem{tHooft:1975uxh}
G.~'t~Hooft, \emph{{The Background Field Method in Gauge Field Theories}},  in
  \emph{{12th Annual Winter School of Theoretical Physics}}, 1975.

\bibitem{Abbott:1980hw}
L.F.~Abbott, \emph{{The Background Field Method Beyond One Loop}},
  \href{http://dx.doi.org/10.1016/0550-3213(81)90371-0}{\emph{Nucl. Phys. B}
  {\bf 185} (1981) 189}.

\bibitem{Abbott:1981ke}
L.F.~Abbott, \emph{{Introduction to the Background Field Method}}, {\emph{Acta
  Phys. Polon. B} {\bf 13} (1982) 33}.

\bibitem{Sarkar:1974ni}
S.~Sarkar and H.~Strubbe, \emph{{Anomalous Dimensions in Background Field
  Gauges}}, \href{http://dx.doi.org/10.1016/0550-3213(75)90633-1}{\emph{Nucl.
  Phys. B} {\bf 90} (1975) 45}.

\bibitem{Nakanishi:1966zz}
N.~Nakanishi, \emph{{Covariant Quantization of the Electromagnetic Field in the
  Landau Gauge}}, \href{http://dx.doi.org/10.1143/PTP.35.1111}{\emph{Prog.
  Theor. Phys.} {\bf 35} (1966) 1111}.

\bibitem{Lautrup:1967zz}
B.~Lautrup, \emph{{Canonical Quantum Electrodynamics in covariant Gauges}}, .

\bibitem{Becchi:1975nq}
C.~Becchi, A.~Rouet and R.~Stora, \emph{{Renormalization of Gauge Theories}},
  \href{http://dx.doi.org/10.1016/0003-4916(76)90156-1}{\emph{Annals Phys.}
  {\bf 98} (1976) 287}.

\bibitem{Tyutin:1975qk}
I.V.~Tyutin, \emph{{Gauge Invariance in Field Theory and Statistical Physics in
  Operator Formalism}},  \href{https://arxiv.org/abs/0812.0580}{{\tt
  arXiv:0812.0580}}.

\bibitem{Bern:2008qj}
Z.~Bern, J.J.M.~Carrasco and H.~Johansson, \emph{{New Relations for
  Gauge-Theory Amplitudes}},
  \href{http://dx.doi.org/10.1103/PhysRevD.78.085011}{\emph{Phys. Rev. D} {\bf
  78} (2008) 085011} [\href{https://arxiv.org/abs/0805.3993}{{\tt
  arXiv:0805.3993}}].

\bibitem{vanRitbergen:1998pn}
T.~van~Ritbergen, A.N.~Schellekens and J.A.M.~Vermaseren, \emph{{Group theory
  factors for Feynman diagrams}},
  \href{http://dx.doi.org/10.1142/S0217751X99000038}{\emph{Int. J. Mod. Phys.
  A} {\bf 14} (1999) 41} [\href{https://arxiv.org/abs/hep-ph/9802376}{{\tt
  hep-ph/9802376}}].

\bibitem{Ruijl:2017dtg}
B.~Ruijl, T.~Ueda and J.~Vermaseren, \emph{{FORM version 4.2}},
  \href{https://arxiv.org/abs/1707.06453}{{\tt arXiv:1707.06453}}.

\bibitem{Pascual:1984zb}
P.~Pascual and R.~Tarrach, \emph{{QCD: Renormalization for the Practitioner}},
  vol.~194 (1984).

\bibitem{Chetyrkin:1982nn}
K.G.~Chetyrkin and F.V.~Tkachov, \emph{{Infrared R Operation and ultraviolet
  Counterterms in the MS scheme}},
  \href{http://dx.doi.org/10.1016/0370-2693(82)90358-6}{\emph{Phys. Lett.} {\bf
  114B} (1982) 340}.

\bibitem{Chetyrkin:1984xa}
K.G.~Chetyrkin and V.A.~Smirnov, \emph{{R* operation corrected}},
  \href{http://dx.doi.org/10.1016/0370-2693(84)91291-7}{\emph{Phys. Lett.} {\bf
  144B} (1984) 419}.

\bibitem{Smirnov:1986me}
V.A.~Smirnov and K.G.~Chetyrkin, \emph{{R* Operation in the Minimal Subtraction
  Scheme}}, \href{http://dx.doi.org/10.1007/BF01017902}{\emph{Theor. Math.
  Phys.} {\bf 63} (1985) 462}.

\bibitem{Chetyrkin:2017ppe}
K.G.~Chetyrkin, \emph{{Combinatorics of $\mathbf{R}$-, $\mathbf{R^{-1}}$-, and
  $\mathbf{R^*}$-operations and asymptotic expansions of feynman integrals in
  the limit of large momenta and masses}},
  \href{https://arxiv.org/abs/1701.08627}{{\tt arXiv:1701.08627}}.

\bibitem{Herzog:2017bjx}
F.~Herzog and B.~Ruijl, \emph{{The R$^{*}$-operation for Feynman graphs with
  generic numerators}},
  \href{http://dx.doi.org/10.1007/JHEP05(2017)037}{\emph{JHEP} {\bf 05} (2017)
  037} [\href{https://arxiv.org/abs/1703.03776}{{\tt arXiv:1703.03776}}].

\bibitem{deVries:2019nsu}
J.~de~Vries, G.~Falcioni, F.~Herzog and B.~Ruijl, \emph{{Two- and three-loop
  anomalous dimensions of Weinberg\textquoteright{}s dimension-six CP-odd
  gluonic operator}},
  \href{http://dx.doi.org/10.1103/PhysRevD.102.016010}{\emph{Phys. Rev. D} {\bf
  102} (2020) 016010} [\href{https://arxiv.org/abs/1907.04923}{{\tt
  arXiv:1907.04923}}].

\bibitem{Beekveldt:2020kzk}
R.~Beekveldt, M.~Borinsky and F.~Herzog, \emph{{The Hopf algebra structure of
  the R*-operation}},
  \href{http://dx.doi.org/10.1007/JHEP07(2020)061}{\emph{JHEP} {\bf 07} (2020)
  061} [\href{https://arxiv.org/abs/2003.04301}{{\tt arXiv:2003.04301}}].

\bibitem{Cao:2021cdt}
W.~Cao, F.~Herzog, T.~Melia and J.R.~Nepveu, \emph{{Renormalization and
  non-renormalization of scalar EFTs at higher orders}},
  \href{http://dx.doi.org/10.1007/JHEP09(2021)014}{\emph{JHEP} {\bf 09} (2021)
  014} [\href{https://arxiv.org/abs/2105.12742}{{\tt arXiv:2105.12742}}].

\bibitem{Gorishnii:1983su}
S.G.~Gorishnii, S.A.~Larin and F.V.~Tkachov, \emph{{The Algorithm for OPE
  Coefficient Functions in the MS scheme}},
  \href{http://dx.doi.org/10.1016/0370-2693(83)91439-9}{\emph{Phys. Lett. B}
  {\bf 124} (1983) 217}.

\bibitem{Gorishnii:1986gn}
S.G.~Gorishnii and S.A.~Larin, \emph{{Coefficient Functions of Asymptotic
  Operator Expansions in Minimal Subtraction Scheme}},
  \href{http://dx.doi.org/10.1016/0550-3213(87)90283-5}{\emph{Nucl. Phys. B}
  {\bf 283} (1987) 452}.

\bibitem{Nogueira:1991ex}
P.~Nogueira, \emph{{Automatic Feynman graph generation}},
  \href{http://dx.doi.org/10.1006/jcph.1993.1074}{\emph{J. Comput. Phys.} {\bf
  105} (1993) 279}.

\bibitem{Baikov:2010hf}
P.A.~Baikov and K.G.~Chetyrkin, \emph{{Four Loop Massless Propagators: An
  Algebraic Evaluation of All Master Integrals}},
  \href{http://dx.doi.org/10.1016/j.nuclphysb.2010.05.004}{\emph{Nucl. Phys.}
  {\bf B837} (2010) 186} [\href{https://arxiv.org/abs/1004.1153}{{\tt
  arXiv:1004.1153}}].

\bibitem{Lee:2011jt}
R.N.~Lee, A.V.~Smirnov and V.A.~Smirnov, \emph{{Master Integrals for Four-Loop
  Massless Propagators up to Transcendentality Weight Twelve}},
  \href{http://dx.doi.org/10.1016/j.nuclphysb.2011.11.005}{\emph{Nucl. Phys. B}
  {\bf 856} (2012) 95} [\href{https://arxiv.org/abs/1108.0732}{{\tt
  arXiv:1108.0732}}].

\bibitem{Georgoudis:2018olj}
A.~Georgoudis, V.~Goncalves, E.~Panzer and R.~Pereira, \emph{{Five-loop
  massless propagator integrals}},
  \href{https://arxiv.org/abs/1802.00803}{{\tt arXiv:1802.00803}}.

\bibitem{Georgoudis:2021onj}
A.~Georgoudis, V.~Gon\c{c}alves, E.~Panzer, R.~Pereira, A.V.~Smirnov and
  V.A.~Smirnov, \emph{{Glue-and-cut at five loops}},
  \href{http://dx.doi.org/10.1007/JHEP09(2021)098}{\emph{JHEP} {\bf 09} (2021)
  098} [\href{https://arxiv.org/abs/2104.08272}{{\tt arXiv:2104.08272}}].

\bibitem{Ruijl:2017cxj}
B.~Ruijl, T.~Ueda and J.A.M.~Vermaseren, \emph{{Forcer, a FORM program for the
  parametric reduction of four-loop massless propagator diagrams}},
  \href{https://arxiv.org/abs/1704.06650}{{\tt arXiv:1704.06650}}.

\bibitem{Freedman:1974gs}
D.Z.~Freedman, I.J.~Muzinich and E.J.~Weinberg, \emph{{On the Energy-Momentum
  Tensor in Gauge Field Theories}},
  \href{http://dx.doi.org/10.1016/0003-4916(74)90448-5}{\emph{Annals Phys.}
  {\bf 87} (1974) 95}.

\bibitem{Freedman:1974ze}
D.Z.~Freedman and E.J.~Weinberg, \emph{{The Energy-Momentum Tensor in Scalar
  and Gauge Field Theories}},
  \href{http://dx.doi.org/10.1016/0003-4916(74)90040-2}{\emph{Annals Phys.}
  {\bf 87} (1974) 354}.

\end{thebibliography}\endgroup
 
\end{document}